\documentclass[11pt,a4paper,english,english]{article}
\usepackage[T1]{fontenc}
\usepackage[utf8]{inputenc}
\usepackage{fancyhdr}
\usepackage{color}
\usepackage{babel}
\usepackage{array}
\usepackage{float}
\usepackage{mathrsfs}
\usepackage{bm}
\usepackage{amsmath}
\usepackage{amssymb}
\usepackage{stmaryrd}
\usepackage{graphicx}
\usepackage[bookmarks=false,
 breaklinks=false,pdfborder={0 0 1},backref=section,colorlinks=true]
 {hyperref}
\hypersetup{
 pdftex,pdfstartview=FitV,linkcolor=linkcolor,citecolor=linkcolor,urlcolor=linkcolor,hyperindex=true,hyperfigures=false,backref=false}

\makeatletter

\pdfpageheight\paperheight
\pdfpagewidth\paperwidth

\providecommand{\tabularnewline}{\\}

\usepackage{graphicx}
\usepackage{import}
\usepackage{amsmath}
\usepackage{amssymb}
\usepackage{wrapfig}
\usepackage{subcaption}
\usepackage[nottoc,numbib]{tocbibind}
\usepackage{color} 
\usepackage{setspace}
\usepackage{ragged2e}
\usepackage{natbib}
\usepackage{hyperref}
\usepackage[table]{xcolor}
\usepackage{adjustbox}

\usepackage{tabularx}

\usepackage{psfrag}
\usepackage{sistyle}

 \usepackage{pdfpages}

\usepackage{eurosym}
\def\€{\euro{}}

\definecolor{linkcolor}{rgb}{0,0,0.6}		

\usepackage{fancyhdr}
\makeatletter
\renewcommand{\sectionmark}[1]{%
  \markright{%
    \ifnum\value{section}=0 \else\thesection.~\fi
    #1%
  }%
}
\renewcommand{\subsectionmark}[1]{%
  \markright{%
    \ifnum\value{subsection}=0 \else\thesubsection.~\fi
    #1%
  }%
}
\makeatother

\fancyheadoffset[LE,RO]{0pt} 

\newlength{\myheadwidth}
\makeatletter
\@ifpackageloaded{babel}{%
}{%
}
\makeatother

\hypersetup{backref=false}
\usepackage{etoolbox}
\AtEndPreamble{\hypersetup{backref=false,pagebackref=false}}
\usepackage{etoolbox}
\AtBeginDocument{%
  \hypersetup{backref=false,pagebackref=false}%
  \renewcommand*{\backref}[1]{}%
  \renewcommand*{\backrefalt}[4]{}%
}

\makeatother

\begin{document}
\title{Dynamical large deviations and long-range correlations for local weak
wave turbulence}

\author{
  Brice Douet \thanks{ENS de Lyon, Laboratoire de Physique, Lyon, France \newline brice.douet@ens-lyon.fr} , 
  Freddy Bouchet \thanks{Laboratoire de Météorologie Dynamique, École Normale Supérieure, Paris, France \newline freddy.bouchet@cnrs.fr}
}

\maketitle
\begin{abstract}
Wave turbulence describes the statistical dynamics of dispersive waves
with weakly nonlinear interactions. While the classical kinetic equation
captures the mean evolution of the spectrum, the study of its fluctuations
due to finite-size effects and intermittency requires a probabilistic
framework for space--time trajectories of the spectrum dynamics.
Following the previous large deviation theories for wave turbulence
\cite{guioth_path_2022,Onuki_2023,Guioth_2024} we develop a simplification
meant for simpler qualitative and numerical predictions of measurable
quantities. In this work we derive a new, simplified large deviation
principle for the spectrum in the case of isotropic and local wave
interaction kernel. It fully characterizes typical and rare fluctuations
of the spectrum. In a joint article \cite{article2_brice_freddy},
we obtain a theory which is a generalised form of Macroscopic Fluctuation
Theory, but with two conservation laws (mass and energy). In this
paper, we use it to explicitly compute the equation for Gaussian fluctuations
around Kolmogorov--Zakharov spectra. In addition to the usual equilibrium
contribution, our results shows long-range correlations in the inertial
range, which can be decomposed into three independent contributions:
one is driven by the flux in the bulk and another is driven by the
forcing and its possible fluctuations. In addition, these contributions
are computed for the first time with a method adapted to boundary
conditions where only the fluxes are fixed. The results provide a
general, replicable method for analyzing wave turbulence in more complex
settings. Finally, the generalization of this theory to the inhomogeneous
wave turbulence provides a possible explanation to the instability
of the Kolmogorov--Zakharov spectra in some 1D inhomogeneous models
with 4-wave interactions such as the Majda--McLaughlin--Tabak (MMT).
By introducing a small random source of spatially-inhomogeneous perturbations,
the large deviations may be the leading source of translation symmetry
breaking at the origin of this instability. This work opens the discussion
regarding universal and non universal properties in two-point correlation
functions. This opens new range of study on the phenomena of intermittency
which is partially developed in this paper.\textbf{ }
\end{abstract}

\newpage{}

\setcounter{page}{1}

\setlength{\parindent}{16pt}

\setcounter{secnumdepth}{-1}   

\section*{Introduction}

\setcounter{secnumdepth}{3} 

\textbf{}

The aim of this paper is to simplify and use the existing theory \cite{guioth_path_2022,Onuki_2023,Guioth_2024}
to describe probabilistically the mesoscopic evolutions of wave fields.
We derive a new large deviation principle that describes completely
the typical and rare fluctuations of the wave spectral density when
the wave interaction kernel is isotropic and local. This simplification
enables to embed it in a theoretical framework extending the Macroscopic
Fluctuation Theory (MFT) for two conservation laws and presented in
another joint article \cite{article2_brice_freddy}. It allows simple
explicit computation for measurable quantities, notably the long-range
correlations that emerge when bulk currents are present. We fully
compute it with a numerical approach. This work relies on different
active fields in theoretical physics: recent mathematical and theoretical
advances in the kinetic theory of wave turbulence, large deviation
theory for kinetic theories and the description of long-range correlations
in nonequilibrium systems.

Wave turbulence describes the statistical dynamics of weakly nonlinear
dispersive waves in systems ranging from surface gravity and capillary
waves \cite{Cavaleri2007WaveModelling,Ardhuin2010Semiempirical,roland2014developments,popinet2012adaptive,falcon_observation_2007,falcon2022experiments}
to internal waves which develop in stratified fluids \cite{lvov2001hamiltonian,sutherland2010internal,davis2020succession},
inertial waves in rotating fluids \cite{caillol2000kinetic,galtier2003weak,yarom2014experimental,yarom2017experimental},
nonlinear optics, Bose--Einstein condensates, bending waves \cite{miquel2013transition,during2006weak,lbz3-nzxn,Muralidhar2024Extreme},
and plasma oscillations \cite{Galtier_2020,ryzhik1996transport}.
When the wave amplitudes are small and the system large, two key simplifications
occur. First, the phases of individual modes rapidly decorrelate,
so that the hierarchy of correlation functions can be asymptotically
closed: higher-order correlations become negligible, and the slow
evolution of the wave spectrum is governed by a wave-kinetic equation
for the average spectral density $\bar{n}\left(\bm{k},t\right)$.
This closure procedure, built on the assumptions of weak coupling
and random phases, explains how nonlinear mode interactions lead to
irreversible transfer of conserved quantities such as energy and wave
action across scales, producing stationary Kolmogorov--Zakharov spectra
\cite{zakharov_kolmogorov_1992,galtier2024wave,Galtier_2020,newell2011wave,nazarenko2011wave,hasselmann1962non}.

The deterministic kinetic equation captures the mean evolution of
the system, but real wave fields display strong fluctuations, finite-size
corrections, and intermittency, which manifest as deviations from
the ideal cascade predictions. These fluctuations originate from finite
numbers of resonant triads, imperfect scale separation, unknown initial
conditions and sporadic bursts of nonlinear terms \cite{zakharov_kolmogorov_1992,zakharov2004one,Onuki_2023,Guioth_2024,FGV,nazarenko_wave_2011}.
To describe such deviations one needs a statistical framework for
the probability of space--time trajectories of the wave spectrum.
Large-deviation theory provides this: it assigns an action functional
$\mathcal{A}\left[n\right]$ to each macroscopic path, whose minimum
corresponds to the kinetic evolution, while the large deviation estimate
$P\left[\left\{ n_{\varepsilon}\left(t\right)=n\left(t\right)\right\} _{0\leq t\leq T}\right]\underset{\varepsilon\to0}{\asymp}\exp\left[-\tfrac{\mathcal{A}\left[n\right]}{\varepsilon}\right]$
quantifies probability of typical as well as rare events. This approach
has recently been developed for homogeneous weak wave turbulence \cite{guioth_path_2022}
and extended to inhomogeneous media and random scattering \cite{Onuki_2023,Guioth_2024}.
When spatial inhomogeneity or randomness is included, the relevant
quantity becomes the local spectral density $n\left(\bm{x},\bm{k},t\right)$
defined through a Wigner transform. Simplified isotropic local models,
in which the spectrum depends mainly on the wavenumber magnitude,
allow explicit calculation of steady states and correlation functions
\cite{zakharov_kolmogorov_1992,FGV,benney1967propagation,newell_wave_2001,newell2011wave}.
Recent analytical and numerical studies have used these settings to
test kinetic predictions, deviations and possibly extreme fluctuations
in numerical \cite{hrabski2024verification,dku2022domain,zhu2022testing,banks2022direct}
or laboratory experiments \cite{falcon2022experiments,michel2022statistics,falcon2020saturation,lbz3-nzxn,Muralidhar2024Extreme}. 

Large deviation theory for wave dynamics has already been used to
study rare events. It has been done theoretically for the evolution
of the empirical spectrum on the kinetic time scale \cite{guioth_path_2022,Onuki_2023,Guioth_2024}
and mathematically for wave amplitude due to short time phase mixing
in the non-linear Schrödinger dynamics \cite{garrido2023large}, but
also numerically for studying the appearance of rogue waves \cite{dematteis2018rogue,dematteis2019experimental,michel2022statistics}.
Instanton structures have been predicted and compared with experimental
data measured in a very large wave basin \cite{dematteis2019experimental}.
Another aim is to reproduce the statistics of the dynamics by numerical
models running simple effective stochastic equations instead of deterministic
ones. The applications are very broad, but for wave turbulence one
could for example hint some of them in ocean weather forecast of the
significant wave height, the flux fluctuations being a limitation
to classical predictions \cite[part 7.2]{Cavaleri2007WaveModelling}
(see also \cite{Cavaleri2007WaveModelling,Ardhuin2010Semiempirical,roland2014developments,popinet2012adaptive}
for classical literature). However the range of validity of wave turbulence
being exceeded when the waves are large, one should add appropriate
ingredients accounting for wave breaking and nonlinear dissipation,
which we will not discuss here. Finally, a very relevant application
of our theory is to predict the transition time to a new state when
the Kolmogorov--Zakharov cascade solution is unstable. As pointed
out in the conclusion of \cite{guioth_path_2022}, the stochastic
noise derived within the large deviation framework is the leading
source of symmetry breaking perturbations, which could be at the origin
of multistability phenomena for the 4-wave Majda--McLaughlin--Tabak
(MMT) model \cite{majda1997one,cai2001dispersive,chibbaro2017weak,newell2012spontaneous}.
This model is unstable to spatially-inhomogeneous perturbations and
a 4-wave dynamical large theory in an inhomogneous framework could
unveil the transition rates and transition paths between possible
different attractors.

In parallel, Macroscopic Fluctuation Theory (MFT) has used large-deviation
ideas for driven diffusive and interacting-particle systems, yielding
a general formulation for the probability of macroscopic fields \cite{bertini2015macroscopic,bodineau2008long,illien2024deankawasakiequationstochasticdensity,bertini2009towards,DLS2}.
In dynamical large deviation theory and in MFT, the quasipotential
(stationary rate functional) satisfies a Hamilton--Jacobi equation
whose derivatives generate the non-equilibrium fluxes and cumulants
of the distribution of the stationary $n$. Determining this quasipotential
or the cumulants is of major importance when one is interested in
the fluctuations around a nonequilibrium steady state. More generally
in kinetic theories, the macroscopic state is described from the motion
of elementary constituents as a law of large numbers. Beyond the average,
one can be interested in their statistical fluctuations starting from
first principles. This work has recently made important progresses
in other contexts than wave turbulence, \textit{e.g. }for stochastic
dynamics with mean field interaction \cite{BOUCHET:2016:D}, for discrete
models that mimic dilute gases and with Boltzmann like behavior \cite{rezakhanlou1998large,leonard1995large},
for dilute gases related to the Boltzmann equation \cite{bouchet2020boltzmann,bodineau2022long},
for the Kac model \cite{heydecker2023large,basile2024asymptotic},
for plasma below the Debye length, related to the Landau equation
\cite{feliachi2021dynamical}, for homogeneous systems with long range
interactions related to the Balescu--Guernsey--Lenard equation \cite{feliachi2022dynamical}.
The large deviation principles describe fluctuations but also reveal
gradient structure for the dynamics \cite{mielke2014relation,bouchet2020boltzmann}.

Finally, large deviation approach enables to compute or estimate fluctuations
around the deterministic limit (interpreted as the most probable state
by a law of large number). These fluctuations are ideally quantified
by the computation of the large deviation functional, but this is
reserved to a handful of simple models. More modestly, close to a
given state, one can look for the second order cumulant (or two-point
correlation function), which is the second derivative of the large
deviation functional. It can give a first hindsight on the dynamical
properties of the system. In particular, knowing if two points macroscopically
distant are strongly correlated or not indicates the large or small
scale propagation of fluctuations, which is an important feature to
describe physical systems. This leads to the study of long-range correlations,
which is a very active topic (see \cite{article2_brice_freddy}) 
In the context of wave turbulence, one can find the expression of
spatial structure functions in \cite[part 5.6]{nazarenko2011wave}
assuming RPA (Random Phase and Amplitude), and experimentally measured
in \cite{nazarenko2010statistics} and in \cite{lvov_noisy_2004}.
They correspond to one-point moment (\textit{i.e.} the diagonal part)
of the Fourier wave modes in momentum space, assuming RPA. These computations
do not measure the possible long-range contribution, which is also
difficult to measure experimentally, although some experimental setups
have tried a few different configurations \cite{campagne2019energy,deike2013etudes,aubourg20173}.

In this work, we aim to go beyond this RPA approximation and analyse
the multi-point correlations. This will be done in the light of the
fluctuations identified by large deviation theory established for
wave turbulence, with Random Phase (RP) hypothesis instead of RPA
\cite{guioth_path_2022,Onuki_2023,Guioth_2024}. Very remarkably,
this RP hypothesis was recently justified, on the mathematical side,
as \cite{deng2311long} proved that correlations played a marginal
role in the dynamics, thus justifying the wave kinetic equation for
arbitrarily long times. The parallel result was also recently proven
for Boltzmann's equation \cite{deng2024long}. When computing these
quantities (wave spectrum, correlations...), a recurrent problem is
the large inertial ranges (bulk) between forcing and dissipation scales,
which are generically considered in wave turbulence. To tackle this
problem, numerical resolutions should be restrained to a part of the
inertial range, because resolving up to the forcing and dissipation
scales is not numerically manageable. In this context, the boundary
conditions cannot simply be treated as if there were particle reservoirs
or thermostats (which lead to fixed-density type boundary conditions),
because the fluctuations of the current at the boundary should mimic
bulk current fluctuations. The only imposed quantities are the fluxes,
and we used to lack a theory for the correlation function with flux-type
boundary conditions. We will treat this problem in the light of a
large deviation framework adapted to two conservation laws (mass and
energy), explained in details in \cite{article2_brice_freddy}. In
this framework, we also explained how to address boundary conditions
which are fixed-flux type instead of fixed-density type. This type
of boundary condition do not specify the value of the density at the
boundary, thus leaving a degree of freedom for the homogeneous solution.
Their solution corresponds to the contribution to the correlations
which are fundamentally created by the currents. Another contribution
is expected if we specify the value of the boundary density, corresponding
to a particular choice of the homogeneous solution. 

The main results of this article are the following. We write a new
simplified large deviation principle compared to \cite{guioth_path_2022,Guioth_2024,Onuki_2023}
which is valid when the wave interaction kernel is isotropic and local.
We obtain a new isotropic local large deviation Hamiltonian. As for
the classical isotropic local theory of the kinetic equation \cite{connaughton_non-stationary_2003},
this leads to simpler equations (PDEs) verified by the key quantities
(average spectrum and the cumulants of its distribution) for the large
deviations. We then consider the task of concretely computing these
quantities and we give a detailed and validated method to compute
the long-range correlations accounting for the Gaussian fluctuations
around a Kolmogorov--Zakharov spectrum. Since the forcing is not
resolved, we need to adopt suitable flux-type boundary conditions
and we use the theoretical framework and the methods developed in
\cite{article2_brice_freddy} for dynamics with two conserved quantities
(wave action and energy). We compute numerically the wave spectrum
and the long-range correlations in inertial ranges, our method being
general and replicable in more complex situations. In this numerical
part, our scientific contribution lies in the numerical method, its
adaptation to flux-type boundary conditions and eventually its application
to the determination of the long-range correlations for wave turbulence.
Finally, extending this theory to inhomogeneous wave turbulence \cite{Guioth_2024}
offers a potential explanation for the instability of Kolmogorov--Zakharov
spectra in certain one-dimensional inhomogeneous models with four-wave
interactions, such as the Majda--McLaughlin--Tabak (MMT) model.
By introducing a small, spatially inhomogeneous random perturbation,
large deviations may become the primary mechanism for breaking translational
symmetry, thereby driving this instability. We observe this instability
through the exponential divergence in time of the correlation function,
which is an indicator of an unstable state.

\textbf{}

This article is organised as follows. First, in Section \ref{sec:Dynamical-theory-WT},
we recall the general framework of large deviations for wave turbulence
introduced in \cite{guioth_path_2022,Onuki_2023,Guioth_2024} and
simplify it when the wave interaction kernel is isotropic and local.
We recall give a general framework adapted to diffusive systems with
conserved quantities in Section \ref{sec:General-framework-for-LD},
as detailed in \cite{article2_brice_freddy}. We explain the sources
of long-range correlations and give the Lyapunov equation they verify.
Eventually, in Section \ref{sec:Application-to-WT}, we use a numerical
scheme established and tested in \cite{article2_brice_freddy} to
compute these correlations for isotropic local wave turbulence. 

\section{Dynamical theory of large deviations for wave turbulence with isotropic
local interactions \label{sec:Dynamical-theory-WT}}

\subsection{Large deviation theory for weak wave turbulence}

In this section, we summarize the main known general results in large
deviations for weak wave turbulence that have emerged recently and
we explain the simplified framework which we will adopt in this article.
The starting point of this work will be the results of \cite{guioth_path_2022}
which derived the first general path large deviation theory for weak
wave turbulence in the kinetic limit. This work was followed by others
\cite{Onuki_2023,Guioth_2024} which generalised this result for spatially
inhomogeneous wave action spectra, respectively due to linear wave
scattering by a random medium and to nonlinear wave-wave scattering.
The first paper \cite{guioth_path_2022} lies in the kinetic regime
of wave turbulence, which assumes weak nonlinearity and large system
size. Then, assuming Random Phase (RP) approximation, the empirical
spectral density is the natural precursor to the spectrum for finite
system sizes and verifies a Large Deviation Principle (LDP) with a
large deviation Hamiltonian which is computed. We give in this section
a concise summary of this paper to set the starting point of our study.
For general mathematical and physical literature about large deviations,
stochastic methods and a few applications, one can for instance refer
to \cite{cramer1938nouveau,Don,Freidlin_Wentzel_1984_book,Gardiner_1994_Book_Stochastic,touchette2009large,bertini2015macroscopic,bouchet2020boltzmann}.

\subsubsection{Wave turbulence notations and hypotheses\label{subsec:Wave-turbulence-notations-hyp}}

Let us now define the wave turbulence notations and hypotheses. First,
we justify the form of the dynamic's Hamiltonian that we will assume,
then we briefly explain the kinetic regime of wave turbulence. 

We consider a space $\mathbb{V}_{L}^{d}=\left[0,L\right]^{d}$ of
size $L$ and dimension $d$. On this space, we consider a nonlinear
wave equation associated with a scalar field $\psi\left(\bm{x},t\right),\left(\;\bm{x}\in\mathbb{V}_{L}^{d},\,t\geqslant0\right)$.
The paradigmatic example is the Non Linear Schrödinger (NLS) equation
\[
i\partial_{t}\psi=-\frac{\mu^{2}}{2}\Delta\psi+\epsilon\left|\psi\right|^{2}\psi.
\]
For simplicity, we will restrict to systems with generic 4-wave interactions
at leading order. If we look at the Fourier amplitude of the field
$A_{\bm{k}}\left(t\right)=\frac{1}{L^{d}}\int_{\mathbb{V}_{L}^{d}}\psi\left(\bm{x},t\right)e^{-i\bm{k}.\bm{x}}\text{d}^{d}\bm{x},\left(\;\bm{k}\in\left(\frac{2\pi}{L}\mathbb{Z}\right)^{d}\right)$,
it follows a Hamitonian dynamics $i\frac{\partial A_{\bm{k}_{1}}}{\partial t}=\frac{\partial\mathcal{H}}{\partial A_{\bm{k}_{1}}^{*}}$
with Hamiltonian 
\begin{equation}
\mathcal{H}={\displaystyle \sum_{\bm{k}}\omega_{\bm{k}}A_{\bm{k}}A_{\bm{k}}^{*}+\sum_{\bm{k}_{1}+\bm{k}_{2}=\bm{k}_{3}+\bm{k}_{4}}W_{\bm{k}_{1}\bm{k}_{2}\bm{k}_{3}\bm{k}_{4}}A_{\bm{k}_{1}}A_{\bm{k}_{2}}A_{\bm{k}_{3}}^{*}A_{\bm{k}_{4}}^{*}}+\ldots\label{eq:dynamicsHamiltonian-4Wave}
\end{equation}
where we will use the notation $\sum_{\bm{k}}\equiv\sum_{\bm{k}\in\left(\frac{2\pi}{L}\mathbb{Z}\right)^{d}}$,
unless stated otherwise. $W_{\bm{k}_{1}\bm{k}_{2}\bm{k}_{3}\bm{k}_{4}}$
is the interaction kernel (\textit{e.g.} for NLS, $W$ is a constant)
and $\omega_{\bm{k}}$ is the dispersion law of waves. Due to the
symmetries in the sum, $W_{\bm{k}_{1}\bm{k}_{2}\bm{k}_{3}\bm{k}_{4}}$
can be considered symmetric under permutation of the wave vectors:
for any permutation $\pi$, $W_{\bm{k}_{\pi\left(1\right)}\bm{k}_{\pi\left(2\right)}\bm{k}_{\pi\left(3\right)}\bm{k}_{\pi\left(4\right)}}=W_{\bm{k}_{1}\bm{k}_{2}\bm{k}_{3}\bm{k}_{4}}$
. It has been shown in \cite{zakharov_kolmogorov_1992} that the
Hamiltonian can always be put in this form up to a non-linear canonical
change of variables which enable to only keep resonant terms. Additional
information on the generality of this form is provided in Appendix
\ref{subsec:Appendix-4Wave}, but this discussion is not central for
the present work (although it obviously affects its generality). The
important information to keep from this appendix is that the computations
presented in parts (\ref{subsec:LD-isotropic-WT}, \ref{subsec:LD-isotropic-local-WT})
are valid if we suppose the dispersion relation $\omega_{\bm{k}}$
to be of the \emph{nondecay} type, that is $\omega_{\bm{k}}$ is a
concave function of $k=\left\Vert \bm{k}\right\Vert $. If this hypothesis
is not valid, one should take the isotropic local form of the large
deviation Hamiltonian (\ref{eq:LD-Hamiltonian-local}) for an effective
approximation which can be relevant, but for which our justification
does not apply.

\paragraph{Range of hypotheses in the kinetic regime of weak wave turbulence}

Let us explain briefly the hypotheses we will make to work in the
kinetic regime in wave turbulence. One can also refer to \cite{guioth_path_2022}
for more details.

In order to work at finite energy density (\ref{eq:dynamicsHamiltonian-4Wave})
in the large $L$ limit, we will rescale the wave amplitudes $A_{\bm{k}}$
and define $a_{\bm{k}}=\left(\frac{L}{2\pi}\right)^{d/2}\epsilon^{-1/2}A_{\bm{k}}$
 where $\epsilon$ is an extra scaling parameter ($\epsilon\ll1$),
\textit{a priori} independent from $L$, which will account for weak
nonlinearity. The dynamics of $a_{\bm{k}}$ reads
\begin{equation}
i\frac{\partial a_{\bm{k}_{1}}}{\partial t}=\frac{\partial\mathcal{\tilde{H}}}{\partial a_{\bm{k}_{1}}^{*}}\label{eq:dyn-ak}
\end{equation}
where 
\begin{equation}
\tilde{\mathcal{H}}=\left(\frac{2\pi}{L}\right)^{d}{\displaystyle \sum_{\bm{k}}\omega_{\bm{k}}a_{\bm{k}}a_{\bm{k}}^{*}+\epsilon\left(\frac{2\pi}{L}\right)^{2d}\sum_{\bm{k}_{1}+\bm{k}_{2}=\bm{k}_{3}+\bm{k}_{4}}W_{\bm{k}_{1}\bm{k}_{2}\bm{k}_{3}\bm{k}_{4}}a_{\bm{k}_{1}}a_{\bm{k}_{2}}a_{\bm{k}_{3}}^{*}a_{\bm{k}_{4}}^{*}}+\ldots\label{eq:dynamics-Hamiltonian-rescaled}
\end{equation}
 is the rescaled energy density. In this expression (\ref{eq:dynamics-Hamiltonian-rescaled}),
the first term corresponds to the linear evolution whereas the second
describes the nonlinear evolution of the amplitudes, which is controlled
by the parameter $\epsilon$. We define the \textit{empirical spectral
density }
\[
\hat{n}(\bm{\xi},t)=\left(\frac{2\pi}{L}\right)^{d}\sum_{\bm{k}}a_{\bm{k}}a_{\bm{k}}^{*}\delta(\bm{k}-\bm{\xi})
\]
where $\delta$ is a Dirac distribution in the $d$-dimensional space
of wavenumbers.

The aim of the previous works was to obtain the probability for path
dynamics of the spectral density $\hat{n}$ in the limit of weak nonlinearity
(small wave amplitude) and large system size. These limits enable
perturbative expansions in the dynamics of $a_{\bm{k}}$ and thus
of $\hat{n}$. This is equivalent to look at the system at a certain
\textit{time scale} defining the kinetic regime, detailed in \cite[part 3.2]{guioth_path_2022}.
The characteristic time scale $\Delta t$ at which the system evolves
is assumed to verify the three following properties:
\begin{enumerate}
\item \textit{Random Phase (RP) hypothesis}: we assume that the phases of
the waves have a distribution which quickly converges to a uniform
distribution within a time $t_{mix}\left(\bm{k},\epsilon\right)\ll\Delta t$
(mixing property). The justification of this property is beyond the
scope of this work requires to assess how the nonlinear dispersion
relation and/or the effects of the chaotic nonlinear dynamics lead
to this convergence, within a very short time compared to the typical
time of the nonlinear evolution of the amplitudes \cite{spohn2006phonon,newell2011wave}.
This entails a Markov property which is crucial for the derivation
of the large deviation theory. There was recently a major advance
in this field, as Deng, Hani and Zaher \cite{deng2311long} managed
to justify mathematically for long times that it holds. The proof
has strong similarities with the one given for the Boltzmann equation
\cite{deng2024long} and shows that the global dynamics is not affected
by the collisions (wave scattering for wave turbulence) and can in
a good approximation be considered as if waves were phase-independent.
This RP property was not proven before, yet widely used in wave turbulence
literature, and one should emphasize that no other property such as
the Random Phase and Amplitude (RPA) is needed.
\item \textit{Kinetic regime of wave turbulence}:
\begin{enumerate}
\item $\Delta t\ll t_{nl}$ where $t_{nl}$ is the characteristic time for
the development of the nonlinearities impacting the spectral density
$\hat{n}$. One can show that this time scales as $t_{nl}\sim\frac{1}{\epsilon^{2}}$.
Thus to compare all time time scales with this nonlinear evolution,
we rescale all times by $\epsilon^{2}$, defining the kinetic time
$\tau=t\epsilon^{2}$, and $\tau_{nl}=t_{nl}\epsilon^{2}$. 
\item The nonlinear term in (\ref{eq:dynamics-Hamiltonian-rescaled}) presents
enough resonances which contribute in the dynamics and verify
\[
\begin{cases}
\bm{k}_{1}+\bm{k}_{2} & =\bm{k}_{3}+\bm{k}_{4}\\
\omega_{\bm{k}_{1}}+\omega_{\bm{k}_{2}} & =\omega_{\bm{k}_{3}}+\omega_{\bm{k}_{4}}
\end{cases}.
\]
More precisely, to contribute within a time interval $\Delta t$,
they should verify a broadened resonance condition $\left(\omega_{\bm{k}_{1}}+\omega_{\bm{k}_{2}}-\omega_{\bm{k}_{3}}-\omega_{\bm{k}_{4}}\right)\Delta t\ll1$
\cite{nazarenko2011wave,guioth_path_2022}. The wave number spacing
is $\frac{2\pi}{L}$, thus in the large $L$ limit, the frequency
differences scale as $\Delta\omega\underset{L\to\infty}{\sim}\frac{2\pi}{L}\left|\frac{\partial\omega}{\partial\bm{k}}\right|$.
This entails that the condition $\frac{2\pi}{L}\left|\frac{\partial\omega}{\partial\bm{k}}\right|\Delta t\ll1$
is sufficient to meet the broadened resonance condition.
\end{enumerate}
\end{enumerate}
Gathering all conditions on the kinetic time increment $\Delta\tau$
leads to the double condition verified by the kinetic times $\Delta\tau$:
\[
\epsilon^{2}t_{mix}\left(\bm{k},\epsilon\right)\ll\Delta\tau\ll\text{min}\left\{ \tau_{nl},\frac{L\epsilon^{2}}{2\pi}\left|\frac{\partial\omega}{\partial\bm{k}}\right|^{-1}\right\} .
\]
In order to have space for small time increments in the ``small $\epsilon$
-- large $L$'' limit, a sufficient condition is to take a joint
limit called \textit{kinetic limit}. It is defined for any function
$\phi_{L,\epsilon}$ parameterised by $L$ and $\epsilon$ as 
\[
\text{Kin Lim }\phi_{L,\epsilon}=\lim_{\begin{array}{c}
L\to\infty,\epsilon\to0\\
\exists C>0,L\epsilon^{2}>C
\end{array}}\phi_{L,\epsilon}.
\]
Now that we explained the hypotheses for the wave system, let us look
at the dynamics of $\hat{n}$ in the kinetic limit. 

\subsubsection{Large deviation Hamiltonian\label{subsec:Large-deviation-Hamiltonian-general-WT}}

We are looking for a large deviation principle on the probability
to observe a trajectory for the empirical spectrum $\left\{ \hat{n}(t)\right\} _{0\leq\tau\leq\tau_{f}}$
close to a prescribed trajectory $\left\{ n(\tau)\right\} _{0\leq\tau\leq\tau_{f}}$,
conditioned on the initial condition $\hat{n}\left(\tau=0\right)=n_{0}$.
We can write it as 
\begin{equation}
P_{n_{0}}\left[\left\{ \hat{n}(t)\right\} _{0\leq\tau\leq\tau_{f}}=\left\{ n(t)\right\} _{0\leq\tau\leq\tau_{f}}\right]\underset{L\to\infty}{\asymp}e^{-\left(\frac{L}{2\pi}\right)^{d}\int_{0}^{T}\text{d}t\text{ sup}_{\lambda}\left\{ \int\text{d}\bm{k}\lambda\dot{n}-H\left[n,\lambda\right]\right\} }\label{eq:LDP-general}
\end{equation}
where the large deviation parameter is $\left(\frac{L}{2\pi}\right)^{d}$.
Due to the Markov property, $\hat{n}$ can be considered as a continuous
time Markov process and the Gärtner-Ellis states that we can look
at the cumulant generating function to prove such a LDP. The large
deviation Hamiltonian can be computed using the formula 
\[
H\left[n,\lambda\right]=\lim_{\Delta\tau\to0}\tfrac{1}{\Delta\tau}\underset{L\to\infty}{\text{Kin Lim}}\left(\tfrac{2\pi}{L}\right)^{d}\ln\left(\mathbb{E}_{n}\left[e^{-\left(\frac{L}{2\pi}\right)^{d}\int\text{d}\bm{k}\lambda\left(\bm{k}\right)\left[\hat{n}(\bm{k},\Delta t)-n(\bm{k})\right]}\right]\right)
\]
 where $\mathbb{E}_{n}$ is conditioned on the initial value $\hat{n}(\bm{k},t=0)=n(\bm{k})$.
$\lambda$ is the conjugated momentum to $\dot{n}$ The time increment
$\hat{n}(\bm{k},\Delta t)-n(\bm{k})$ is estimated by perturbative
expansion in $\epsilon$ using (\ref{eq:dyn-ak}) and the Markov property.
The paper \cite{guioth_path_2022} proves the result
\begin{equation}
H\left[n,\lambda\right]=\int_{\mathcal{R}_{4}^{d}}|W_{1234}|^{2}\left(\lambda_{1}+\lambda_{2}-\lambda_{3}-\lambda_{4}\right)n_{1}n_{2}n_{3}n_{4}\left[\frac{1}{n_{1}}+\frac{1}{n_{2}}-\frac{1}{n_{3}}-\frac{1}{n_{4}}+\lambda_{1}+\lambda_{2}-\lambda_{3}-\lambda_{4}\right]\label{eq:LD-hamiltonian-general}
\end{equation}
where for compactness we denoted 
\[
\mathcal{R}_{4}^{d}=\left\{ \bm{k}_{1},\bm{k}_{2},\bm{k}_{3},\bm{k}_{4}\;|\;\bm{k}_{1}+\bm{k}_{2}=\bm{k}_{3}+\bm{k}_{4}\ \&\ \omega_{\bm{k}_{1}}+\omega_{\bm{k}_{2}}=\omega_{\bm{k}_{3}}+\omega_{\bm{k}_{4}}\right\} \subset\mathbb{R}^{4d}
\]
the resonant manifold, $W_{1234}=W_{\bm{k}_{1}\bm{k}_{2}\bm{k}_{3}\bm{k}_{4}}$,
$n_{i}=n\left(\bm{k}_{i}\right)$ and $\lambda_{i}=\lambda\left(\bm{k}_{i}\right)$,
$\lambda$ being the conjugated momentum to $\dot{n}$. This expression
enables to deduce some important properties which are consistent with
the classical kinetic theory but also enable to go further. For instance,
an equivalent fluctuating dynamics with small noise for the empirical
spectrum generalises the kinetic equation (which is the kinetic limit,
or law of large numbers, of this fluctuating dynamics). We find consistency
with the conserved quantities (total wave action number, energy and
momentum), and we can compute the stationary action for large times
at equilibrium, called \textit{equilibrium quasipotential}. This is
detailed in the original paper \cite{guioth_path_2022} and summarised
in Appendix \ref{subsec:Properties-dyn-a,d-eq-quasipot-general}. 

\subsubsection{Problematics for computing measurable quantities for applications}

The preceding results are very insightful. However, this formalism
is abstract and the dynamical equations are complicated to solve.
For instance the kinetic equation is an integro-differential equation
\begin{equation}
\partial_{t}n_{1}=4\int_{\mathcal{R}_{3}^{d}\left(\bm{k}_{1}\right)}|W_{1234}|^{2}n_{1}n_{2}n_{3}n_{4}\left[\frac{1}{n_{1}}+\frac{1}{n_{2}}-\frac{1}{n_{3}}-\frac{1}{n_{4}}\right].\label{eq:kin-eq-classical}
\end{equation}
The solution is known when there is a large scale separation between
forcing and dissipation (\textit{large inertial range}). In this case,
the Kolmogorov--Zakharov (KZ) solutions \cite{zakharov_kolmogorov_1992}
yield power laws with explicit exponents and prefactors (called \textit{Kolmogorov
constants}). Beyond these assumptions, the general solution requires
numerical methods, and even in stationary regimes, the large deviation
rate function (quasipotential) remains unknown out of equilibrium.
To capture the essential physics, simpler effective models are therefore
valuable. A common simplification is isotropy, often combined with
additional hypotheses to replace integro-differential equations with
partial differential equations, leading to models such as:
\begin{equation}
\partial_{t}n=\partial_{\omega}^{2}\mathcal{K}+\left(s-d\right)\label{eq:iso-local-model-intro}
\end{equation}
where the function $\mathcal{K}$ depends on $n\left(\omega\right)$
and its derivatives (and possibly $\omega$), and $s$ and $d$ are
possible source and dissipation terms (see \cite{dyachenko1992optical}
and section \ref{subsec:LD-isotropic-local-WT}). Such models often
predict the spectra with correct power-law exponents but not the prefactors
(see section \ref{subsec:Out-of-equilibrium-solutions-kinetic-eq-wide-inertial-ranges}
and Appendix \ref{subsec:Appendix:-Computation-of-kolmogorov-constant}).
They are widely used in practice, for instance in ocean wave forecasting
where coefficients are adjusted to observations, rather than derived
from first principles. The first contribution of this work is to derive
such models as (\ref{eq:iso-local-model-intro}) from first principles
within the large deviation framework, although under stronger assumptions
(isotropy and local interactions). Not only do we obtain this kinetic
equation, but also the large deviation principle. In Sections \ref{sec:General-framework-for-LD}
and \ref{sec:Application-to-WT}, we study their general properties,
which remain relevant even when these assumptions are relaxed by treating
the model as effective.

\subsection{Large deviations for isotropic wave turbulence \label{subsec:LD-isotropic-WT}}

In this section, we suppose that the state $n$ is and remains isotropic
at any time: $n(\bm{k})=n(k)$. For simplicity, we also suppose a
simple power law for the dispersion relation $\omega\left(\bm{k}\right)=\gamma k^{\alpha}$.
This dispersion relation is important to obtain explicit expressions
for prefactors. Yet, another isotropic dispersion relation could also
be considered. Under these hypotheses, we provide a new expression
of the large deviation Hamiltonian for isotropic wave turbulence. 

First, with isotropy hypothesis, we will use the variable $\omega$
instead of $k$ and try to rewrite a LDP 
\begin{equation}
P\left[\left\{ \hat{N}(\tau)\right\} _{0\leq\tau\leq\tau_{f}}=\left\{ N(\tau)\right\} _{0\leq\tau\leq\tau_{f}}\right]\underset{L\to\infty}{\asymp}e^{-\left(\frac{2\pi}{L}\right)^{d}\int_{0}^{\tau_{f}}\text{d}\tau\text{ sup}_{\Lambda}\left\{ \int\text{d}\omega\Lambda\dot{N}-H_{iso}\left[N,\Lambda\right]\right\} }\label{eq:LDP-isotropic}
\end{equation}
We introduced $N\left(\omega\right)$ the frequency distribution (spectrum
integrated over the solid angle $\Omega$), corresponding to the number
of waves in the frequency band $\left[\omega,\omega+\text{d}\omega\right]$.
It is defined by demanding that for any test function $\phi$, $\int\text{d}\omega\phi\left(\omega\right)N\left(\omega\right)\text{d}\omega=\int\phi\left(\omega\left(\bm{k}\right)\right)n\left(\bm{k}\right)\text{d}^{d}\bm{k}$,
giving 
\begin{equation}
N(\omega)=\Omega_{d}k^{d-1}\frac{dk}{d\omega}n(k(\omega)).\label{eq:n-to-N}
\end{equation}
 where $\frac{dk}{d\omega}=\gamma^{-1/\alpha}\alpha^{-1}\omega^{1/\alpha-1}$.
For the rest of this paper, we will sometimes use $N$ and sometimes
write $n$. This notation is an intentional convention to avoid heavy
notations involving $\left(\Omega_{d}k^{d-1}\frac{dk}{d\omega}\right)^{-1}N$.
When $n$ is written, one should understand it as $\left(\Omega_{d}k^{d-1}\frac{dk}{d\omega}\right)^{-1}N$.
For convenience, we will also denote 
\begin{equation}
\Omega\left(\omega\right)=\Omega_{d}k^{d-1}\frac{dk}{d\omega}=\Omega_{d}\frac{\omega^{\frac{d}{\alpha}-1}}{\gamma^{d/\alpha}\alpha}.\label{eq:def_Omega_prefactor}
\end{equation}
In the large deviation principle (\ref{eq:LDP-isotropic}), $\Lambda$
is the conjugated moment to $\dot{N}$ and $H_{iso}$ is the corresponding
large deviation Hamiltonian. Here, $\Omega_{d}$ is the surface of
the unit sphere in dimension $d$. In order to change the variable,
we simply perform a contraction principle to the LDP (see Appendix
\ref{subsec:Contraction-principle} for a definition). Since there
is an exact correspondence between $N$ and $n$, both LDP (\ref{eq:LDP-general})
and (\ref{eq:LDP-isotropic}) should coincide, meaning
\[
\begin{cases}
H_{iso}\left[N,\Lambda\right] & =H\left[n,\lambda\right]\\
\int\text{d}^{d}\bm{k}\lambda\dot{n} & =\int\text{d}\omega\Lambda\dot{N}
\end{cases}\iff\begin{cases}
H_{iso}\left[N,\Lambda\right] & =H\left[n,\lambda\right]\\
\int\text{d}k\int\text{d}\Omega k^{d-1}\lambda\left(\bm{k}\right)\dot{n}\left(\bm{k}\right) & =\int\Lambda\left(\omega\right)\text{d}\omega\frac{dk}{d\omega}\int\text{d}\Omega k^{d-1}\dot{n}\left(k\right).
\end{cases}
\]
Thus the transformation is 
\[
\begin{cases}
H_{iso}\left[N,\Lambda\right] & =H\left[n,\lambda\right]\\
N(\omega) & =\Omega\left(\omega\right)n(k(\omega))\\
\Lambda(\omega) & =\lambda(k(\omega))
\end{cases}
\]
We can simply check the consistency of this transformation by computing
and identifying the functional variation: $\delta H=\delta H_{iso}$
and deduce the functional derivatives $\frac{\delta H}{\delta n}$,
$\frac{\delta H}{\delta\lambda}$, $\frac{\delta H_{iso}}{\delta N}$,
$\frac{\delta H_{iso}}{\delta\Lambda}$. With them, we can check that
$N$, $\Lambda$ are indeed conjugated variables as the Hamilton's
equations they verify are:
\[
\begin{cases}
\dot{N} & =\frac{\delta H_{iso}}{\delta\Lambda}\\
\dot{\Lambda} & =-\frac{\delta H_{iso}}{\delta N}
\end{cases}.
\]
Now we can proceed to the simplification of the Hamiltonian $H_{iso}$.
Starting from the expression (\ref{eq:LD-hamiltonian-general}) with
isotropic $n\left(k\right)$ we perform an angle integration and denote
\begin{equation}
S_{\omega_{1}\omega_{2}\omega_{3}\omega_{4}}={\displaystyle 24\pi\frac{\left(\omega_{1}\omega_{2}\omega_{3}\omega_{4}\right)^{\frac{d}{\alpha}-1}}{\left(\gamma^{d/\alpha}\alpha\right)^{4}}\int\delta\left(\bm{k}_{1}+\bm{k}_{2}-\bm{k}_{3}-\bm{k}_{4}\right)|W_{\bm{k}_{1},\bm{k}_{2},\bm{k}_{3},\bm{k}_{4}}|^{2}\text{d}\Omega_{1}\text{d}\Omega_{2}\text{d}\Omega_{3}\text{d}\Omega_{4}}\label{eq:S-formula}
\end{equation}
the interaction kernel in the space of $\omega$, where we have $\frac{\omega^{\frac{d}{\alpha}-1}}{\gamma^{d/\alpha}\alpha}=k^{d-1}\frac{dk}{d\omega}$.
$S$ has natural symmetries inherited from $W$ under the transformations
$1\leftrightarrow2,3\leftrightarrow4\text{ and }\left(1,2\right)\leftrightarrow\left(3,4\right)$.
If $W$ is homogeneous of degree $\gamma_{W}$, then $S$ is homogeneous
of degree $\gamma_{S}=\frac{2\gamma_{W}+3d}{\alpha}-4$. With this
definition, we obtain 
\begin{align}
{\tiny H_{iso}\left[N,\Lambda\right]} & ={\scriptstyle \int S_{\omega_{1}\omega_{2}\omega_{3}\omega_{4}}\delta\left(\omega_{1}+\omega_{2}-\omega_{3}-\omega_{4}\right)\left(\lambda_{1}+\lambda_{2}-\lambda_{3}-\lambda_{4}\right)n_{1}n_{2}n_{3}n_{4}\left[\tfrac{1}{n_{1}}+\tfrac{1}{n_{2}}-\tfrac{1}{n_{3}}-\tfrac{1}{n_{4}}+\lambda_{1}+\lambda_{2}-\lambda_{3}-\lambda_{4}\right]}\label{eq:LD-Hamiltonian-isotropic}
\end{align}
where $n(k(\omega))=\Omega\left(\omega\right)^{-1}N(\omega)$ and
$\lambda(k(\omega))=\Lambda(\omega)$. As for the general case, one
can quickly check the properties of the dynamics which we can deduce
from this expression. This is detailed in Appendix \ref{subsec:Properties-dyn-a,d-eq-quasipot-isotropic}.
In particular, we obtain a new fluctuating dynamics with small noise
(\ref{eq:fluctuating-dyn-isotropic}), the kinetic equation as a law
of large number, the conservation of total wave action and energy
(the total momentum is null due to isotropy, so this quantity is no
longer interesting in isotropic framework, unlike the classical one
\cite{guioth_path_2022}).

\subsection{Large deviations for isotropic local wave turbulence \label{subsec:LD-isotropic-local-WT}}

In this section, we assume the previous isotropy hypothesis and we
suppose in addition that the interaction kernel $S$ has a \textit{locality}
property. We also use the previous hypothesis that the leading contribution
to the dynamics are 4-wave interactions that can be reduced to $2\to2$
interactions. A sufficient condition for this second hypothesis, as
explained in Section \ref{subsec:Wave-turbulence-notations-hyp} and
Appendix \ref{subsec:Appendix-4Wave}, is that the dispersion relation
$\omega\left(\bm{k}\right)=\gamma k^{\alpha}$ is of the nondecay
type (valid if and only if $\alpha<1$). Using a classical perturbative
expansion \cite{dyachenko1992optical}, we obtain the first main results
of this work: an isotropic local version of the large deviation Hamiltonian,
and the equivalent fluctuating dynamics. 

The local hypothesis considers $S_{\omega_{1}\omega_{2}\omega_{3}\omega_{4}}$,
defined in (\ref{eq:S-formula}), localised around $\omega_{1}=\omega_{2}=\omega_{3}=\omega_{4}$.
In other words, we suppose that the dynamics is mainly driven by interactions
between modes with close momenta. Physically, it means that the transport
of wave action number and energy is ``diffusive'', that is to say
that waves having momenta of the same order interact strongly and
this ``diffusive'' effect drives the behaviour of the second member
in the kinetic equation. This regime is called ``locality'' for
wave turbulence. One should be aware that the two hypotheses are independent:
we could do local approximation without isotropy or vice-versa. However,
for the sake of simplicity we will use both simultaneously. In the
local approximation, the integro-differential equations (\ref{eq:kin-eq-classical})
simplify to PDEs, as shown in \cite{dyachenko1992optical}. The computation
is analogous as in this paper, see part 3.5 there. The idea is to
write for $i=2,3,4$ : $\omega_{i}=\omega_{1}\left(1+p_{i}\right)$,
use the homogeneity of $S$ and Taylor expand the terms $\frac{1}{n_{1}}+\frac{1}{n_{2}}-\frac{1}{n_{3}}-\frac{1}{n_{4}}$
and $\lambda_{1}+\lambda_{2}-\lambda_{3}-\lambda_{4}$. Here we use
the hypothesis that the dynamics are 4-wave interactions that can
be reduced to $2\to2$ interactions. 

We obtain a new LDP
\[
P\left[\left\{ \hat{N}(\tau)\right\} _{0\leq\tau\leq\tau_{f}}=\left\{ N(\tau)\right\} _{0\leq\tau\leq\tau_{f}}\right]\underset{L\to\infty}{\asymp}e^{-\left(\frac{2\pi}{L}\right)^{d}\int_{0}^{\tau_{f}}\text{d}\tau\text{ sup}_{\Lambda}\left\{ \int\text{d}\omega\Lambda\dot{N}-H_{\ell}\left[N,\Lambda\right]\right\} }
\]
with isotropic local large deviation Hamiltonian:
\begin{equation}
H_{\ell}\left[N,\Lambda\right]=\int\text{d}\omega\mu\left[N\right]\left[\partial_{\omega}^{2}\lambda\partial_{\omega}^{2}\left(\frac{1}{n}+\lambda\right)\right]\label{eq:LD-Hamiltonian-local}
\end{equation}
where $\mu\left[N\right]=S_{0}n^{4}\omega^{s}$ is a mobillity, $s=6+\gamma_{S}$
is an exponent containing physical parameters and $S_{0}$ is a real
number given by the integral 
\begin{equation}
S_{0}=\frac{1}{4}\int\text{d}p_{2}\text{d}p_{3}\text{d}p_{4}S_{1,1+p_{2},1+p_{3},1+p_{4}}\delta\left(p_{2}-p_{3}-p_{4}\right)\left(p_{2}^{2}-p_{3}^{2}-p_{4}^{2}\right)^{2}\label{eq:S0}
\end{equation}
The range of validity of this formula is discussed in Appendix \ref{subsec:Appendix:-Local-approx-computation}.
As an order of magnitude, if we look at $S_{1,1+p_{2},1+p_{3},1+p_{4}}$
as a distribution in the space of $p_{i}$'s, we can consider the
approximation valid if its characteristic width is much smaller than
1. Appendix \ref{subsec:Appendix:values-W-exponents} gives a review
of some values for the interaction kernels $W$ and the main exponents
$\gamma_{S}$ and $s$. As we will see in section \ref{subsec:Equilibirum-quasipotential-local},
the term $\frac{1}{n}$ appearing in the expression (\ref{eq:LD-Hamiltonian-local})
can be interpreted as the derivative of the equilibrium quasipotential
$Q_{eq}$: $\frac{1}{n}=\frac{\delta Q_{eq}}{\delta n}$.

Now that we have a new large deviation theory in the isotropic local
framework, let us review its main properties.

\subsection{Properties of the isotropic local large deviation theory for wave
turbulence}

In this paragraph, we review the properties derived from the isotropic
local large-deviation Hamiltonian (\ref{eq:LD-Hamiltonian-local}).
We construct a fluctuating dynamics with two conservation laws, determine
the equilibrium distribution, quasipotential, and cumulants, and then
introduce a forcing that drives the system out of equilibrium. This
generates two currents, associated to the wave action and energy.
Their properties and fluctuations are analyzed, followed by the characterization
of solutions for the nonequilibrium spectrum.

\subsubsection{Properties of the isotropic local large deviation Hamiltonian\label{subsec:properties-local-H}}

The large deviation Hamiltonian (\ref{eq:LD-Hamiltonian-local}) is
quadratic in the response field $\Lambda$ which means the statistics
of the time increments $\dot{N}\text{d}t$ is gaussian. It takes the
form of a Freidlin-Wentzell large deviation Hamiltonian for a weak-noise
diffusion process
\[
H_{\ell}\left[N,\Lambda\right]=b_{\ell}\left[N\right].\Lambda+\Lambda.a_{\ell}\left[N\right]\Lambda
\]
where $b_{\ell}\left[N\right]=\partial_{\omega}^{2}\left[\mu\left[N\right]\partial_{\omega}^{2}\left(\frac{1}{n}\right)\right]\ $
gives the classical wave kinetic equation verified by the average
state $\bar{N}$. It is convenient to introduce a ``current'' function
\begin{equation}
\mathcal{K}\left[N\right]=\mu\left[N\right]\partial_{\omega}^{2}\left(\frac{1}{n}\right)\label{eq:def-K-(deterministic part)}
\end{equation}
such that the wave kinetic equation writes $\partial_{t}\bar{N}=\partial_{\omega}^{2}\mathcal{K}\left[\bar{N}\right]$.
In this equation, the double derivative is synonym of a double conservation
law of wave action and energy, as in the isotropic case. The noise
covariance $a_{\ell}\left[N\right](\cdot)=\partial_{\omega}^{2}\left[\mu\left[N\right]\left(\partial_{\omega}^{2}\cdot\right)\right]$
is a nonnegative-definite, self-adjoint operator. We can furthermore
write the equivalent fluctuating dynamics $\text{d}N=b_{\ell}\left[N\right]\text{d}t+\sqrt{2\left(\frac{2\pi}{L}\right)^{d}}\sigma_{\ell}\left[N\right]\text{d}W$
where the operator $\sigma_{\ell}\left[N\right]\ \cdot=\partial_{\omega}^{2}\left(\sqrt{\mu\left[N\right]}\ \cdot\right)$
verifies $\sigma_{\ell}\sigma_{\ell}^{T}=a_{\ell}$ and $W$ is a
Wiener process. This stochastic kinetic equation can thus be written
in the form
\begin{align}
\text{d}N & =\partial_{\omega}^{2}\left(\mathcal{K}\left[N\right]\text{d}t+\sqrt{2\left(\frac{2\pi}{L}\right)^{d}\mu\left[N\right]}\text{d}W\right)\label{eq:fluctuating-dyn-local}
\end{align}
The part $\sqrt{2\left(\frac{2\pi}{L}\right)^{d}\mu\left[N\right]}\text{d}W$
can be regarded as a fluctuating part of $\mathcal{K}$, with a small
amplitude of the order $\left(\frac{2\pi}{L}\right)^{d/2}$. We introduce
the notation $K\left[N\right]$:
\begin{equation}
K\left[N\right]\text{d}t=\mathcal{K}\left[N\right]\text{d}t+\sqrt{2\left(\frac{2\pi}{L}\right)^{d}\mu\left[N\right]}\text{d}W\label{eq:def-K-(deterministic+fluctuating part)}
\end{equation}
which contains both the deterministic part and the fluctuating part.
As we will see in Section \ref{subsec:Fluxes-conservation-laws},
this equation conserves the total wave action number and energy..
Interestingly, this equation is now a much simpler model than the
initial one (\ref{eq:fluctuating-dyn-general}). Instead of a stochastic
integro-differential equation, we have a stochastic PDE, so the second
member is much simpler to evaluate as it is local in $\omega$. Although
the general solution is not analytically solvable, we can find it
in particular cases (some specific stationary regimes) and otherwise
solve the equation numerically. 

\subsubsection{Equilibirum quasipotential and equilibirum cumulants for the isotropic
local theory\label{subsec:Equilibirum-quasipotential-local} }

In order to know the fluctuations at equilibrium\footnote{Equilibrium is defined as a state verifying detailed balance},
one can compute the quasipotential $Q_{\mathcal{N},\mathcal{E}}$,
exactly analog to \cite[Appendix B]{guioth_path_2022}. The method
is to express the equilibrium pdf in the kinetic limit using the microcanonical
distribution and is presented in Appendix \ref{subsec:Appendix:-equil-quasipot}.
The result is the following:
\begin{equation}
Q_{\mathcal{N},\mathcal{E}}\left[N\right]=\begin{cases}
\frac{\Omega_{d}}{\gamma^{d/\alpha}\alpha}\int\text{d}\omega\omega^{\frac{d}{\alpha}-1}\left(\frac{n}{n^{*}}-1-\ln\left(\frac{n}{n^{*}}\right)\right) & \text{if }\mathcal{E}=\int\text{d}\omega N\omega\text{ and }\mathcal{N}=\int\text{d}\omega N\\
+\infty & \text{otherwise}
\end{cases}\label{eq:eq-quasipot}
\end{equation}
where $\mathcal{E},\mathcal{N}$ are the two conserved quantities
(total energy and wave action), $n^{*}\left(\omega\right)=\frac{1}{A\omega+B}$
is the Rayleigh-Jeans distribution and where $A,B$ are fixed so that
$\mathcal{E}=\int\text{d}\omega N^{*}\omega\text{ and }\mathcal{N}=\int\text{d}\omega N^{*}$
are verified. This comes using the same method as \cite[Appendix B]{guioth_path_2022},
just taking care of the conserved quantities, which are for us the
energy and wave action. We then use a contraction principle to express
it in variable $\omega$ as in eq. (\ref{eq:eq-quasipot}). We can
check that this functional solves the stationary Hamilton-Jacobi equation
$H\left[n,\frac{\delta Q_{\mathcal{N},\mathcal{E}}}{\delta n}\right]=0$,
which is a necessary (but not sufficient) condition for $Q_{\mathcal{N},\mathcal{E}}$
to be the quasipotential.

\paragraph{Detailed balance at equilibrium}

At equilibrium, we expect the detailed balance property to hold \cite{bouchet2020boltzmann}.
If $Q_{\mathcal{N},\mathcal{E}}$ is the equilibrium quasipotential
and $H$ the large deviation Hamiltonian, the detailed balance property
at equilibrium reads
\[
\forall N,\Lambda\qquad H\left[N,-\Lambda\right]=H\left[N,\Lambda+\frac{\delta Q_{\mathcal{N},\mathcal{E}}}{\delta N}\right].
\]
Since we have $\frac{\delta Q_{\mathcal{N},\mathcal{E}}}{\delta N}=-\Omega_{0}\frac{\omega^{\frac{d}{\alpha}-1}}{\gamma^{d/\alpha}\alpha N}$,
one can easily show that this is true for $H_{\ell}$.

\subsubsection{Conservation laws and fluxes\label{subsec:Fluxes-conservation-laws}}

As we had conservation laws in the dynamics for the general dynamics
(\ref{eq:fluctuating-dyn-general}), we expect the isotropic local
dynamics (\ref{eq:fluctuating-dyn-local}) to inherit them. As in
the isotropic case, we expect two conservation laws, for wave action
and energy.

The isotropic local dynamics has the particular form of a conservation
equation with two conservation laws. Indeed, if we denote $E=N\omega$
the local energy, we have two conservation equations $\partial_{t}N=-\partial_{\omega}j_{N}$
and \textit{$\partial_{t}E=-\partial_{\omega}j_{E}$}, where 
\begin{equation}
\begin{cases}
j_{N}\left[N\right]\text{d}t & =-\partial_{\omega}K\text{\ensuremath{\text{d}}}t\\
 & =-\partial_{\omega}\mathcal{K}\left[N\right]\text{\ensuremath{\text{d}}}t-\sqrt{2\left(\frac{2\pi}{L}\right)^{d}}\partial_{\omega}\left[\sqrt{\mu\left[N\right]}\text{d}W\right]\\
j_{E}\left[N\right]\text{\ensuremath{\text{d}}}t & =\left(K+\omega j_{N}\right)\text{\ensuremath{\text{d}}}t=\left(K-\omega\partial_{\omega}K\right)\text{\ensuremath{\text{d}}}t\\
 & =\left(\mathcal{K}-\omega\partial_{\omega}\mathcal{K}\right)\text{\ensuremath{\text{d}}}t+\sqrt{2\left(\frac{2\pi}{L}\right)^{d}}\left[\sqrt{\mu\left[N\right]}\text{d}W-\omega\partial_{\omega}\left[\sqrt{\mu\left[N\right]}\text{d}W\right]\right]
\end{cases}\label{eq:jn-je}
\end{equation}
are two currents with a deterministic part (their average) and a fluctuating
part. We can also write $K$ as a function of the currents:
\[
K=j_{E}-j_{N}\omega
\]
We deduce that the total wave action number $\mathcal{N}=\int\text{d}\omega N$
and the total energy $\mathcal{E}=\int\text{d}\omega N\omega$ are
conserved for this dynamics as soon as there are appropriate boundary
conditions (\textit{i.e.} the total net entering fluxes are zero).
Besides, we can check that the Hamiltonian $H_{\ell}\left[N,\Lambda\right]=\int\text{d}\omega\mu\left[N\right]\left[\partial_{\omega}^{2}\lambda\partial_{\omega}^{2}\left(\frac{1}{n}+\lambda\right)\right]$
has the two symmetries
\begin{align*}
\forall N,\Lambda,\text{ and }\forall\alpha,\beta\in\mathbb{R}^{2},\qquad H_{\ell}\left[N,\Lambda+\alpha\underset{=1}{\underbrace{\frac{\delta\mathcal{N}}{\delta N}}}\right] & =H_{\ell}\left[N,\Lambda+\beta\underset{=\omega}{\underbrace{\frac{\delta\mathcal{E}}{\delta N}}}\right]=H_{\ell}\left[N,\Lambda\right]
\end{align*}
which is expected in the presence of these conserved quantities \cite{bouchet2020boltzmann}.

\paragraph{Out-of-equilibrium dynamics induced by a forcing}

As explained in paragraph \ref{par:Out-of-equilibrium-dynamics},
physical systems are often out-of-equilibrium when forcing terms are
added, to amount for sources and dissipation of $N$. This can be
done by adding terms for source $s\left[N,\omega,t\right]\geq0$ and
dissipation $d\left[N,\omega,t\right]\geq0$ to the dynamics. Importantly,
these functions must be isotropic as well in order to keep isotropy
of the state $N$. We write the fluctuating dynamics:
\[
\partial_{t}N=\partial_{\omega}^{2}K+\left(s-d\right).
\]
This has the property to break the conservation laws since we create
and dissipate $N$ with the terms $s$ and $d$. Consider a separation
between forcing scales $\omega_{f}$ at which $s\neq0$, and dissipation
scales $\omega_{d}^{+},\omega_{d}^{-}$ at which $d\neq0$, such that
$\omega_{d}^{+}>\omega_{f}>\omega_{d}^{-}$. Then we can compute explicitly
the (constant) values of currents $j_{N}^{(1)},j_{E}^{(1)}$ in the
first interval $\left[\omega_{d}^{+},\omega_{f}\right]$ and $j_{N}^{(2)}j_{E}^{(2)}$
in the second interval $\left[\omega_{f},\omega_{d}^{-}\right]$.
We call these intervals \textit{inertial ranges}. The currrents are
a function of the forcing amplitude, denoted $S_{f}$, and the forcing
and dissipation scales $\omega_{d}^{+},\omega_{f},\omega_{d}^{-}$.
We can easily show that when the dissipation is localized close to
the scales $\omega_{d}^{\pm}$, we have 
\begin{equation}
\begin{cases}
j_{N}^{(1)} & =-S_{f}\left(\frac{\omega_{d}^{+}-\omega_{f}}{\omega_{d}^{+}-\omega_{d}^{-}}\right)\\
j_{N}^{(2)} & =S_{f}\left(\frac{\omega_{f}-\omega_{d}^{-}}{\omega_{d}^{+}-\omega_{d}^{-}}\right)\\
j_{E}^{(1)} & =-S_{f}\omega_{d}^{-}\left(\frac{\omega_{d}^{+}-\omega_{f}}{\omega_{d}^{+}-\omega_{d}^{-}}\right)\\
j_{E}^{(2)} & =S_{f}\omega_{d}^{+}\left(\frac{\omega_{f}-\omega_{d}^{-}}{\omega_{d}^{+}-\omega_{d}^{-}}\right)
\end{cases}\label{eq:je-jn-with-forcingAmpllitude-and-frequencies-1}
\end{equation}
Further explanation is provided in Appendix \ref{subsec:Out-of-equilibrium-dynamics-induced-by-forcing}.
In addition, the classical Fjørtoft argument (see \textit{e.g.} \cite{nazarenko_wave_2011})
is recovered when taking the limits $\omega_{d}^{+}\gg\omega_{f}\gg\omega_{d}^{-}$
in eq. (\ref{eq:je-jn-with-forcingAmpllitude-and-frequencies-1}).
It states that in the first inertial range ($\omega_{d}^{-}\ll\omega\ll\omega_{f}$)
there is an inverse wave-action cascade with constant negative current
$J_{N}<0$ and no energy current, while in the second range ($\omega_{f}^ {}\ll\omega\ll\omega_{d}^{+}$)
there is a direct energy cascade $J_{E}>0$ with constant positive
current and no wave-action current.

\subsubsection{Out-of-equilibrium solutions to the kinetic equation with wide inertial
ranges\label{subsec:Out-of-equilibrium-solutions-kinetic-eq-wide-inertial-ranges}}

In this section, we present the well-known kinetic equation solutions
in the presence of large inertial ranges. In both the general nonlocal
case (Kolmogorov--Zakharov solutions, \cite{zakharov_kolmogorov_1992})
and under the local hypothesis, the solutions are power laws with
the same exponents. Here, they are treated as the mean values of the
previously derived fluctuating dynamics.

In the nonlocal case, as shown in \cite{zakharov_collapse_nodate},
we can find zeros of the collision integral 
\begin{equation}
b_{iso}\left[N\right]\left(\omega_{1}\right)=4{\displaystyle \int\text{d}\omega_{1}\text{d}\omega_{2}\text{d}\omega_{3}\text{d}\omega_{4}}S_{\omega_{1}\omega_{2}\omega_{3}\omega_{4}}\delta\left(\omega_{1}+\omega_{2}-\omega_{3}-\omega_{4}\right)n_{1}n_{2}n_{3}n_{4}\left[\frac{1}{n_{1}}+\frac{1}{n_{2}}-\frac{1}{n_{3}}-\frac{1}{n_{4}}\right]\label{eq:b_iso}
\end{equation}
 by a judicious splitting of the integration region and changes of
variables (Zakharov transformations \cite{zakharov_kolmogorov_1992,zakharov1967weak}).
We are thus able to find power law solutions in the form $n\sim\omega^{-x}$
where $x\in\left\{ 0,1,\frac{s-2}{3},\frac{s-3}{3}\right\} $. The
two first ones correspond to thermodynamical equilibrium and the two
last ones are Kolmogorov--Zakharov (KZ) solutions \cite{zakharov_kolmogorov_1992}.
In the local theory, we can look at zeros of $b_{\ell}\left[N\right]=S_{0}\partial_{\omega}^{2}\left[n^{4}\omega^{s}\partial_{\omega}^{2}\left(\frac{1}{n}\right)\right]$
by integrating twice the equation $b_{\ell}=0$. We find
\begin{equation}
\left(J_{E}-J_{N}\omega\right)\omega^{-s}n^{-4}=\partial_{\omega}^{2}\left(\frac{1}{n}\right)\label{eq:stationary-kinetic-eq-twofluxes}
\end{equation}
where the two integration constants $J_{E},J_{N}$ are clearly interpreted
as energy and wave action currents. The equation (\ref{eq:stationary-kinetic-eq-twofluxes})
is nonlinear but admits a unique two-dimensional solution by the Cauchy--Lipschitz
theorem (the function is locally Lipschitz). 

While the general case, with two nonzero currents, has no known analytical
solution, it becomes tractable when one current vanishes, as in the
large scale-separation regime. If both currents vanish, the solution
reduces to the Rayleigh--Jeans distribution. For power-law solutions,
one recovers the KZ forms $n\sim\omega^{-x}$ where $x\in\left\{ \frac{s-2}{3},\frac{s-3}{3}\right\} $,
with the corresponding exponents and fixed Kolmogorov constants. For
$J_{E}=0$, we have 
\begin{align}
n_{N} & =A_{N}\omega^{-x_{N}},\quad x_{N}=\frac{s-3}{3},A_{N}=\left(-\frac{J_{N}}{x_{N}\left(x_{N}-1\right)}\right)^{1/3}\label{eq:n_N-sol-powerlaw}
\end{align}
and for $J_{N}=0$:
\begin{equation}
n_{E}=A_{E}\omega^{-x_{E}},\quad x_{E}=\frac{s-2}{3},A_{E}=\left(\frac{J_{E}}{x_{E}\left(x_{E}-1\right)}\right)^{1/3}.\label{eq:n_E-sol-powerlaw}
\end{equation}
These exponents $x_{N}$ and $x_{E}$ exactly match those predicted
by the general (nonlocal) theory\footnote{In these solutions (\ref{eq:n_N-sol-powerlaw}-\ref{eq:n_E-sol-powerlaw}),
one can remark that the sign of $J_{N},J_{E}$ is imposed by the sign
of $x_{N}\left(x_{N}-1\right)$ and $x_{E}\left(x_{E}-1\right)$,
such that $A_{E},A_{N}$ are positive. We deduce that the solutions
are physically acceptable only if the physical current are in accordance
with that. In practice, in the physical cases listed in table \ref{tab:exponents},
this is always verified.}. In addition, in the general (non-local) case, the Kolmogorov constants
can also be computed. They are dimensionally similar as they are proportional
to $J_{E}^{1/3}$ and $J_{N}^{1/3}$ but the non-dimensional prefactor
writes \textit{e.g.} $A_{E}=\left(J_{E}/\mathcal{G}I'\left(x_{E}\right)\right)^{1/3}$
where\footnote{The proof of this (known) result is provided in Appendix \ref{subsec:Appendix:-Computation-of-kolmogorov-constant}}
$I$ is the nondimensionalised collision integral (\ref{eq:b_iso})
and $\mathcal{G}$ is a constant depending on $\alpha,d,\gamma_{S}$.
The exact Kolmogorov constant is thus not as simple as those derived
in our isotropic model (\ref{eq:n_N-sol-powerlaw}-\ref{eq:n_E-sol-powerlaw})
\cite{pushkarev2000turbulence,deike2014direct}. 

We can also go beyond the KZ solutions by examining corrections that
arise when two fluxes are present but one is small compared to the
other. To our knowledge, this analytical generalization is new, although
numerical resolutions of the kinetic equation with two fluxes are
possible and more general. The computation is presented in Appendix
\ref{subsec:perturbative corr-to-n}, and enables to treat situations
where the inertial ranges are quite large, but not sufficiently to
ensure validity of KZ solutions. The main result is that we obtain
corrections to the spectrum which are in general negative, linear
in $\omega$ or $\tfrac{1}{\omega}$ (depending on which flux is small
compared to the other), and larger when we approach the dissipation
frequencies $\omega_{d}^{+},\omega_{d}^{-}$. We will not use them
directly in this work, but this property remains interesting when
we would like refinements of solutions. 

\subsection{Inhomogeneous wave turbulence}

In the line of the possible generalizations for this work, one should
mention the case when spatial inhomogeneities cause a space dependence
of the spectrum $n\left(\bm{k},\bm{x},t\right)$. This case has been
studied in \cite{Guioth_2024,Onuki_2023} and is summarised in Appendix
\ref{subsec:Appendix-inhomogeneous-WT}. In short, one should keep
in mind that all of the present work can be generalised to inhomogeneous
dynamics, up to an additional space dependency in $\bm{x}$ and an
additional transport term in the large deviation Hamiltonian. In the
following however, we will write everything in the homogeneous framework
for simplicity.\\

Up to now, we have been interested in the dynamics of the wave action
spectrum $n$ and we have written a large deviation principle and
an associated fluctuating dynamics. The main interest of the large
deviation theory we derived is to study the fluctuations of $n$ because
the dynamics of the average is already well known and does not require
such a theoretical framework. We know these fluctuations at equilibrium
because we know the quasipotential. Therefore, in the next sections,
we will compute the fluctuations of the dynamics brought out of equilibrium
by a source and a dissipation. 

\section{Summary of the general framework for path large deviation theory
with two conservation laws \label{sec:General-framework-for-LD}}

Wave turbulence as a driven diffusive system with two conservation
laws, is an example of a dynamics with an equilibrium structure forced
by non equilibrium boundary conditions. Most transport phenomena in
physics share this property. This equilibrium structure forced by
non equilibrium boundary conditions is a very important property.
In this section we will explain its importance by the general framework
of transverse decomposition for non linear and linearised dynamics,
introduced in \cite{article2_brice_freddy}, which we summarize. A
key consequence and the main result of this part is that the two point
correlation function can be decomposed in three parts : an equilibrium
diagonal part, and two parts which are each proportional to the non-equilibrium
currents. In the case of wave turbulence, we have two conservation
laws, two currents $J_{N}$ and $J_{E}$, and a differential operator
of order 2. As a consequence, the nonequilibrium parts of the two
point correlation function will be non local, continuous, and strongly
affected by boundary conditions. This will be one of the main results
of this section.

Building on \cite{article2_brice_freddy}, which analyzes Gaussian
nonequilibrium fluctuations via the Lyapunov equation, key results
include clear identification of the sources of long-range correlations
and the impact of boundary conditions. Since wave turbulence involves
boundary fluxes, fixed-flux conditions are considered, leading to
distinct long-range behaviours compared to fixed-density conditions.
We focus on the second-order cumulant $C_{2}\left(\omega_{1},\omega_{2}\right)$
in diffusive systems with small noise, over the frequency domain $\mathcal{D}=\left[\omega_{\text{min}},\omega_{\text{max}}\right]$
where $0\leq\omega_{\text{min}}<\omega_{\text{max}}\leq+\infty$. 

While \cite{article2_brice_freddy} studies the cases with one or
two conservation laws, we exclusively focus on the two conservation
laws case, with a general dynamics of diffusive systems of the type
\begin{equation}
\text{d}N=\partial_{\omega}^{2}\left(\mathcal{K}\left[N\right]\text{d}t+\sqrt{2\varepsilon\mu\left[N\right]}\text{d}W\right)+\left(s-d\right)\text{d}t\label{eq:PB2_2_CL}
\end{equation}
where $\mathcal{K}$ is a functional of $N$ and possibly $\omega$,
$\mu$ (which we call \textit{mobility}) is a function of $N$ and
possibly $\omega$, $\varepsilon\ll1$ is the large deviation parameter
and $W$ a Wiener process. $s$ and $d$ stand for source and dissipation
terms and will possibly break the conservation law in certain frequency
ranges. For simplicity, we will suppose they vanish at the boundary
$\partial\mathcal{D}$. As explained in section \ref{subsec:Fluxes-conservation-laws},
the total wave action number (or mass) $\mathcal{N}=\int\text{d}\omega N$
and the total energy $\mathcal{E}=\int\text{d}\omega N\omega$ are
conserved for this dynamics as soon as there are appropriate boundary
conditions. They are associated to two currents $j_{N}\left[N\right]=-\partial_{\omega}K$,
$j_{E}\left[N\right]=K-\omega\partial_{\omega}K$ where $K=\mathcal{K}\left[N\right]+\sqrt{2\varepsilon\mu\left[N\right]}\frac{\text{d}W}{\text{d}t}$
is a fluctuating quantity. For the simplicity of notations, for the
rest of this section we will drop the $N$-dependencies which are
implicitly assumed. We will denote the functions with a bar when they
depend on the average value $\bar{N}$: for instance we will denote
$\mu\left(\omega\right)$ (resp. $\bar{\mu}\left(\omega\right)$)
instead of $\mu\left[N\right]\left(\omega\right)$ (resp. $\mu\left[\bar{N}\right]\left(\omega\right)$).

For this dynamics, we consider two types of configurations. In the
first one, the domain $\mathcal{D}$ contains all the source and dissipation
terms, such that beyond dissipation scales nothing happens and the
boundary conditions are not a real matter of interest. In the second
configuration, the domain $\mathcal{D}$ excludes all source and dissipation
scales, such that the term $s-d$ vanishes everywhere in $\mathcal{D}$.
This regime is called \textit{inertial}, and is particularly relevant
for numerical studies where the numerical cost of resolving the whole
range is determinant. In this configuration, we have to define precisely
the boundary conditions which are not trivial. In general, the boundary
conditions are expected to depend on the dynamics and be non universal.
However, in order to study the initial range, we choose to study fixed
currents boundary conditions, as the simplest and most natural case.
For this case, $j_{N}\left[N\right],j_{E}\left[N\right]$ are fixed
at the boundary $\partial\mathcal{D}$, such that incoming and outgoing
fluxes are equal. Other intermediate configurations, for instance
when the forcing is included in $\mathcal{D}$ but not the dissipation,
could easily be studied with our methods, but for the sake of simplicity
we will omit them first, and mention the necessary modifications as
a remark in section \ref{subsec:forcing}. In this section, we start
with the general configuration and then specify the boundary conditions
to study the inertial configuration.

\subsection{Transverse decomposition in systems with two conservation laws\label{subsec:Transverse-decomposition-1CL-2CL}}

In the dynamics (\ref{eq:PB2_2_CL}), a transverse decomposition with
respect to the quasipotential $Q$ corresponds to a splitting of the
function $\mathcal{K}\left[N\right]$ into a gradient contribution
$\mathcal{K}_{\nabla}=\mu\left[N\right]\partial_{\omega}^{2}\frac{\delta Q}{\delta N}$
and a transverse contribution:
\[
\mathcal{K}=\mathcal{K}_{\nabla}+\mathcal{K}_{\perp}
\]
with the transverse condition $\frac{\delta Q}{\delta N}.\partial_{\omega}^{2}\mathcal{K}_{\perp}=0$.
This transverse structure also declines at the linearised level as
we look at small density fluctuations, see next section \ref{subsec:Lyapunov-equation-for-longrange-correlations}. 

Now let us consider a dynamics with \textit{statistical equilibrium
structure} (as defined in \cite{article2_brice_freddy}: a \textit{statistical
equilibrium structure} shows a transverse decomposition with respect
to the equilibrium quasipotential $Q_{eq}$). In a second step, we
consider the particular case $\mathcal{K}$ of the type 
\begin{equation}
\mathcal{K}\left[N\right]=\mathscr{D}\left[N\right]\partial_{\omega}^{2}g\left[N\right]+\mu\left[N\right]E\label{eq:K-transverse-decomposition-particular-form}
\end{equation}
where $E$ is an external field depending on $\omega$, $g$ is a
function of $N$ and possibly $\omega$, as well as $\mu$ and $\mathscr{D}$
. In this expression, no source and dissipation $s-d$ are present
as in (\ref{eq:PB2_2_CL}) but one could in principle add some. This
general dynamics corresponds to a statistical equilibrium dynamics
if $g$ can be interpreted as the functional derivative of the equilibrium
quasipotential $Q_{eq}$: $g=\frac{\delta Q_{eq}}{\delta N}$, and
if we have the relation 
\begin{equation}
\mathscr{D}=\mu.\label{eq:Einstein-relation-2CL}
\end{equation}
This identity is the analog of the Einstein relation for two conservation
laws and arises, for instance, in isotropic local wave turbulence.
While the dynamics formally has a statistical equilibrium structure,
it may describe nonequilibrium states when boundary conditions enforce
nonzero fluxes, such that the stationary state $\bar{N}$ is not the
equilibrium state. In such cases (\textit{e.g.}, wave turbulence with
imposed currents, where the transverse term $\mathcal{G}$ vanishes)
the stationary state differs from equilibrium despite the equilibrium-like
form of the dynamics.

\subsection{Gaussian fluctuations at equilibrium and out-of-equilibrium - General
formulation of the Lyapunov equation and boundary conditions for two
conservation laws\label{subsec:General-Lyapunov-and-BCs-1} }

\subsubsection{Lyapunov equation for the long-range correlations\label{subsec:Lyapunov-equation-for-longrange-correlations}}

Within a dynamics of the type (\ref{eq:PB2_2_CL}), we call $\delta N=\frac{N-\bar{N}}{\sqrt{\varepsilon}}$
the rescaled density fluctuation with $\varepsilon$ the large deviation
parameter. At leading order in $\varepsilon$, it verifies a linear
dynamics with a linear operator $\mathcal{L}\cdot$ acting on $\delta N$.
General considerations over this linearised dynamics are recalled
in Appendix \ref{subsec:Intro-Linearised-dynamics+Lyapunov}, where
is also explained how to obtain the general equation over the (rescaled)
second order cumulant $C_{2}\left(\omega_{1},\omega_{2}\right)=\mathbb{E}\left(\delta N\left(\omega_{1}\right)\delta N\left(\omega_{2}\right)\right)$,
called \textit{Lyapunov equation}:
\[
\partial_{t}C_{2}=\mathcal{L}_{\omega_{1}}C_{2}+\mathcal{L}_{\omega_{2}}C_{2}+2A
\]
where $\mathcal{L}_{\omega_{1}}\cdot$ is the operator of the linearised
dynamics and $A$ is the noise covariance, defined in Appendix \ref{subsec:Intro-Linearised-dynamics+Lyapunov},
equation (\ref{eq:A-intro}).

When the core dynamics inherit a statistical equilibrium structure
relative to the equilibrium quasipotential $Q_{eq}$ and the boundary
conditions are compatible with this structure, we have an equilibrium
dynamics. Then the quasipotential $Q_{eq}\left[N\right]:=\int\text{d}\omega\:q_{eq}\left(N\left(\omega\right)\right)$
is known, with $q_{eq}$ a simple function of $N\left(\omega\right)$.
The solution of the Lyapunov equation is simply $C_{2}=C_{2,eq}$,
where due to the local expression of $Q_{eq}$, $C_{2,eq}$ is diagonal:
$C_{2,\text{eq}}=\frac{1}{q_{eq}''}\delta\left(\omega_{1}-\omega_{2}\right)$. 

However, in general the boundary conditions are not compatible with
the equilibrium quasipotential and we have a non-equilibrium dynamics,
with a dynamical structure in the bulk that inherit the equilibrium
structure. It is then natural to define $B$, the nonequilibrium contribution
to the two point-correlation function, by the decomposition $C_{2}=C_{2,eq}+B$.
Then we can easily derive an equation for $B$:
\begin{equation}
\partial_{t}B=\mathcal{L}_{\omega_{1}}B+\mathcal{L}_{\omega_{2}}B+2\tilde{A}\label{eq:lyapunov-B-general}
\end{equation}
where $\mathcal{L}_{\omega_{1}}\cdot=-\partial_{\omega_{1}}^{2}\left(\mu'\partial_{\omega_{1}}^{2}q_{eq}\left(\bar{N}\left(\omega_{1}\right)\right)\cdot\right)-\partial_{\omega_{1}}^{2}\left(\mu\partial_{\omega_{1}}^{2}\left[q_{eq}'\left(\bar{N}\left(\omega_{1}\right)\right)\cdot\right]\right)$
is the operator of the linearised dynamics and $\tilde{A}\left(\omega_{1},\omega_{2}\right)=-\partial_{\omega_{1}}^{2}\left[\left(\mu'\partial_{\omega_{1}}^{2}q_{eq}\left(\bar{N}\left(\omega_{1}\right)\right)\right)C_{2,eq}\left(\omega_{1},\omega_{2}\right)\right]$.

We consider the simpler case where $\mathcal{K}$ writes as (\ref{eq:K-transverse-decomposition-particular-form})
with Einstein relation (\ref{eq:Einstein-relation-2CL}). We also
consider stationary solutions of the Lyapunov equation (\ref{eq:lyapunov-B-general}).
Then the equilibrium contribution writes $C_{2,\text{eq}}=-\frac{1}{\bar{g}'}\delta\left(\omega_{1}-\omega_{2}\right)$
and the nonequilibrium contribution $B$ verifies in the stationary
regime 
\begin{equation}
\mathscr{L}B=\left(\mathcal{K}^{*}\left(\omega_{1}\right)+\mathcal{K}^{*}\left(\omega_{2}\right)\right)\delta''\left(\omega_{1}-\omega_{2}\right)\label{eq:Lyapunov-B-2CL}
\end{equation}
where in the first equation $\mathscr{L}\cdot=\partial_{\omega_{1}}^{2}\left(L\cdot\right)+\partial_{\omega_{2}}^{2}\left(\cdot L^{T}\right)$
with $L\cdot=\frac{\delta\mathcal{K}}{\delta N}\left[\bar{N}\right]\cdot=\bar{\mathscr{D}}'\partial_{\omega_{1}}^{2}\bar{g}\cdot+\bar{\mathscr{D}}\partial_{\omega_{1}}^{2}\left(\bar{g}'\cdot\right)+\bar{\mu}'E$
and $\mathcal{K}^{*}=\frac{\bar{\mathcal{\mu}}'\bar{\mathcal{K}}}{\bar{\mathcal{\mu}}\bar{g}'}$.
To be clear, $L$ is a linear operator acting on the first variable
$\omega_{1}$ whereas $L^{T}$ is its adjoint, acting on the second
variable $\omega_{2}$. 

We can make several more comments on the equation for stationary $B$
(\ref{eq:Lyapunov-B-2CL}). First, we note that $\mathscr{L}$ is
an elliptic operator of the fourth order. Second, in the right-hand-side,
we have source terms. Then the solution will depend on the boundary
conditions. Still some general conclusions can be drawn. On the right
hand side, the bulk term $\left(\mathcal{K}^{*}\left(\omega_{1}\right)+\mathcal{K}^{*}\left(\omega_{2}\right)\right)\delta''\left(\omega_{1}-\omega_{2}\right)$
is proportional to the average ``flux'' function $\bar{\mathcal{K}}$,
the solution will then be proportional to $\bar{\mathcal{K}}$ and
zero for equilibrium solutions when $\bar{\mathcal{K}}=0$ and more
generally when $\mathcal{K}^{*}=0$. Moreover this source is diagonal
and a distribution of order two (the second derivative of a Dirac
distribution). Because the elliptic operator is of order 4, and the
equation forced by a source term of order 2, the elliptic regularity
theorem entails that the solution is at most of order $2-4=-2$. In
other words it is a continuous function (with possibly a discontinuous
first derivative with discontinuities located on the diagonal $\omega_{1}=\omega_{2}$).
Moreover, the solution is smooth over any neighborhood not containing
the diagonal $\omega_{1}=\omega_{2}$, strongly affected by the boundary
conditions and thus non local. Hence, this smooth contribution of
the correlation function $B$ describes the long-range correlations
in our system. 

\subsubsection{Impact of a forcing in the domain $\mathcal{D}$\label{subsec:forcing}}

It is worth mentioning that we can show additional interesting properties
if we relax the hypothesis of a vanishing forcing $s-d$ inside the
domain $\mathcal{D}$. If this forcing is \textit{deterministic},
\textit{i.e.} it only depends on the density $N$ and possibly $\omega$,
we can show that the previous results are unchanged up to a modification
of the second member by adding a source term $-2C_{\text{eq}}\frac{\delta}{\delta N}\left(s-d\right)\left[\bar{N}\right]\delta\left(\omega_{1}-\omega_{2}\right)$,
and a modification of the operator $\mathcal{L}$ by $\frac{\delta}{\delta N}\left(s-d\right)\left[\bar{N}\right]$if
$s-d$ depends on $N$. In addition, adding a fluctuation to the forcing
modeled by a random noise might also impact the correlations, but
this contribution is independent and additive, since it only adds
a source term to the second member. These forcing terms will possibly
create two additional components to the long-range correlation function
$B$, one coming from the deterministic part and one coming from the
stochastic part. The properties of these terms are discussed in \cite{article2_brice_freddy}
but will not be used in our simple numerical study presented in the
next parts, although it can be applied to wave turbulence.\\

\subsubsection{Two point correlation functions for the inertial configuration}

Let us now specify the inertial configuration. This case of inertial
configuration will be very interesting for its simplicity as the two
point correlation can be easily decomposed in two components. 

We define the inertial configuration as the case when the source and
dissipation vanish, and the dynamics is determined by fixed flux boundary
conditions. For this case, $j_{N}\left[N\right],j_{E}\left[N\right]$
are fixed at the boundary $\partial\mathcal{D}$, such that incoming
and outgoing fluxes are equal to respectively to the two constant
values $J_{N}$ and $J_{E}$. In this regime, the currents of wave-action
and energy $J_{N},J_{E}$ are constant in the whole domain and $\mathcal{\bar{K}}=J_{N}-\omega J_{E}$.
In addition, we have to specify the boundary conditions. The stationarity
Lyapunov equation with fixed-flux boundary conditions then reads \cite{article2_brice_freddy}:
\begin{equation}
\begin{cases}
\mathscr{L}B=\left(\mathcal{K}^{*}\left(\omega_{1}\right)+\mathcal{K}^{*}\left(\omega_{2}\right)\right)\delta''\left(\omega_{1}-\omega_{2}\right)\\
\begin{cases}
LB\left(\omega_{1},\omega_{2}\right) & =\mathcal{K}^{*}\left(\omega_{1}\right)\delta\left(\omega_{1}-\omega_{2}\right)\\
\partial_{\omega_{1}}\left(LB\left(\omega_{1},\omega_{2}\right)\right) & =\partial_{\omega_{1}}\left(\mathcal{K}^{*}\delta\left(\omega_{1}-\omega_{2}\right)\right)
\end{cases} & \text{ for }\omega_{1}\in\partial\mathcal{D}\\
\text{ same symmetrized boundary conditions } & \text{ for }\omega_{2}\in\partial\mathcal{D}
\end{cases}\label{eq:Lyapunov-B-2CL+BCs}
\end{equation}

Because the stationary solution is a inhomogeneous linear equation
and $\mathcal{\bar{K}}=J_{N}-\omega J_{E}$, the solution of (\ref{eq:Lyapunov-B-2CL})
is called \textit{flux-driven} contribution. In general, the operator
$\mathscr{L}$ depends on $N$ and thus also on $J_{N}$ and $J_{E}$,
so the contributions linked to both fluxes are intertwined. However,
in the two particular cases where $J_{N}=0$ or $J_{E}=0$, we will
see later that there is a simple proportionality property of the solution
of (\ref{eq:Lyapunov-B-2CL}) with some power of the flux value. \\

In the following, we will focus on the flux-driven contribution of
the correlations for wave turbulence. We will compute these long-range
correlations for the isotropic local wave turbulence model. 

\section{The flux-driven two point correlation function for isotropic and
local wave turbulence \label{sec:Application-to-WT}}

Studying long range correlation in transport phenomena in general,
and in wave turbulence is very interesting. For instance the two-point
correlation function that we study in the frequency domain $C_{2}\left(\omega_{1},\omega_{2}\right)$,
which is function of order 4 in the variable $\psi$, is directly
related to quantities which are often measured in experiments. For
instance the measured the 4th order structure function $S_{4}\left(\bm{x},\bm{x}+\bm{r}\right)=\mathbb{E}\left[\left|\psi\left(\bm{x}\right)-\psi\left(\bm{x}+\bm{r}\right)\right|^{4}\right]$,
or the 4th order spatial correlation function can be computed can
be computed from $C_{2}$ \cite{campagne2019energy,deike2013etudes,aubourg20173}
by Fourier transforms and corrections with lesser order terms.

In this part we study the two point correlation for the case of isotropic
local wave turbulence. For simplicity, we consider the inertial case,
as defined in the previous section: boundary conditions with fixed
fluxes. We will compute and discuss the stationary solution of the
flux-driven part $B\left(\omega_{1},\omega_{2}\right)$ of the two
point correlation function.

We recall $B$ obeys the Lyapunov equation (\ref{eq:Lyapunov-B-2CL}),
which is a linear fourth order inhomogeneous elliptic PDE with nonconstant
coefficients. As a consequence the solution has a regularity property:
it is smooth outside the diagonal and continuous (with possibly a
discontinuous derivative with discontinuities located on the diagonal
$\omega_{1}=\omega_{2}$). The analytical computation is generally
not possible when we are out of equilibrium. To solve it, we need
to use a numerical approach, which has been explained and tested in
\cite{article2_brice_freddy} with the example of the Symmetric Exclusion
Process (SEP) model. In the following, we use the same approach to
compute the correlation function for the isotropic local wave turbulence
model. As a particularly interesting example, we will look at the
fluctuations around the Kolmogorov--Zakharov (KZ) spectrum (\ref{eq:n_E-sol-powerlaw}). 

\subsection{Numerical computation of the flux-driven two point correlation function
for isotropic and local wave turbulence \label{subsec:Computation-3-contrib}}

We consider the isotropic local model for wave turbulence. The stochastic
kinetic equation can be written in the form (\ref{eq:PB2_2_CL}).
The flux-driven contribution to the long-range correlations is $B\left(\omega_{1},\omega_{2}\right)=C_{2}\left(\omega_{1},\omega_{2}\right)-C_{eq}\delta\left(\omega_{1}-\omega_{2}\right)$
and verifies the Lyapunov equation (\ref{eq:Lyapunov-B-2CL}). For
isotropic wave turbulence, we recall that for simplicity of the mathematical
expressions, we work with $n$ rather than $N$, where $N=\Omega\left(\omega\right)n$
(see eq. (\ref{eq:def_Omega_prefactor})). Therefore the linearised
dynamics equation reads
\[
\Omega\left(\omega\right)\text{d}\delta n=\mathcal{L}\left[\bar{n}\right]\delta n\text{d}t+\sqrt{2}\sigma_{\ell}\left[\bar{n}\right]\text{d}W.
\]
We write the Lyapunov equation for two-point correlation for $n$:
$C_{n,2}\left(\omega_{1},\omega_{2}\right)=\mathbb{E}\left[\delta n\left(\omega_{1}\right)\delta n\left(\omega_{2}\right)\right]=\Omega^{-1}\left(\omega_{1}\right)\Omega^{-1}\left(\omega_{2}\right)\mathbb{E}\left[\delta N\left(\omega_{1}\right)\delta N\left(\omega_{2}\right)\right]$
and modify the definition of $B$ by the same prefactor: $B_{n}\left(\omega_{1},\omega_{2}\right)=\Omega^{-1}\left(\omega_{1}\right)\Omega^{-1}\left(\omega_{2}\right)B\left(\omega_{1},\omega_{2}\right)$.
We see that $C_{n,2}$ and $B_{n}$ are rescaled versions of $C_{2}$
and $B$. In the following, as we will always discuss $C_{n,2}$ and
$B_{n}$ only, we will drop the index $n$ without risk of confusion
and simply denote $B_{n}$ by $B$. 

With this rescaling, the rescaled $B$ verifies a Lyapunov equation
of the same form (\ref{eq:Lyapunov-B-2CL}). However, the coefficients
in this equation have to be modified and we have the following expression:
\begin{equation}
\begin{cases}
g\left[N\right] & =\frac{1}{n}\\
\mu\left[N\right]=\mathscr{D}\left[N\right] & =\omega^{s}n^{4}\\
E & =0
\end{cases}\quad\Longrightarrow\quad\begin{cases}
\mathcal{K} & =\mu\left[N\right]\partial_{\omega}^{2}\left(\frac{1}{n}\right)\\
 & =\omega^{s}n\left[2\left(\partial_{\omega}n\right)^{2}-n\partial_{\omega}^{2}n\right]\\
\mathcal{K}^{*} & =-4n^{2}\omega^{s}\left[2\left(\partial_{\omega}n\right)^{2}-n\partial_{\omega}^{2}n\right]\\
L\cdot & =\frac{\delta\mathcal{K}}{\delta n}\cdot=\partial_{\omega_{1}}^{2}\left(\frac{1}{n}\right)\partial_{n}\mu\cdot+\mu\left[N\right]\partial_{\omega_{1}}^{2}\left(-\frac{1}{n^{2}}\cdot\right)\cdot\\
 & =u\left(\omega_{1}\right)\partial_{\omega_{1}}^{2}+v\left(\omega_{1}\right)\partial_{\omega_{1}}+w\left(\omega_{1}\right)\cdot\\
\text{where } & \begin{cases}
u\left(\omega\right) & =-\omega^{s}n^{2}\\
v\left(\omega\right) & =\left(4\omega^{s}n\partial_{\omega}n\right)\\
w\left(\omega\right) & =\left(2\omega^{s}\left(n\partial_{\omega}^{2}n-\left(\partial_{\omega}n\right)^{2}\right)\right)
\end{cases}
\end{cases}\label{eq:WT-expressions-g-mu-E}
\end{equation}
where we recall $s$ is a known physical exponent (\ref{eq:S-formula}).
A review of the possible values of $s$ on some physical cases is
provided in table \ref{tab:exponents}. 

To solve equation (\ref{eq:Lyapunov-B-2CL+BCs}), for a given configuration,
the first task is to compute the most probable state $\bar{N}$, which
strongly influence the results. We then deduce the associated $\bar{n}$
and the functions defined in equation (\ref{eq:WT-expressions-g-mu-E})
evaluated at $\bar{n}$. Although a necessary and quite technical
step, the computation of the stationary spectrum is not the core result
of this work, so the method is left in Appendix \ref{subsec:Computation-spectrum-WT}.
In the following, we implement the Lyapunov equation and elaborate
a numerical method to obtain the result for $B$. 

\subsubsection{Numerical method for computing the flux-driven two point correlation
\label{subsec:Computation-long-correl-WT}}

In this paragraph, we explain the numerical computation of the flux-driven
long-range correlations $B$ for isotropic local wave turbulence.
To do so, we adopt a discretization method for the operator of the
linearised dynamics with appropriate boundary conditions. Then we
employ existing numerical schemes to solve for equation (\ref{eq:Lyapunov-B-2CL+BCs})
in the space of discrete frequencies $\omega$, by using the same
method as for the SEP model used in \cite{article2_brice_freddy}.
In brief, in this paper, a numerical method was developed. It is able
to solve the Lyapunov equation for one conservation law and validated
on the SEP model which is one of the simplest models with long-range
correlations. In the following section, we generalise this method
to two conservation laws (mass and energy) and apply our numerical
method to find the solution of the Lyapunov equation for wave turbulence.

\paragraph{Numerical protocol}

First, let us adapt the numerical method presented in \cite{article2_brice_freddy}
for two conservation laws. As for the kinetic equation, we discretise
the space of $\omega$ in $I$ values $\left(\omega_{1},\ldots,\omega_{I}\right)$
with a regular spacing $\Delta\omega$. First, we would like to discretise
the Lyapunov equation (\ref{eq:Lyapunov-B-2CL+BCs}). To do it, we
need to correctly discretise the second member and the linear operator
$\mathscr{L}$ with correct boundary conditions. We would like to
do it so that the operator $\mathscr{L}$ automatically verifies the
boundary conditions. Since these conditions are inhomogeneous in the
corners of the domain, they cannot simply be written as such within
a linear operator.We already met this problem for the SEP model in
\cite{article2_brice_freddy}. To solve it, we define $\tilde{B}$
by a change of variable 
\[
\tilde{B}_{ij}=\begin{cases}
B_{ij} & \text{if }1\leq i\leq I\\
B_{-1,j}-\frac{2\Delta\omega^{3}h_{j}^{\left(0\right)}}{u_{-1}} & \text{if }i=-1\\
B_{0,j}+\frac{\Delta\omega^{2}g_{j}^{\left(0\right)}}{u_{0}-\frac{\Delta\omega v_{0}}{2}} & \text{if }i=0\\
B_{I+1,j}+\frac{\Delta\omega^{2}g_{j}^{\left(1\right)}}{u_{I+1}+\frac{\Delta\omega v_{I+1}}{2}} & \text{if }i=I+1\\
B_{I+2,j}+\frac{2\Delta\omega^{3}h_{j}^{\left(1\right)}}{u_{I+2}} & \text{if }i=I+2
\end{cases}
\]
where we define $g_{j}^{\left(0\right)},g_{j}^{\left(1\right)},h_{j}^{\left(0\right)},h_{j}^{\left(1\right)}$
as the functions defining the boundary conditions for $B$:
\begin{align}
 & \begin{cases}
\left(LB\right)_{i=1,j} & =g_{j}^{\left(0\right)}\\
\left(LB\right)_{i=I,j} & =g_{j}^{\left(1\right)}\\
\left(\partial_{\omega_{1}}LB\right)_{i=1j} & =h_{j}^{\left(0\right)}\\
\left(\partial_{\omega_{1}}LB\right)_{i=I,j} & =h_{j}^{\left(1\right)}
\end{cases}\label{eq:LB boundary 2CL-1}
\end{align}
that is to say, $g_{j}^{\left(0\right)}:=\mathcal{K}^{*}\left(\omega_{\text{min}}\right)\frac{\delta_{j,1}}{\Delta\omega}$,
$g_{j}^{\left(1\right)}:=\mathcal{K}^{*}\left(\omega_{\text{max}}\right)\frac{\delta_{j,I}}{\Delta\omega}$
and $h_{j}^{\left(0\right)}:=\mathcal{K}^{*}\left(\omega_{j}\right)\frac{\delta'_{j,1}}{\Delta\omega^{2}}$,
$h_{j}^{\left(1\right)}:=\mathcal{K}^{*}\left(\omega_{j}\right)\frac{\delta'_{j,1}}{\Delta\omega^{2}}$
are the discretisations of the boundary functions $\mathcal{K}^{*}\left(\omega_{1}\right)\delta\left(\omega_{1}-\omega_{2}\right)$
and $\mathcal{K}^{*}\left(\omega_{2}\right)\delta'\left(\omega_{1}-\omega_{2}\right)$
respectively. In these expressions, we wrote $\frac{\delta_{j,1}}{\Delta\omega}$
as the discretisation of the Dirac delta function $\delta$ and $\frac{\delta'_{j,1}}{\Delta\omega^{2}}$
as the discretisation of $\delta'$ as vectors of size $I$, namely
$\delta'_{j,1}=\left(\begin{array}{ccccc}
-1 & 1 & 0 & \ldots & 0\end{array}\right)_{j}$ and $\delta'_{j,I}=\left(\begin{array}{ccccc}
0 & \ldots & 0 & -1 & 1\end{array}\right)_{j}$.

By doing this change of variable, we obtain equation obtain the discretised
Lyapunov equation in the form 
\begin{equation}
\mathfrak{L}\tilde{B}+\tilde{B}\mathfrak{L}^{T}=Q\label{eq:discretized-Lyapunov-Btilde-2CL}
\end{equation}
where $\mathfrak{L}\in\mathscr{M}_{I,I}\left(\mathbb{R}\right)$ is
the discretised linear operator incorporating the boundary conditions\footnote{$\mathscr{M}_{I,J}\left(\mathbb{R}\right)$ denotes the set of $I\times J$
real matrices}, $Q\in\mathscr{M}_{I,I}\left(\mathbb{R}\right)$ is the matrix discretizing
the second member and $\tilde{B}\in\mathscr{M}_{I,I}\left(\mathbb{R}\right)$
is the discretization of the solution. In practice: 
\begin{enumerate}
\item The discretised operator $\mathfrak{L}$ reads 
\[
\mathfrak{L}=D_{I\times I-2}^{2}\cdot\tilde{L}\in\mathscr{M}_{I,I}\left(\mathbb{R}\right)
\]
 where $D_{I\times I-2}^{2}\in\mathscr{M}_{I,I-2}\left(\mathbb{R}\right)$
is the matrix standing for the discretised second order derivative
with appropriate boundary conditions (\ref{eq:LB boundary 2CL-1})
and $\tilde{L}\in\mathscr{M}_{I-2,I}\left(\mathbb{R}\right)$ is a
discretization of the operator $L$ given in (\ref{eq:WT-expressions-g-mu-E}):
\begin{equation}
L=u\left(\omega_{1}\right)\partial_{\omega_{1}}^{2}+v\left(\omega_{1}\right)\partial_{\omega_{1}}+w\left(\omega_{1}\right)\label{eq:L-operator-continuous-1}
\end{equation}
where in wave turbulence:
\[
\begin{cases}
u\left(\omega\right) & =-\omega^{x_{2}}n^{2}\\
v\left(\omega\right) & =\left(4\omega^{x_{2}}n\partial_{\omega}n\right)\\
w\left(\omega\right) & =\left(2\omega^{x_{2}}\left(n\partial_{\omega}^{2}n-\left(\partial_{\omega}n\right)^{2}\right)\right).
\end{cases}
\]
Further explanations and explicit expressions for these matrices,
with details of the computation are gathered in Appendix \ref{subsec:Appendix:-computation-discretized-Lyapunov-2CL}.
We define the (discretized) mass and energy of the matrix $\tilde{B}$
as the vectors $\left(1...1\right).\tilde{B}$ and $\left(\omega_{1},\ldots,\omega_{I}\right).\tilde{B}$
respectively. One can easily check that this operator automatically
conserves the mass and the energy:
\begin{align}
\left(1...1\right).\mathfrak{L} & =0\nonumber \\
\left(\omega_{1},\ldots,\omega_{I}\right).\mathfrak{L} & =0\label{eq:mass-energy-conservation-discretized}
\end{align}
\item The matrix for the second member $Q$ is defined as $Q=A+A^{T}+S+S^{T}$.
The first term of this second member is the matrix 
\[
A:=\mathcal{K}_{I-2\times I}^{*}\cdot\delta''_{I}\in\mathscr{M}_{I,I}\left(\mathbb{R}\right)
\]
 with $\mathcal{K}_{I-2\times I}^{*}\in\mathscr{M}_{I,I-2}\left(\mathbb{R}\right)$
and $\delta''_{I}=\in\mathscr{M}_{I-2,I}\left(\mathbb{R}\right)$
stand for discretisations of $\omega\mapsto\mathcal{K}^{*}\left(\omega\right)$
and $\omega\mapsto\delta''\left(\cdot-\omega\right)$ respectively.
Eventually, the last matrix $S$ standing at the second member is
the counterpart of the change of variable $B\mapsto\tilde{B}$.
\end{enumerate}

Solving (\ref{eq:L-operator-continuous-1}) gives us a solution to
the Lyapunov equation with the correct boundary conditions. 

\subparagraph{Mathematical comments}

Before solving the equation (\ref{eq:discretized-Lyapunov-Btilde-2CL}),
we can make some mathematical comments.
\begin{enumerate}
\item This discretisation procedure has a precision $\mathcal{O}\left(\Delta\omega^{2}\right)$
for the evaluation of the differential operators and $\mathcal{O}\left(\Delta\omega\right)$
for the method imposing the boundary conditions, thus leading to
overall $\mathcal{O}\left(\Delta\omega\right)$ precision for the
result. 
\item This equation is a matrix equation used in the stability analysis
of linear dynamical systems. It is a known result (see the mathematics
of Bartels-Stewart algorithm) that a unique solution to (\ref{eq:discretized-Lyapunov-Btilde-2CL})
exists whenever the eigenvalues of $\mathfrak{L}$ are distinct from
the eigenvalues of $\mathfrak{-L}$. Of course, this assumption implies
that $\mathfrak{L}$ is invertible. Actually this assumption is not
verified in our case, indeed equation (\ref{eq:mass-energy-conservation-discretized})
shows that $\mathfrak{L}$ has a kernel of dimension 2. In such a
case, the solution is not unique. The null eigenvalues are associated
to mass and energy conservation. Choosing the mass and energy values
for the solution then ensure uniqueness. As for the SEP model in \cite{article2_brice_freddy},
we will use a method to get the solution with zero mass and zero energy.
The choice of this mass and energy values are somewhat arbitrary (see
more complete discussion in \cite[section 2.3.1]{article2_brice_freddy}),
but they are compatible with the structure of the equation and can
be modified with a different choice of solution to the homogeneous
equation, as will be discussed in section \ref{subsec:Scaling-properties-of-homogeneous-solution}.
\item Various algorithms enable to solve it. The most popular is the Bartels-Stewart
algorithm with various versions (see \textit{e.g.} \cite{10.1145/361573.361582,golub1979hessenberg})
which uses the Schuhr decomposition of $\mathfrak{L}$ to find the
exact solution. This is the method employed by the function \textit{solve\_continuous\_lyapunov}
in Python or Matlab for instance. The uniqueness of the solution found
by this algorithm requires the matrix to have eigenvalues with strictly
negative real parts. For the operator $\mathfrak{L}_{\tau_{N}}$ used
later, this will be true. In the case of large matrices, the solution
can be approximated by a low rank solution using the Alternating-Direction-Implicit
method (ADI, see \textit{e.g.} \cite{lu1991solution,li2002low}).
This method relies on finite difference discretization and decouples
the matrix operator into simpler subproblems, typically by sequentially
solving triangular systems associated with alternating directions.
This approach reduces computational cost by avoiding direct inversion
of large matrices, making it efficient for high-dimensional problems.
In our case where the matrix $\mathfrak{L}$ is large but sparse,
this method can be very efficient.
\end{enumerate}

\subparagraph{Consequence: Resolution of the system with mass and energy dissipation\label{par:Consequence:-Resolution-with-mass-energy-dissipation}}

As remarked before, the solution of the Lyapunov equation is not unique
unless we fix a ``mass and energy condition''. This problem is analysed
in \cite{article2_brice_freddy} for one conservation law. In our
situation, the boundary conditions enable to fix a ``null mass and
energy'' for the solution, that is $\mathbf{1}^{T}.\tilde{B}=\bm{\omega}^{T}.\tilde{B}=0$.
with $\mathbf{1}=\left(1,\ldots,1\right)^{T}$ and $\bm{\omega}=\left(\omega_{1},\ldots,\omega_{I}\right)^{T}$.
This is the choice we will make. To find the solution with null mass
and energy, we will proceed similar to \cite{article2_brice_freddy},
by using mass and energy damping times $\tau_{N},\tau_{E}$, and applying
the following steps:
\begin{enumerate}
\item Diagonalize $\mathfrak{L}$: We denote $\lambda_{i}\leq0$ the eigenvalues
of $\mathfrak{L}$ (with $\lambda_{1}=\lambda_{2}=0$) respectively
associated to eigenvectors $v_{i}$. We denote $P$ a diagonalization
matrix such that we have $P=\left[\mathbf{1},\bm{\omega},v_{2},\ldots v_{I}\right]$
and $P^{-1}\mathfrak{L}P=\Lambda$ with $\Lambda=\text{Diag}\left[0,0,\lambda_{3},\ldots,\lambda_{I}\right]$.
Now we denote $\Lambda_{\tau_{N},\tau_{E}}=\text{Diag}\left[-\frac{1}{\tau_{N}},-\frac{1}{\tau_{E}},\lambda_{3},\ldots,\lambda_{I}\right]$
for any $\tau_{N},\tau_{E}>0$ and 
\[
\mathfrak{L}_{\tau_{N},\tau_{E}}=P\Lambda_{\tau_{N},\tau_{E}}P^{-1}.
\]
\item Solve the equation 
\begin{equation}
\mathfrak{L}_{\tau_{N},\tau_{E}}\tilde{B}+\tilde{B}\mathfrak{L}_{\tau_{N},\tau_{E}}^{T}=Q.\label{eq:discretized-Lyapunov-WT-tauN-tauE}
\end{equation}
Indeed, we can easily show that if $\tilde{B}$ is solution of $\mathfrak{L}\tilde{B}+\tilde{B}\mathfrak{L}^{T}=Q$
and if $\tilde{B}.\mathbf{1}=\mathbf{1}^{T}\tilde{B}=\bm{\omega}^{T}.\tilde{B}=\tilde{B}.\bm{\omega}=0$,
then $\tilde{B}$ is solution of $\mathfrak{L}_{\tau_{N},\tau_{E}}\tilde{B}+\tilde{B}\mathfrak{L}_{\tau_{N},\tau_{E}}^{T}=Q$
for any $\tau_{N},\tau_{E}$. Since the solution of such an equation
is unique because $\mathfrak{L}_{\tau_{N},\tau_{E}}$ has only strictly
negative eigenvalues, then we have found the unique solution with
null mass of $\mathfrak{L}\tilde{B}+\tilde{B}\mathfrak{L}^{T}=Q$.
We have to choose wisely the values of $\tau_{N},\tau_{E}$, as illustrated
in \cite{article2_brice_freddy} for the SEP model. As pointed out
in \cite{article2_brice_freddy}, it seems relevant to choose the
characteristic times of the order of magnitude of minus the inverse
of the smallest nonzero eigenvalue. This enables to have a correct
solution with weak mass and energy. We will choose it as a first guess
and then we could refine the resolution by doing the same type of
analysis as in \cite{article2_brice_freddy}. We start by fixing the
relevant values $\tau_{N}=\tau_{E}=0.4$ since this is the correct
order of magnitude since we have $\lambda_{3}=-2.8$. 
\end{enumerate}
More precisely, $\mathfrak{L}_{\tau_{N},\tau_{E}}$ is the operator
of a linearised dynamics $\partial_{t}X=\mathfrak{L}_{\tau_{N},\tau_{E}}X$
which has the same stationary solution as $\partial_{t}\delta n=\mathfrak{L}\delta n$
if the initial condition is chosen with zero mass. It dissipates mass
at rate $\tau_{N}$ and energy at rate $\tau_{E}$, without interfering
on the other modes. To recover the original solution $\delta n$,
we can simply solve $\partial_{t}X=\Lambda_{\tau_{N},\tau_{E}}X$
and transform with $P$ : $\delta n=P^{-1}X$.

\subsubsection{Nondimensional Lyapunov equation and universal scaling of the solution}

In the following, we aim at resolving the Lyapunov equation to amount
for the Gaussian fluctuations around the paradigmatic Kolmogorov-Zakharov
spectra. To solve the equation (\ref{eq:Lyapunov-B-2CL+BCs}) in the
most general way as possible, we write it with dimensionless parameters.
We introduce $\omega_{0}$ a characteristic frequency, which we will
set to the inertial range width: $\omega_{0}=\omega_{\text{max}}-\omega_{\text{min}}$.
We denote $\tilde{\omega}=\omega/\omega_{0}$. We denote $n_{E}\left(\omega\right)=A_{E}\omega^{-x_{E}}$
and $n_{N}\left(\omega\right)=A_{N}\omega^{-x_{N}}$the Kolmogorov-Zakharov
spectra, respectively with a constant energy flux $J_{N}$ alone (case
1) and a constant wave action current $J_{N}$ alone (case 2). We
recall that we have 
\begin{align*}
x_{E} & =\frac{s-2}{3},A_{E}=\left(\frac{J_{E}}{x_{E}\left(x_{E}-1\right)}\right)^{1/3}\\
x_{N} & =\frac{s-3}{3},A_{N}=\left(-\frac{J_{N}}{x_{N}\left(x_{N}-1\right)}\right)^{1/3}
\end{align*}
and we can easily show that a nondimensional version of the Lyapunov
equation (\ref{eq:Lyapunov-B-2CL+BCs}) with parameters (\ref{eq:WT-expressions-g-mu-E})
is 
\[
\tilde{\mathscr{L}}\tilde{B}=\left(\tilde{\mathcal{K}^{*}}\left(\tilde{\omega_{1}}\right)+\tilde{\mathcal{K}^{*}}\left(\tilde{\omega_{2}}\right)\right)\delta''\left(\tilde{\omega_{1}}-\tilde{\omega}_{2}\right)
\]
where $\tilde{\mathscr{L}}=\partial_{\tilde{\omega}_{1}}^{2}\left(\tilde{L}_{\tilde{\omega}_{1}}\cdot\right)+\partial_{\tilde{\omega}_{2}}^{2}\left(\tilde{L}_{\tilde{\omega}_{2}}\cdot\right)$
and we introduce different nondimensionalizations for the case 1 with
energy flux and the case 2 with wave action flux. For the case 1,
we have
\[
\begin{cases}
\tilde{B}\left(\tilde{\omega_{1}},\tilde{\omega_{2}}\right) & =A_{E}^{-2}\omega_{0}^{2x_{E}+1}B\left(\omega_{1},\omega_{2}\right)\\
\tilde{L}_{\tilde{\omega}} & =A_{E}^{-2}\omega_{0}^{-x_{E}}L_{\omega_{0}\tilde{\omega}}\\
\tilde{\mathcal{K}}^{*}(\tilde{\omega}) & =-4x_{E}(x_{E}-1)\,\tilde{\omega}^{-x_{E}}
\end{cases}
\]
while for the case 2, we have
\[
\begin{cases}
\tilde{B}\left(\tilde{\omega_{1}},\tilde{\omega_{2}}\right) & =A_{N}^{-2}\omega_{0}^{2x_{N}+1}B\left(\omega_{1},\omega_{2}\right)\\
\tilde{L}_{\tilde{\omega}} & =A_{N}^{-2}\omega_{0}^{-x_{N}-1}L_{\omega_{0}\tilde{\omega}}\\
\tilde{\mathcal{K}}^{*}(\tilde{\omega}) & =-4x_{N}(x_{N}-1)\,\tilde{\omega}^{-(x_{N}+1)}
\end{cases}
\]
 give the nondimensional functions. We observe the following universal
scaling due to this dimensional analysis:
\begin{itemize}
\item We have 
\begin{equation}
B\sim A_{E}^{-2}\omega_{0}^{2x_{E}+1}\text{ or }B\sim A_{N}^{-2}\omega_{0}^{2x_{N}+1}\label{eq:asymp-dependencies-B}
\end{equation}
 respectively for case 1 and case 2. In other words, it scales with
the current at the power $-2/3$ and with the inertial range's width
$\omega_{0}$ to the power $2x_{E}+1$ or $2x_{E}+1$, respectively. 
\item The solution $\tilde{B}$ is invariant by inertial range dilatation
$\left[\omega_{\text{min}},\omega_{\text{max}}\right]\mapsto\left[\lambda\omega_{\text{min}},\lambda\omega_{\text{max}}\right]$
for any constant $\lambda$.
\end{itemize}
In the following, we resolve nondimensional solutions for both case
1 with energy flux and case 2 with wave action flux. 

\subsubsection{Numerical results \label{subsec:Computation-long-correl-WT-R=0000E9sultats}}

\paragraph{}

\begin{figure}[H]
\centering{}%
\begin{tabular}{@{}cc}
\includegraphics[totalheight=0.4\textwidth]{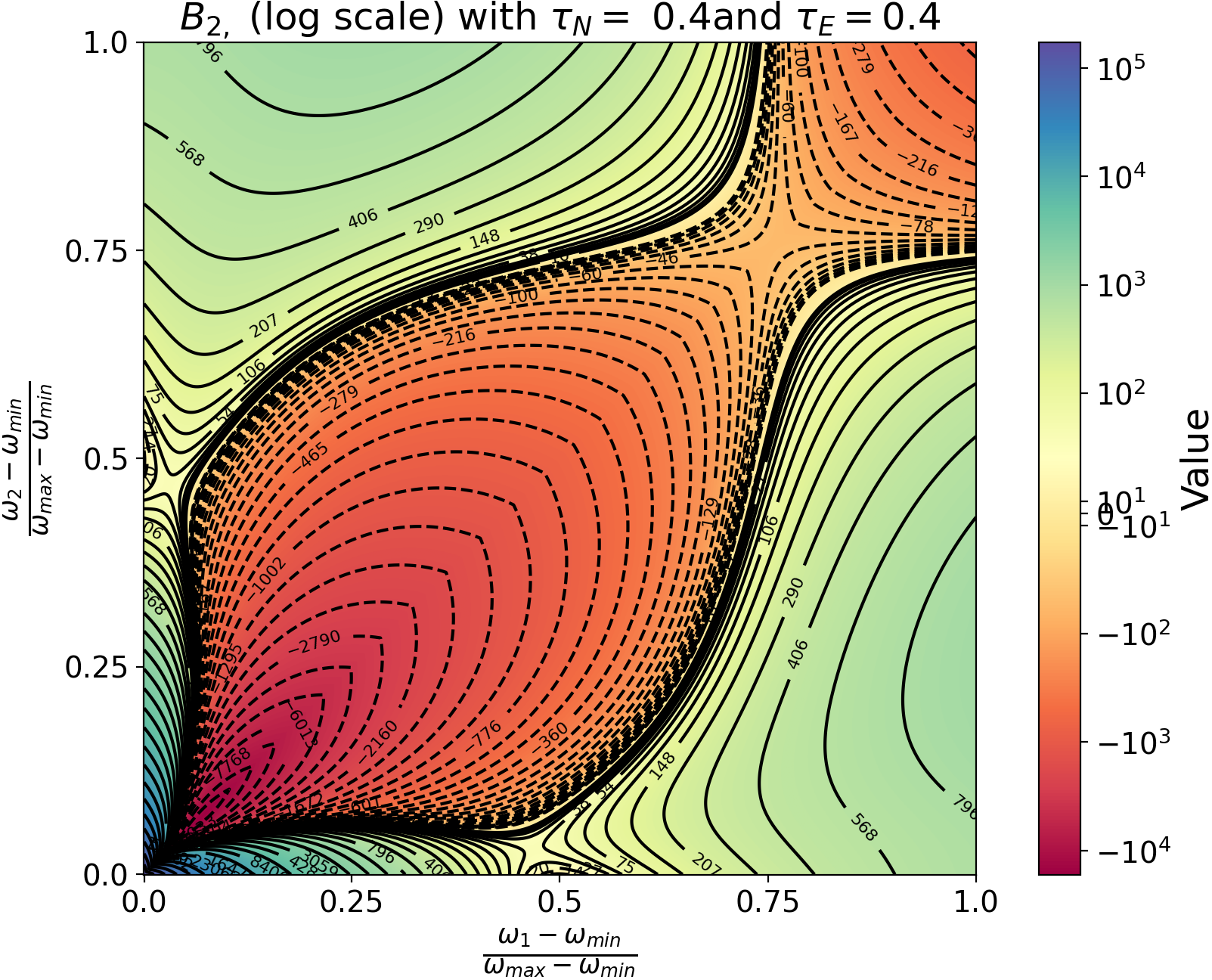} & \includegraphics[totalheight=0.4\textwidth]{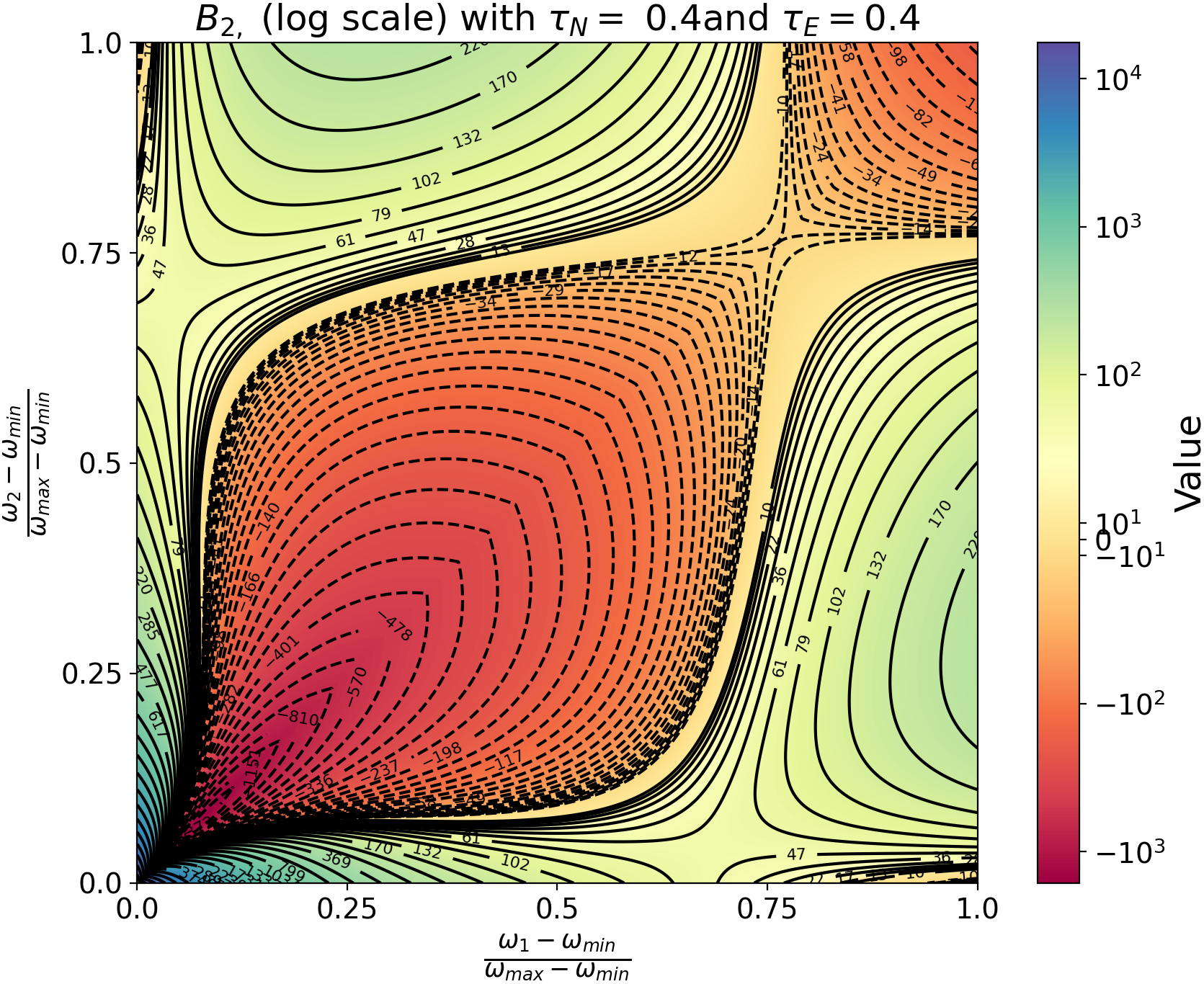}\tabularnewline
\textbf{(a)} & \textbf{(b)}\tabularnewline
\end{tabular}\caption{{\small\textbf{Numerical resolution for the flux-driven long-range
correlations in wave turbulence model with fixed-flux boundary conditions}}{\small{}
}\protect \\
{\small In this figure, we show the numerical solutions for the flux-driven
long-range correlations $B$ for wave turbulence (symlog color scale)
with the method explained in paragraph \ref{par:Consequence:-Resolution-with-mass-energy-dissipation}.
An arbitrary choice of parameters (exponent $s=7$ and an inertial
range $\left[\omega_{\text{min}},\omega_{\text{max}}\right]$ with
$\omega_{\text{min}}/\omega_{\text{max}}=0.1$) has been made. The
figure (a) (resp. (b)) corresponds to Gaussian fluctuations computed
around the Kolmogorov-Zakharov spectrum with an energy current $J_{E}$
alone (resp a wave action $J_{N}$ current alone). \label{fig:WT-B-numerical}}}
\end{figure}

\begin{figure}[H]
\centering{}\includegraphics[totalheight=8cm]{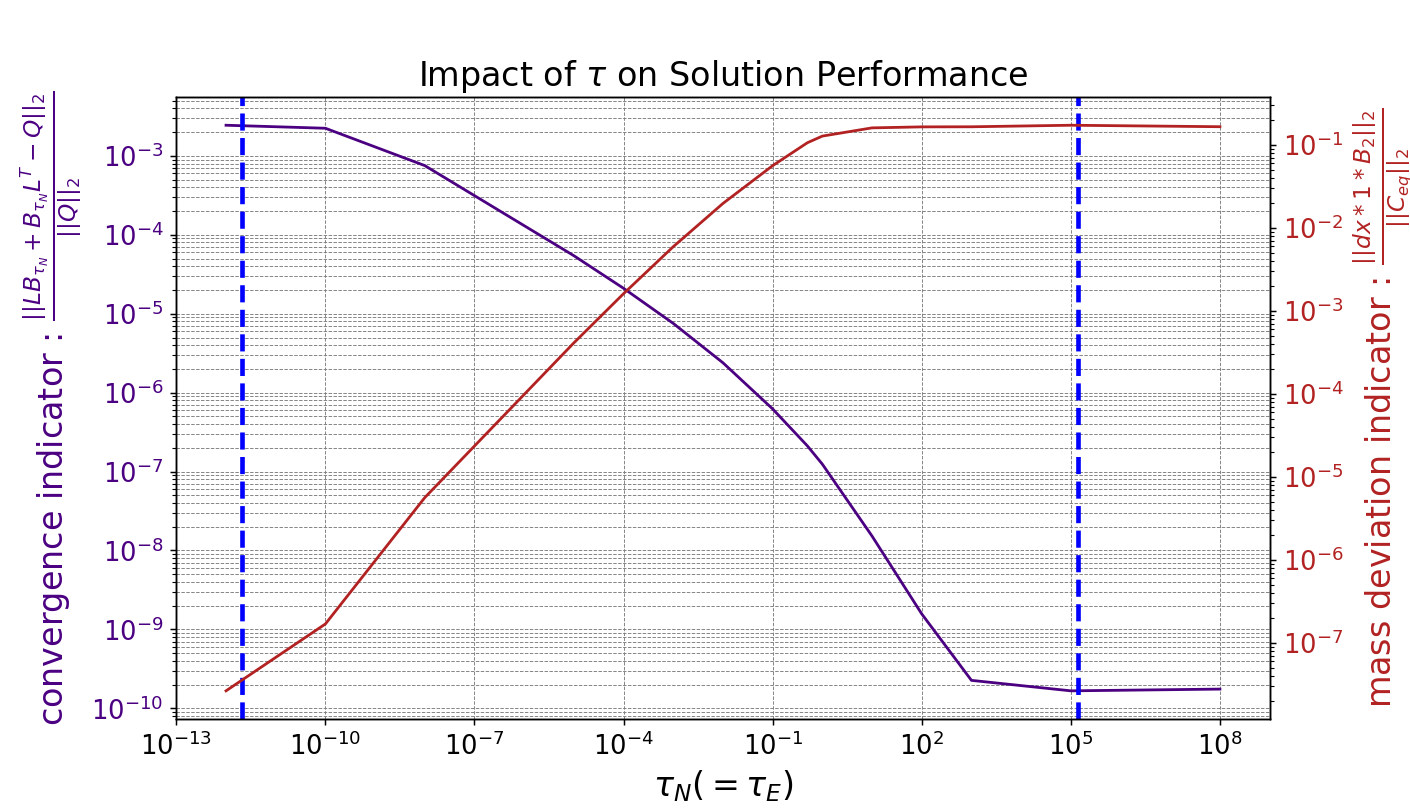}\\
\textbf{(a)}\\
\includegraphics[totalheight=8cm]{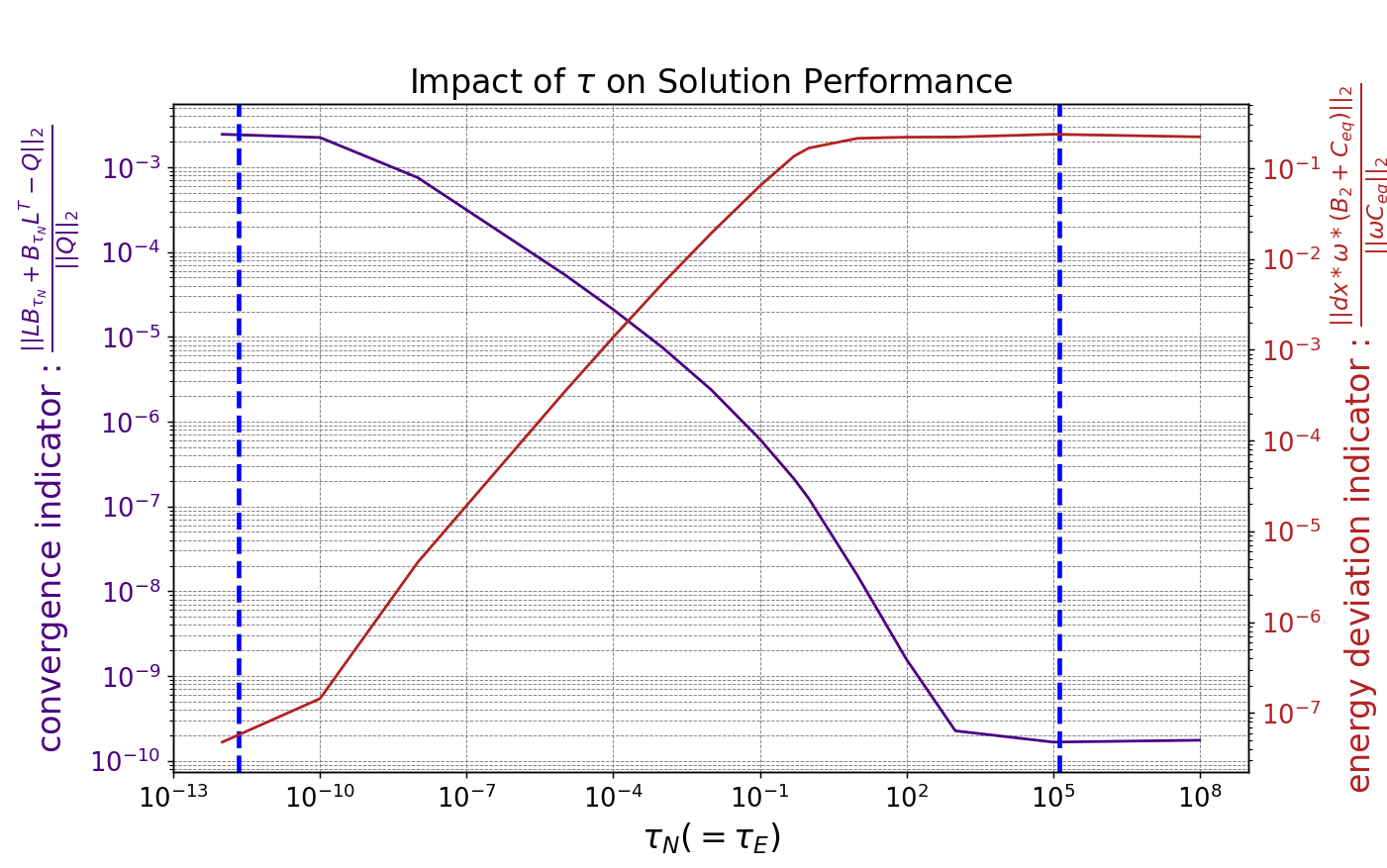}\\
\textbf{(b)}\caption{{\small\textbf{Evolution of the convergence indicator and mass indicator
with dissipation rates $\tau_{N},\tau_{E}$}}{\small{} }\protect \\
{\small In this figure, we show the evolution of convergence index
$c\left(B\right)$, the mass index $m\left(B\right)$ (top figure
(a)) as well as the energy index $e\left(B\right)$ (bottom figure
(b)) with $\tau_{N}$ ($\tau_{E}$ is chosen equal to $\tau_{N}$
for simplicity, although these two parameters could be chosen independently).
The blue dotted vertical lines represent the values of minus the inverse
of the minimum and maximum eigenvalues of the matrix $\mathfrak{L}$.
We observe that the convergence index drops when $\tau_{N}$ approaches
the maximum eigenvalue. A good value for $\tau_{N}$ seems to be approximately
$10^{-1}$, where the convergence indicator becomes very small ($<10^{-7}$)
and the mass and energy indicators still remains limited ($\sim5.10^{-2}$
for both). The choice of $\tau_{N}$ is in any case a compromise between
good convergence and weak mass. We remarked that if we tried to diminish
the mass indicator further by decreasing $\tau_{N}$, the shape of
the correlations (Figure \ref{fig:WT-B-numerical}) is visually deformed,
so it seemed us preferable to keep very small values for convergence
indicator instead of demanding a very small mass and energy. \label{fig:WT-B-numerical-indicators-vs-tauN-tauE}}}
\end{figure}

The flux-driven long-range correlation function $B$, solution of
the preceding section, is presented in Figure \ref{fig:WT-B-numerical}.
We can first check that this function is close to the solution of
the discretised Lyapunov equation by verifying that the residual of
the equation (\ref{eq:discretized-Lyapunov-WT-tauN-tauE}) takes small
values. To do it, we monitor the normalized norm of the residual 
\[
c\left(B\right):=\frac{\left\Vert \mathfrak{L}B+B\mathfrak{L}^{T}-Q\right\Vert _{2}}{\left\Vert Q\right\Vert _{2}}.
\]
As it is normalized, the convergence is satisfactory when it is much
smaller than 1. We can also check that the mass and energy indices
\begin{align*}
m\left(B\right) & :=\frac{\left\Vert \left[\begin{array}{ccccc}
\sum_{i}B_{i1}\Delta\omega+C_{eq}\left(\omega_{1}\right) &  & ,\ldots, & \sum_{i}B_{iI}\Delta\omega+C_{eq}\left(\omega_{I}\right)\end{array}\right]\right\Vert _{2}}{\left\Vert \left[C_{eq}\left(\omega_{i}\right)\right]_{1\leq i\leq I}\right\Vert _{2}}\\
e\left(B\right) & :=\frac{\left\Vert \left[\begin{array}{ccccc}
\sum_{i}\omega_{i}B_{i1}\Delta\omega+\omega_{1}C_{eq}\left(\omega_{1}\right) &  & ,\ldots, & \sum_{i}\omega_{i}B_{iI}\Delta\omega+\omega_{I}C_{eq}\left(\omega_{I}\right)\end{array}\right]\right\Vert _{2}}{\left\Vert \left[\omega_{i}C_{eq}\left(\omega_{i}\right)\right]_{1\leq i\leq I}\right\Vert _{2}}
\end{align*}
take values much smaller than 1. These two indices should be as close
to 0 as possible. With the choice $\tau_{N}=\tau_{E}=0.4$, we find
a solution $B$ verifying 
\begin{align*}
c\left(B\right) & =3.4\times10^{-7}\\
m\left(B\right) & =8.2\times10^{-2}\\
e\left(B\right) & =1.0\times10^{-1}
\end{align*}
which is quite satisfactory as they are all much smaller than 1. We
note that the mass and energy indicators are improved by an order
of magnitude compared to the solution without dissipation. The solution
is always a compromise between weak mass/energy and weak convergence
indicator (see \cite{article2_brice_freddy}). We can investigate
the optimal choice for $\tau_{N},\tau_{E}$. The joint evolution of
the three indices are presented in Figure \ref{fig:WT-B-numerical-indicators-vs-tauN-tauE}. 


\begin{figure}[H]
\centering{}%
\begin{tabular}{@{}cc}
\includegraphics[totalheight=0.4\textwidth]{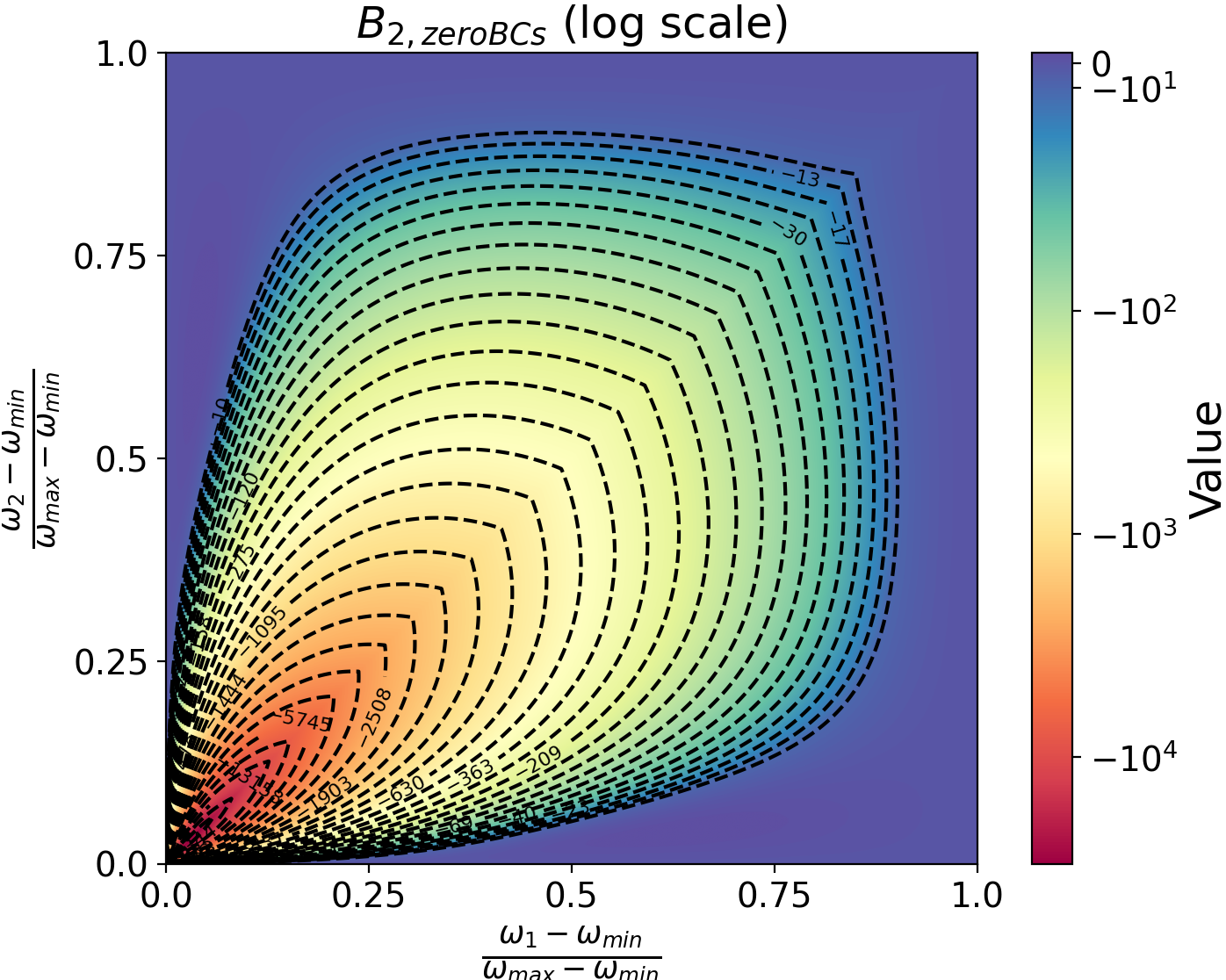} & \includegraphics[totalheight=0.395\textwidth]{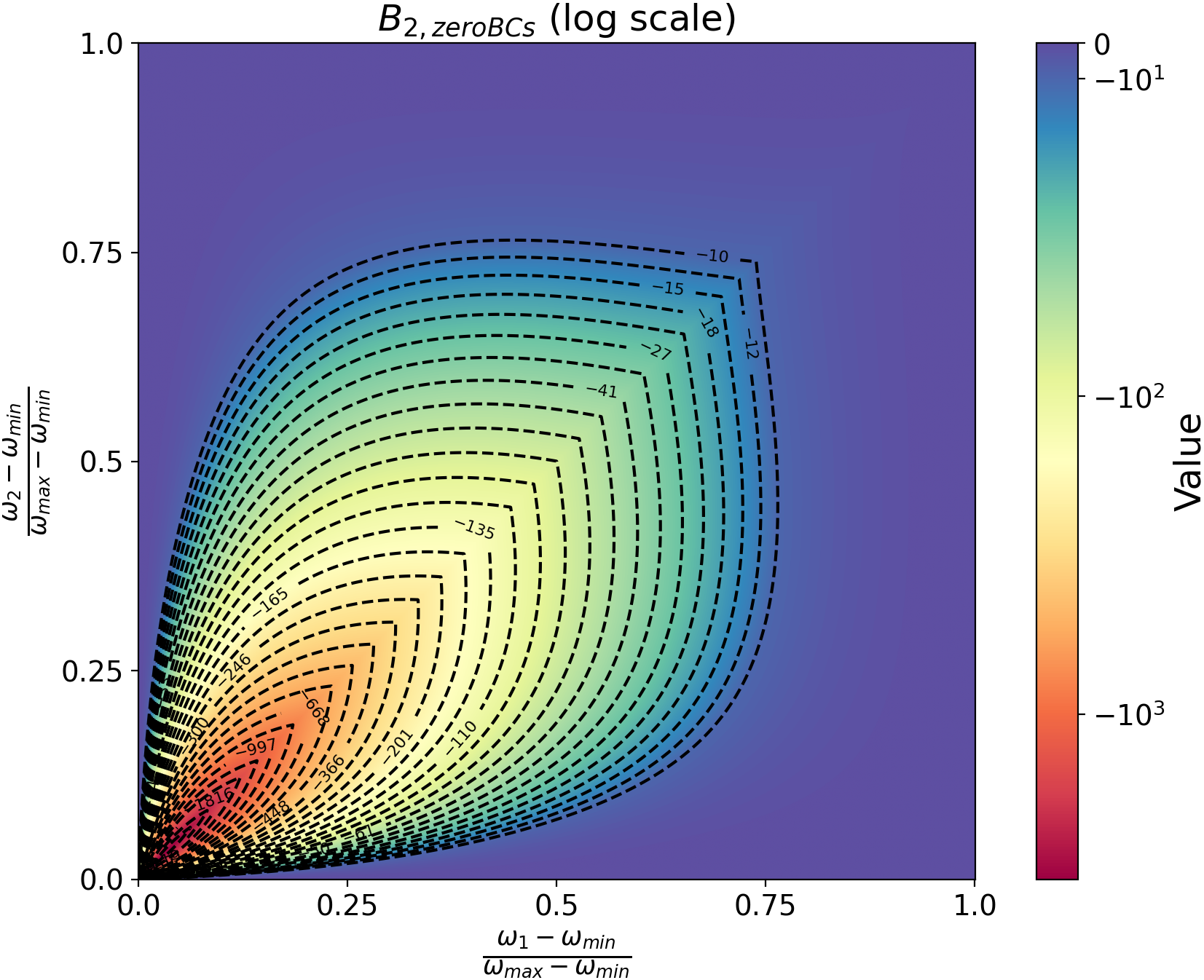}\tabularnewline
\textbf{(a)} & \textbf{(b)}\tabularnewline
\end{tabular}\caption{{\small\textbf{Numerical resolution for the flux-driven long-range
correlations in wave turbulence model with fixed-density boundary
conditions}}{\small{} }\protect \\
{\small In this figure, we show the solution of the method explained
in paragraph \ref{par:Consequence:-Resolution-with-mass-energy-dissipation}
with null boundary conditions for $B$. The color scale is in log
scale. The figure (a) (resp. (b)) corresponds to Gaussian fluctuations
computed around the Kolmogorov-Zakharov spectrum with an energy current
$J_{E}$ alone (resp a wave action $J_{N}$ current alone). \label{fig:WT-B-numerical-zeroBCs}}}
\end{figure}

Finally, as a complement, we can try other boundary conditions, as
done in the SEP model. We remarked for SEP that the boundary conditions
greatly impact the shape of the flux-driven correlation function $B$,
as might be expected. For wave turbulence, although the boundary conditions
are fundamentally flux-type, we can try to fix the density at the
boundary and see the impact on the correlation function. It can be
done easily by changing the operator $\mathfrak{L}$ so that it reflects
null boundary values for $B$. This is presented in Figure \ref{fig:WT-B-numerical-zeroBCs}.
We see that the shape of the correlations are impacted dramatically,
all the more as we get further off-diagonal. 

In this section, we have given an example of resolution for the Lyapunov
equation in wave turbulence. We have chosen a set of parameters, because
the aim is not an exhaustive review of the different shapes which
are possible, but rather an illustration of the methodology we have
developed. Let us make a few discussions on the key characteristics
we have observed.

\subsubsection{Qualitative properties of the flux-driven two point correlation function\label{subsec:Qualitative-properties-B}}

In this section, we discuss the results and figures obtained in previous
section \ref{subsec:Computation-long-correl-WT-R=0000E9sultats}.
We share their key physical insights. 

First, we do observe flux-driven long-range correlations with a nontrivial
structure. As pointed out in \cite{article2_brice_freddy}, this structure
is different depending on the boundary conditions, be it for one or
for two conservation laws. This is one of our key results. In the
light of the figures we obtain, we emphasize the importance of boundary
conditions when computing bulk fluctuations and correlation functions
in physical systems. To our knowledge, this phenomenon has not been
pointed out even for the SEP model, where the differences between
both figures of \cite{article2_brice_freddy} are striking even close
to the diagonal. For wave turbulence, these differences appear even
when considering large inertial ranges (in some unshowed numerical
examples, we used a ratio up to $\omega_{\text{min}}/\omega_{\text{max}}=2\times10^{-4}$,
and ), where we could intuitively think that the correlation structure
depends weakly on the boundary conditions. Although this difference
is interesting to note when the inertial range is wide, our numerical
method is all the more relevant as the inertial range is narrow. This
method is indeed particularly useful to resolve correlations in a
part of the whole inertial range, where the natural boundary conditions
are very naturally flux-type.

Second, both figures \ref{fig:WT-B-numerical} and \ref{fig:WT-B-numerical-zeroBCs}
show a concentration of the correlations for smaller values of $\omega$
(bottom-left corner). This part of the figures \ref{fig:WT-B-numerical}
and \ref{fig:WT-B-numerical-zeroBCs} remain quite similar although
the rest differ significantly.

Third, the non-uniqueness of the solution of the Lyapunov equation
with flux-type boundary conditions forces to fix a value for component
of the solution in the direction of the mass and the energy vectors
$\bm{1}$ and $\bm{\omega}$. This is the discretised counterpart
to the fact of fixing the value of the integral of the solution, as
pointed out in the ``mass and energy'' constraints for the cumulant,
(see \cite{article2_brice_freddy}). 

Fourth, the sign of the solution is known for fixed-density boundary
conditions. The operator of the Lyapunov equation being elliptic,
it has a negative kernel and thus the solution has the opposite sign
as the second member. This property does not apply to the fundamental
solution with flux-type boundary conditions. In this case, the sign
depends on the global mass and energy conditions fixing the value
of the integral of the solution.

Fifth, the solution is a continuous function (the level lines are
continuous) and has discontinuous partial derivatives on the diagonal
$\omega_{1}=\omega_{2}$ (the level lines are not differentiable on
the diagonal). 

Lastly, as already mentioned, the solution has a known scaling with
the physical constants: $B\sim A_{E}^{-2}\omega_{0}^{2x_{E}+1}$ or
$B\sim A_{N}^{-2}\omega_{0}^{2x_{N}+1}$ respectively for case 1 with
an energy current only and case 2 with a wave action current only.
In other words, it scales with the current at the power $-2/3$ and
with the inertial range's width $\omega_{0}$ to the power $2x_{E}+1$
or $2x_{E}+1$, respectively.

\subsection{Scaling properties of the homogeneous flux-driven two-point correlation
function \label{subsec:Scaling-properties-of-homogeneous-solution}}

In this section, we briefly discuss the solution of (\ref{eq:Lyapunov-B-2CL+BCs})
for wave turbulence and its scaling properties. We obtain the analytic
solution of the homogeneous equation and its scaling for $\omega_{1}\gg\omega_{2}$
as power laws. As already mentioned, we computed the particular solution
with a null mass and null energy. For a more complete resolution of
the Lyapunov equation and to find more general solutions with arbitrary
mass and energy, we can add an homogeneous solution which we compute
explicitly in this section.

Our goal is to find the solution $B_{h}$ of the homogeneous Lyapunov
equation and the scaling of the solution for $\omega_{1}\gg\omega_{2}$.
We consider the homogeneous Lyapunov equation over a subdomain $\tilde{\mathcal{D}}\subseteq\mathcal{D}$:
\[
\begin{cases}
\partial_{\omega_{1}}^{2}\left(L_{\omega_{1}}B_{h}\right)+\partial_{\omega_{2}}^{2}\left(L_{\omega_{2}}B_{h}\right)=0\\
\begin{cases}
L_{\omega_{1}}B_{h}\left(\omega_{1},\omega_{2}\right) & =0\\
\partial_{\omega_{1}}\left(L_{\omega_{1}}B_{h}\left(\omega_{1},\omega_{2}\right)\right) & =0
\end{cases} & \text{ for }\omega_{1}\in\partial\tilde{\mathcal{D}}\\
\text{ same symmetrized boundary conditions } & \text{ for }\omega_{2}\in\partial\tilde{\mathcal{D}}
\end{cases}
\]
We can find all the solutions by separation of variables, obtain an
Euler differential equation which is analytically resoluble, and finally
get the homogeneous solution around KZ spectra $n=A\omega^{-x}$:
\begin{equation}
B_{h}\left(\omega_{1},\omega_{2}\right)=n^{2}\left(\omega_{1}\right)n^{2}\left(\omega_{2}\right)\sum_{i,j=1}^{2}a_{ij}\omega_{1}^{r_{i}}\omega_{2}^{r_{j}}\label{eq:solution-homogeneous-lyapunov-B}
\end{equation}
where $\left(a_{ij}\right)_{1\leq ij\leq2}$ are four constants and
$r_{1}=-r_{2}=\sqrt{x\left(x-1\right)}$.\footnote{Actually the case of two complex-conjugate roots ($x\left(x-1\right)<0$),
the sum of $\omega^{\pm i\theta}$ should be interpreted as $e^{\pm i\theta\ln\omega}$
leading to solutions of the form $a_{1}\cos\left(\theta\ln\omega\right)+a_{2}\sin\left(\theta\ln\omega\right)$
because the constants should be chosen to vanish imaginary parts.
The square root should be considered in the sense $r_{k}=\pm i\sqrt{-4x\left(x-1\right)}$
if $x\left(x-1\right)<0$.} The homogeneous Lyapunov equation is a function space with four
parameters, with all the solutions thus given by (\ref{eq:solution-homogeneous-lyapunov-B}).
Since we only look for symmetric solutions in $\omega_{1}\leftrightarrow\omega_{2}$,
we have $a_{21}=a_{12}$, $a_{11}=a_{22}$ and thus only two free
parameters. These solutions (\ref{eq:solution-homogeneous-lyapunov-B})
bear a mass and an energy, thus the constants $a_{11}$ and $a_{12}$
are determined by the value of the mass and energy of the solution. 

This result (\ref{eq:solution-homogeneous-lyapunov-B}) also leads
to asymptotic power laws for $\omega_{1}\gg\omega_{2}$:
\begin{equation}
B_{h}\left(\omega_{1},\omega_{2}\right)\underset{\omega_{1}\gg\omega_{2}}{\sim}\omega_{1}^{-2x+2\sqrt{x\left(x-1\right)}}A^{4}\omega_{2}^{-2x}\left(a_{11}\omega_{2}^{2\sqrt{x\left(x-1\right)}}+a_{12}\omega_{2}^{-2\sqrt{x\left(x-1\right)}}\right).\label{eq:asymp-estimates-B_h}
\end{equation}

\subsection{Analysis of the Majda--McLaughlin--Tabak (MMT) model instability
with the inhomogeneous large deviation theory}

In the conclusion of \cite{guioth_path_2022}, the Majda--McLaughlin--Tabak
(MMT) model \cite{majda1997one,cai2001dispersive,chibbaro2017weak,newell2012spontaneous}
was mentioned. In particular, the article \cite{newell2012spontaneous}
showed that for some 1D models with 4-wave interactions such as the
MMT model, the KZ solutions were unstable to spatially-inhomogeneous
perturbations. The inhomogeneous kinetic equation
\[
\partial_{t}n\left(\bm{x},\bm{k},t\right)+v_{g}\left(\bm{k}\right)\cdot\nabla_{\bm{x}}n\left(\bm{x},\bm{k},t\right)=\mathcal{I}\left[n\right]\left(\bm{x},\bm{k},t\right)
\]
with 
\[
\mathcal{I}\left[n\right]\left(\bm{x},\bm{k}_{1},t\right)=4\int_{\mathcal{R}_{3}^{d}\left(\bm{k}_{1}\right)}|W_{1234}|^{2}n_{1}n_{2}n_{3}n_{4}\left[\frac{1}{n_{1}}+\frac{1}{n_{2}}-\frac{1}{n_{3}}-\frac{1}{n_{4}}\right]
\]
was reported unstable close to the KZ power law solutions $n_{KZ}\sim\omega^{-x_{KZ}}$.
The instability does not exist in the homogeneous case and is fundamentally
related to the transport term $v_{g}\left(\bm{k}\right)\cdot\nabla_{\bm{x}}n\left(\bm{x},\bm{k},t\right)$.
The emerging solution in the numerical computations is breaking the
translation symmetry and wave trains with high amplitude merge and
form coherent wave packets traveling with the group velocity \cite{newell2012spontaneous}.
Previous works have shown a linearly unstable mode for the inhomogeneous
kinetic theory. In this section, we show that this leads to an instability
in the Lyapunov equation for the inhomogeneous large deviation theory
\cite{Guioth_2024}, thus showing that the inhomogeneous framework
is able to explain this instability. Indeed, in a real system, there
exist a microscopic dynamics, which in the inhomogeneous framework
leads to inhomogeneous random fluctuations triggering this instability.

Let us denote the operator of the linearised dynamics $\mathcal{L}_{\text{in}}=\mathcal{L}\left[n\right]+\mathcal{L}_{\text{tr}}$,
where
\[
\mathcal{L}\left[n\right]=\frac{\delta}{\delta n}\mathcal{I}\left[n\right]
\]
 and
\[
\mathcal{L}_{\text{tr}}\cdot=-v_{g}\left(\bm{k}\right)\cdot\nabla_{\bm{x}}\cdot
\]
is the transport part of the linearised dynamics. $v_{g}=\nabla_{\bm{k}}\omega$
is the group velocity. By definition, if we linearise the dynamics
of $n$ close to a given value $n_{0}$ and denote $\delta n=n-n_{0}$
we have 
\[
\partial_{t}\delta n\left(\bm{x},\bm{k},t\right)=\mathcal{L}_{\text{in}}\left[n_{0}\right]\delta n+\mathcal{O}\left(\delta n^{2}\right).
\]
 The article \cite{newell2012spontaneous} showed that close to the
homogeneous KZ solution $n_{0}=n_{KZ}$, there exist unstable directions
to this equation, \textit{i.e.} $\mathcal{L}_{\text{in}}$ can be
linearly unstable and has eigenvalues with positive real part. Let
us denote it $z\in\mathbb{C}$ such that $\text{Re}\left(z\right)>0$.
We also denote $u$ the associated eigenvector, such that
\[
\mathcal{L}_{\text{in}}\left[n_{KZ}\right]u\left(\bm{x},\bm{k},t\right)=zu\left(\bm{x},\bm{k},t\right).
\]
Since $n_{KZ}$ is a stationary solution to the homogeneous kinetic
equation $\mathcal{L}\left[n_{KZ}\right]$ is independent of time.
In addition, $\mathcal{L}_{\text{tr}}\cdot=-v_{g}\left(\bm{k}\right)\cdot\nabla_{\bm{x}}$
is also independent on time. We deduce that $\mathcal{L}_{\text{in}}\left[n_{KZ}\right]$
is independent on time and that we can choose time-independent eigenvectors:
$u\left(\bm{x},\bm{k}\right)$ does not depend on time. 

This explains the possible development of instabilities. However,
since no model accounting for how spatially-inhomogeneous perturbations
could arise, the origin of this spontaneous symmetry breaking was
still unclear. In the following, we show that spontaneous fluctuations
in the inhomogeneous theory could be the origin of this symmetry breaking.
As presented in \cite{guioth_path_2022}, even corrections to the
leading-order term in the kinetic theory would preserve the symmetries
of the system: if the initial conditions are spatially homogeneous,
the average dynamics would preserve this homogeneity. However, the
inhomogeneous large deviation theory derived in \cite{Guioth_2024}
exhibits a large deviation principle with a quadratic large deviation
Hamiltonian. The result is that the wave spectrum $n$ behave as in
the effective dynamics in the form of a Langevin equation with small
noise
\[
\text{d}n\left(\bm{x},\bm{k},t\right)+v_{g}\left(\bm{k}\right)\cdot\nabla_{\bm{x}}n\left(\bm{x},\bm{k},t\right)\text{d}t=\mathcal{I}\left[n\right]\left(\bm{x},\bm{k},t\right)\text{d}t+\sqrt{2\varepsilon}\sigma\left[n\right]\text{d}W\left(\bm{x},\bm{k},t\right)
\]
where $\varepsilon$ is the large deviation parameter, $\sigma$ is
an operator giving the noise covariance, acting on $\bm{k}$. $W$
is a Wiener process such that 
\[
\mathbb{E}\left[\text{d}W\left(\bm{x}_{1},\bm{k}_{1},t_{1}\right)\text{d}W\left(\bm{x}_{2},\bm{k}_{2},t_{2}\right)\right]=\delta\left(\bm{x}_{1}-\bm{x}_{2}\right)\delta\left(\bm{k}_{1}-\bm{k}_{2}\right)\delta\left(t_{1}-t_{2}\right)\text{d}\bm{x}_{1}\text{d}\bm{k}_{1}\text{d}t_{1}.
\]
The realization of the noise $\text{d}W$ is thus potentially a source
of translation symmetry breaking. This source, although small with
amplitude of the order $\sqrt{\varepsilon}$, is actually the leading
order term exhibiting a symmetry breaking.

If this is true, we should observe that the action of the dynamics
has an unstable extremum when evaluated in the KZ solutions. Since
the second order cumulant is the inverse of the second derivative
of the action, we should find values where this second order cumulant
diverges. Let us investigate the dynamical Lyapunov equation verified
by the second order cumulant. To simplify, we consider the inhomogeneous
isotropic local model exposed in Appendix \ref{subsec:Appendix-inhomogeneous-WT}.
We define the second order cumulant $C_{2}\left(\omega_{1},\omega_{2},\bm{x},t\right)=\mathbb{E}\left[\delta N\left(\omega_{1},\bm{x},t\right)\delta N\left(\omega_{2},\bm{x},t\right)\right]$,
where $\delta N=\frac{N-N_{KZ}}{\sqrt{\varepsilon}}$ with $N_{KZ}$
the KZ solution. As shown in Appendix \ref{subsec:Appendix-inhomogeneous-WT},
the Lyapunov equation reads 
\begin{equation}
\partial_{t}C_{2}=-\mathscr{L}C_{2}+2A\left(\omega_{1},\omega_{2},\bm{x},t\right)\label{eq:Lyapunov-B-2CL-2-1}
\end{equation}
where the source term is $A\left(\omega_{1},\omega_{2},\bm{x},t\right)=\partial_{\omega_{1}}^{2}\partial_{\omega_{2}}^{2}\left[\mu\left[N_{KZ}\right]\left(\omega_{1},\bm{x}\right)\delta\left(\omega_{1}-\omega_{2}\right)\right]$
with $\mu\left[N\right]=n^{4}\omega^{s}$. The operator $\mathscr{L}$
reads $\mathscr{L}\cdot=\mathcal{L}_{\text{in}}+\mathcal{L}_{\text{in}}^{T}$
and $\mathcal{L}_{\text{in}}=\mathcal{L}\left[N\right]+\mathcal{L}_{\text{tr}}$
where
\[
\mathcal{L}_{\text{tr}}\cdot=-v_{g}\left(\bm{k}\right)\cdot\nabla_{\bm{x}}\cdot
\]
is the transport part of the linearised dynamics. $v_{g}=\nabla_{\bm{k}}\omega$
is the group velocity. As before, we denote $z\in\mathbb{C}$ its
eigenvalue with positive real value. We also denote $U$ the associated
eigenvector, such that
\[
\mathcal{L}_{\text{in}}\left[N_{KZ}\right]U\left(\bm{x},\omega\right)=zU\left(\bm{x},\omega\right)
\]
 and 
\[
U\left(\bm{x},\omega\right)\mathcal{L}_{\text{in}}^{T}\left[N_{KZ}\right]=z^{*}U\left(\bm{x},\omega\right).
\]
 Let us project $C_{2}$ along $U$: we denote 
\[
c_{U}\left(t\right):=\left\langle C_{2}U,U\right\rangle =\int\text{d}\omega_{1}\text{d}\omega_{2}\text{d}\bm{x}U\left(\bm{x},\omega_{2}\right)C_{2}\left(\bm{x},\omega_{1},\omega_{2},t\right)U\left(\bm{x},\omega_{1}\right)
\]
 such that, using Lyapunov equation:
\begin{align}
\partial_{t}c_{U} & =\left(z+z^{*}\right)c_{U}+2a_{U}\nonumber \\
 & =2\text{Re}\left(z\right)c_{U}+2a_{U}\label{eq:dt-cu-diff-eq}
\end{align}
 where $a_{U}$ is a constant given by
\[
a_{U}:=\left\langle AU,U\right\rangle =\int\text{d}\omega_{1}\text{d}\omega_{2}\text{d}\bm{x}U\left(\bm{x},\omega_{2}\right)\partial_{\omega_{1}}^{2}\partial_{\omega_{2}}^{2}\left[\mu\left[N_{KZ}\right]\left(\omega_{1},\bm{x}\right)\delta\left(\omega_{1}-\omega_{2}\right)\right]U\left(\bm{x},\omega_{1}\right).
\]
Since $A$ is a self-adjoint nonnegative operator, we know that $a_{U}\geq0$,
and $a_{U}=0$ is possible if and only if $U$ is in the kernel of
$A$, \textit{i.e.} if $U$ is in one of the eigenspaces associated
to the conserved quantities. The solution of (\ref{eq:dt-cu-diff-eq})
is asymptotically driven by the exponential rate $\text{Re}\left(z\right)$:
\[
c_{U}=\alpha e^{\text{Re}\left(z\right)t}-\frac{a_{U}}{\text{Re}\left(z\right)}
\]
where $\alpha\in\mathbb{C}$. Actually, the solution depends on the
initial condition. If we choose the initial wave spectrum to be equal
to $N_{KZ}$, we have $\delta N\left(t=0\right)=0$, $C_{2}\left(t=0\right)=0$,
thus $c_{U}\left(t=0\right)=0$ and 
\[
c_{U}\left(t\right)=\frac{a_{U}}{\text{Re}\left(z\right)}\left(e^{\text{Re}\left(z\right)t}-1\right)\underset{t\to\infty}{\sim}\frac{a_{U}}{\text{Re}\left(z\right)}e^{\text{Re}\left(z\right)t}.
\]
Since $\text{Re}\left(z\right)>0$, the solution can be diverging
at an exponential rate given by $\text{Re}\left(z\right)$. In this
case, we conclude that one of the components of $C_{2}$ is exponentially
diverging in time, at a rate given by $\text{Re}\left(z\right)$.
Hence, the KZ spectrum cannot be stable. 

In previous works, a linearly unstable mode for the inhomogeneous
kinetic theory. We have shown that In this section, we show that this
leads to an instability in the Lyapunov equation for the inhomogeneous
large deviation theory \cite{Guioth_2024}, thus showing that the
inhomogeneous framework is able to explain this instability. Indeed,
in a real system, there exist a microscopic dynamics, which in the
inhomogeneous framework leads to inhomogeneous random fluctuations
triggering this instability.

As a result, the kinetic theory for some models such as the MMT model
have been reported unstable to spatially-inhomogeneous perturbations.
In this section, we have shown that the large deviation theory adapted
to the inhomogeneous framework \cite{Guioth_2024} provides a theoretical
explanation to the emergence of such spatially-inhomogeneous fluctuations.
In particular, we have shown that the existence of an unstable mode
for the dynamics entails an unstable mode in the (inhomogeneous) Lyapunov
equation. Divergences are exponential with rate given by $\text{Re}\left(z\right)$,
where $z$ is the eigenvalue of $\mathcal{L}\left[N_{KZ}\right]$
with the largest positive real part. This phenomenon is caused by
the microscopic dynamics which is not homogeneous and leads at large
scale to intrinsic small random fluctuations which are spatially inhomogeneous.
Thus in real systems, the instability is doomed to develop anyway.

However, identifying the real behaviour of the spectrum and finding
the dynamical correlations, possible metastable states and transition
rates, would require to go beyond this linear stability analysis.
Indeed, the identified divergences of this paragraph are only valid
if the density deviations $\delta n$ are small. This would require
the computation of the full quasipotential, which is beyond the scope
of this work.

\setcounter{secnumdepth}{-1}   

\section{Conclusion and perspectives}

\setcounter{secnumdepth}{3} 

All in all, in this work, we have u sed new theoretical tools of statistical
physics to analyse fluctuations in wave turbulence. We have developed
a simplified large deviation principle for wave turbulence in the
case of isotropic and local interaction kernels, providing a complete
probabilistic description of the spectral density fluctuations. By
leveraging an extension of Macroscopic Fluctuation Theory (MFT) to
systems with two conservation laws (wave action and energy), we established
a theoretical framework that enables explicit computation of long-range
correlations in inertial ranges. The simplification of the large deviation
Hamiltonian for isotropic and local kernels leads to tractable partial
differential equations for the average spectrum and its cumulants,
facilitating both analytical and numerical analysis. In this respect,
it would be interesting to use this kind of approach to nonlocal models
that lead to effective diffusion models, which have been shown relevant
for waves which are dominated by highly nonlocal interactions, such
as internal waves \cite{lanchon2023energy}.

In this paper, we focused on the computation of the second-order cumulant,
but the employed method could be enlarged to higher-order cumulants.
In this model, similar to the second order case, out of equilibrium,
higher-order cumulants can be decomposed into a completely diagonal
contribution, and a part verifying elliptic partial-differential equations
of the fourth order. The solution regularity could similarly be examined
with the order of the second member.

A central result is the derivation of a new isotropic local large
deviation Hamiltonian, which allows for the explicit calculation of
Gaussian fluctuations around Kolmogorov--Zakharov spectra. We addressed
the numerical challenge posed by unresolved forcing and dissipation
scales by implementing flux-type boundary conditions, a method inspired
by the theoretical framework introduced in paper \cite{article2_brice_freddy}.
This approach enables the identification of flux-driven contributions
to correlations. Furthermore, we provided a detailed and validated
numerical method to compute long-range correlations, demonstrating
its applicability and replicability in more complex wave turbulence
scenarios. An insightful use of this method would be to explore further
the diversity of the solutions for the Gaussian fluctuations around
the nonequilibrium steady states. We could vary \textit{e.g. }the
reference steady state, the value of fluxes, the physical exponents
and constants, in order to check the homogeneity and asymptotic estimates
(\ref{eq:asymp-dependencies-B}), (\ref{eq:asymp-estimates-B_h})
for instance. Although these long-range correlations are for the moment
difficult to measure experimentally, some have tried in different
configurations \cite{campagne2019energy,deike2013etudes,aubourg20173},
and it would be valuable to compare quantitative refinements of the
present work with experiments. These necessary refinements include
quantitative adjustment of the physical coefficients (prefactors,
exponents), proper nondimensional expression, realistic modelling
of the source and dissipation terms, in addition to adapted inertial
ranges. This method could probably be made more efficient in solving
wider inertial ranges and avoiding numerical approximation of very
small numbers, by a direct resolution in log scale (setting $x=\log\omega$
and $\tilde{n}=\log n$). Additionally, the dimensional analysis in
order to find an \textit{a priori} form for the solution and simplify
the Lyapunov equation has also been considered. For example, an ansatz
of the type $B\left(\omega_{1},\omega_{2}\right)=\omega_{2}^{z}f\left(\frac{\omega_{1}}{\omega_{2}}\right)$
with $f$ a continuous function and $z$ a constant set by dimensional
analysis should be valid when $B$ is supposed independent of any
forcing processes. We claim that such an ansatz immediately turns
the Lyapunov equation into an ordinary linear differential equation
with nonconstant coefficients, which simplifies the resolution. Since
the aim of the numerical part of this work was more methodological
as a proof of concept, we did not pursue efficiency improvements too
far, but some substantial gains are achievable in different steps
of the process.

Finally, we showed that the instability of the Kolmogorov--Zakharov
solutions in some 1D models with 4-wave interactions such as the Majda--McLaughlin--Tabak
(MMT) model could be explained using the extension of this theory
to spatially-inhomogeneous wave turbulence. In this extension, the
primary source of translation-symmetry breaking is the weak noise
induced by the inhomogeneous large deviation principle. Actually,
even corrections to the kinetic equation at next-to-leading order
would by construction preserve the translation symmetry of the solution.
Yet, the realization of the Wiener process predicted by the large
deviation principle would however induce a small symmetry breaking
that could trigger the instability. This instability is conspicuous
using the Lyapunov equation over the second-order cumulant, whose
observed divergence in time is an indicator of the instability of
the Kolmogorov--Zakharov spectrum. One of the central insights is
that homogeneous wave turbulence theory, despite the absence of explicit
sources of inhomogeneity, sometimes fails to fully account for the
behavior of the wave spectrum. Adopting a framework that incorporates
inhomogeneous fluctuations \cite{Onuki_2023,Guioth_2024} therefore
appears essential. Further work would however be necessary to identify
the spatially-inhomogeneous steady states, and analyse the possible
multistability phenomenon along with the associated transition times
predicted by large deviation theory. In addition, other models exhibit
instabilities and remain to be studied, for example \cite{vladimirova2021turbulence}
where introducing an angular anisotropy is identified as a source
of instability. To study it, we should generalise our work to anisotropic
theory, which is accessible. All these analyses remain to be done
and require a more complete understanding of the quasipotential in
the inhomogeneous configuration. This analysis could be done, for
instance by numerical estimation of the quasipotential.

Other perspectives include the application of the inhomogeneous theory
to simple models of weather forecast models, where these small random
effects could be included to improve predictions in terms of variability.
Operational wave prediction models such as WAVEWATCH III and ECWAM
solve Hasselmann's kinetic equation for the mean spectrum, incorporating
empirical parameterisations of wave breaking and wind forcing, but
containing no term describing fluctuations of the spectrum around
its mean value. Our theory provides precisely this missing link: a
stochastic equation for the empirical spectrum whose drift is the
kinetic equation and whose noise amplitude is determined by the large
deviation principle.. In the inhomogeneous framework, this stochastic
correction is given by the large deviation parameter, defined as the
ratio of the typical wavelength by the typical spatial extension of
wavepackets in the inhomogeneous theory \cite{Guioth_2024}. It becomes
negligible for a single deterministic forecast, however may become
important when one is interested in the tails of the distribution
of spectral trajectories. In an operational context, this opens the
way to probabilistic forecasting of extreme spectral events of exceptionally
energetic or spatially focused sea states arising from anomalous spectral
evolution. Given an initial spectral condition, the action quantifies
the cost of a fluctuation path leading to a target rare sea state,
and the most probable path to that state can be identified directly,
without running large ensembles of realisations. A concrete implementation
would require solving the associated instanton equations for a prescribed
target sea state. This represents one of the most operationally relevant
applications of this framework. 

To describe more quantitatively the sea states, it is however necessary
to build a theoretical model going beyond the wave turbulence regime
to describe effectively the statistics of breaking waves. The article
\cite{wu2023breaking} has notably used multi-layer numerical models
to reproduce the statistics of breaking waves, thus adapting the tools
of shallow layer flows to describe greater wave amplitudes. Its quantitative
agreement between simulations and experimental measurements in regimes
where wave turbulence regime is not valid anymore makes it an interesting
idea to extend the existing theories to larger waves. 

\paragraph{Acknowledgement}

Brice Douet is partly funded by the Simons Foundation Award ID: 651475

\bibliographystyle{unsrt}
\bibliography{bibliography/Fbouchet_2024_10,bibliography/All-2024-10,bibliography/bibliography-all-bdouet,\string"/Users/bdouet/Sdrive/Partages_recus/Recherche Brice Freddy/Turbulence d'onde Brice/Notes de Brice/article/bibliography/bibliography-all-bdouet\string"}

\begin{thebibliography}{10}

\bibitem{guioth_path_2022}
Jules Guioth, Freddy Bouchet, and Gregory~L. Eyink.
\newblock Path large deviations for the kinetic theory of weak turbulence.
\newblock {\em Journal of Statistical Physics}, 189(2):20, November 2022.

\bibitem{Onuki_2023}
Yohei Onuki, Jules Guioth, and Freddy Bouchet.
\newblock Dynamical large deviations for an inhomogeneous wave kinetic theory:
  Linear wave scattering by a random medium.
\newblock {\em Annales Henri Poincar{\'e}}, 25(1), 2023.

\bibitem{Guioth_2024}
Jules Guioth, Yohei Onuki, Brice Douet, and Freddy Bouchet.
\newblock Path large deviations for inhomogeneous weak wave turbulence.
\newblock {\em To be submitted to J.Stat. Phys}, 2024.

\bibitem{article2_brice_freddy}
Brice Douet and Freddy Bouchet.
\newblock Dynamical large deviations for systems with two conservation laws and
  their long-range correlations.
\newblock {\em to be published}, 2026.

\bibitem{Cavaleri2007WaveModelling}
L.~Cavaleri, J.-H. G.~M. Alves, F.~Ardhuin, Alexander Babanin, M.~Banner,
  K.~Belibassakis, M.~Benoit, M.~Donelan, J.~Groeneweg, T.~H.~C. Herbers, P.~A.
  E.~M. Hwang, P.~A. E.~M. Janssen, T.~Janssen, I.~V. Lavrenov, R.~Magne, Jaak
  Monbaliu, Miguel Onorato, V.~Polnikov, D.~Resio, W.~E. Rogers, A.~Sheremet,
  J.~McKee Smith, H.~L.~van Tolman, G.~Van~Vledder, Judith Wolf, Ian Young, and
  WISE Group.
\newblock Wave modelling--the state of the art.
\newblock {\em Progress in Oceanography}, 75(4):603--674, 2007.

\bibitem{Ardhuin2010Semiempirical}
Fabrice Ardhuin, Erick Rogers, Alexander~V Babanin, Jean-Fran{\c{c}}ois
  Filipot, Rudy Magne, Aaron Roland, Andre Van Der~Westhuysen, Pierre
  Queffeulou, Jean-Michel Lefevre, Lotfi Aouf, et~al.
\newblock Semiempirical dissipation source functions for ocean waves. part i:
  Definition, calibration, and validation.
\newblock {\em Journal of Physical Oceanography}, 40(9):1917--1941, 2010.

\bibitem{roland2014developments}
Aron Roland and Fabrice Ardhuin.
\newblock On the developments of spectral wave models: numerics and
  parameterizations for the coastal ocean.
\newblock {\em Ocean Dynamics}, 64(6):833--846, 2014.

\bibitem{popinet2012adaptive}
St{\'e}phane Popinet.
\newblock Adaptive modelling of long-distance wave propagation and fine-scale
  flooding during the tohoku tsunami.
\newblock {\em Natural Hazards and Earth System Sciences}, 12(4):1213--1227,
  2012.

\bibitem{falcon_observation_2007}
E.~Falcon, S.~Fauve, and C.~Laroche.
\newblock Observation of {Intermittency} in {Wave} {Turbulence}.
\newblock {\em Physical Review Letters}, 98(15):154501, April 2007.
\newblock Number: 15.

\bibitem{falcon2022experiments}
Eric Falcon and Nicolas Mordant.
\newblock Experiments in surface gravity--capillary wave turbulence.
\newblock {\em Annual Review of Fluid Mechanics}, 54(1):1--25, 2022.

\bibitem{lvov2001hamiltonian}
Yuri~V Lvov and Esteban~G Tabak.
\newblock Hamiltonian formalism and the garrett-munk spectrum of internal waves
  in the ocean.
\newblock {\em Physical review letters}, 87(16):168501, 2001.

\bibitem{sutherland2010internal}
Bruce~R Sutherland.
\newblock {\em Internal gravity waves}.
\newblock Cambridge university press, 2010.

\bibitem{davis2020succession}
G{\'e}raldine Davis, Timoth{\'e}e Jamin, Julie Deleuze, Sylvain Joubaud, and
  Thierry Dauxois.
\newblock Succession of resonances to achieve internal wave turbulence.
\newblock {\em Physical Review Letters}, 124(20):204502, 2020.

\bibitem{caillol2000kinetic}
Ph~Caillol and V~Zeitlin.
\newblock Kinetic equations and stationary energy spectra of weakly nonlinear
  internal gravity waves.
\newblock {\em Dynamics of atmospheres and oceans}, 32(2):81--112, 2000.

\bibitem{galtier2003weak}
S{\'e}bastien Galtier.
\newblock Weak inertial-wave turbulence theory.
\newblock {\em Physical Review E}, 68(1):015301, 2003.

\bibitem{yarom2014experimental}
Ehud Yarom and Eran Sharon.
\newblock Experimental observation of steady inertial wave turbulence in deep
  rotating flows.
\newblock {\em Nature Physics}, 10(7):510--514, 2014.

\bibitem{yarom2017experimental}
Ehud Yarom, Alon Salhov, and Eran Sharon.
\newblock Experimental quantification of nonlinear time scales in inertial wave
  rotating turbulence.
\newblock {\em Physical Review Fluids}, 2(12):122601, 2017.

\bibitem{miquel2013transition}
Benjamin Miquel, Alexandros Alexakis, Christophe Josserand, and Nicolas
  Mordant.
\newblock Transition from wave turbulence to dynamical crumpling in vibrated
  elastic plates.
\newblock {\em Physical review letters}, 111(5):054302, 2013.

\bibitem{during2006weak}
Gustavo D{\"u}ring, Christophe Josserand, and Sergio Rica.
\newblock Weak turbulence for a vibrating plate: Can one hear a kolmogorov
  spectrum?
\newblock {\em Physical review letters}, 97(2):025503, 2006.

\bibitem{lbz3-nzxn}
Murukesh Muralidhar, S\'ebastien Auma\^{\i}tre, and Antoine Naert.
\newblock Extreme events in a random set of nonlinear elastic bending waves.
\newblock {\em Phys. Rev. E}, 112:034212, Sep 2025.

\bibitem{Muralidhar2024Extreme}
Murukesh Muralidhar, Antoine Naert, and S{\'e}bastien Auma{\^{\i}}tre.
\newblock Extreme events in a set of elastic bending waves.
\newblock In {\em APS Division of Fluid Dynamics Meeting Abstracts}, 2024.
\newblock Bibcode: 2024APS..DFDT35001M.

\bibitem{Galtier_2020}
S{\'e}bastien Galtier.
\newblock Wave turbulence: the case of capillary waves.
\newblock {\em Geophysical \& Astrophysical Fluid Dynamics}, 115(3), 2020.

\bibitem{ryzhik1996transport}
Leonid Ryzhik, George Papanicolaou, and Joseph~B Keller.
\newblock Transport equations for elastic and other waves in random media.
\newblock {\em Wave motion}, 24(4):327--370, 1996.

\bibitem{zakharov_kolmogorov_1992}
Vladimir~E. Zakharov, Victor~S. L’vov, and Gregory Falkovich.
\newblock {\em Kolmogorov {Spectra} of {Turbulence} {I}}.
\newblock Springer {Series} in {Nonlinear} {Dynamics}. Springer Berlin
  Heidelberg, Berlin, Heidelberg, 1992.

\bibitem{galtier2024wave}
Sebastien Galtier.
\newblock Wave turbulence: a solvable problem applied to the navier--stokes
  equations.
\newblock {\em Comptes Rendus. Physique}, 25(G1):433--455, 2024.

\bibitem{newell2011wave}
Alan~C Newell and Benno Rumpf.
\newblock Wave turbulence.
\newblock {\em Annual review of fluid mechanics}, 43(1):59--78, 2011.

\bibitem{nazarenko2011wave}
Sergey Nazarenko.
\newblock {\em Wave turbulence}, volume 825.
\newblock Springer Science \& Business Media, 2011.

\bibitem{hasselmann1962non}
Klaus Hasselmann.
\newblock On the non-linear energy transfer in a gravity-wave spectrum part 1.
  general theory.
\newblock {\em Journal of Fluid Mechanics}, 12(4):481--500, 1962.

\bibitem{zakharov2004one}
Vladimir Zakharov, Fr{\'e}d{\'e}ric Dias, and Andrei Pushkarev.
\newblock One-dimensional wave turbulence.
\newblock {\em Physics Reports}, 398(1):1--65, 2004.

\bibitem{FGV}
G.~{Falkovich}, K.~{Gaw\c{e}dzki}, and M.~{Vergasola}.
\newblock {Particles and fields in fluid turbulence}.
\newblock {\em Rev. Mod. Phys.}, 73:913--975, 2001.

\bibitem{nazarenko_wave_2011}
Sergey Nazarenko.
\newblock {\em Wave {Turbulence}}, volume 825 of {\em Lecture {Notes} in
  {Physics}}.
\newblock Springer Berlin Heidelberg, Berlin, Heidelberg, 2011.

\bibitem{benney1967propagation}
DJ~Benney and Alan~C Newell.
\newblock The propagation of nonlinear wave envelopes.
\newblock {\em Journal of mathematics and Physics}, 46(1-4):133--139, 1967.

\bibitem{newell_wave_2001}
Alan~C. Newell, Sergey Nazarenko, and Laura Biven.
\newblock Wave turbulence and intermittency.
\newblock {\em Physica D: Nonlinear Phenomena}, 152-153:520--550, May 2001.

\bibitem{hrabski2024verification}
Alexander Hrabski and Yulin Pan.
\newblock Verification of wave turbulence theory in the kinetic limit.
\newblock {\em Physical Review Research}, 6(2):023184, 2024.

\bibitem{dku2022domain}
Ryan~Sh{\.e}ji{\'e} D{\k{u}} and Oliver B{\"u}hler.
\newblock Domain dependence of wave turbulence theory for the
  majda-mclaughlin-tabak (mmt) model.
\newblock In {\em 23rd Conference on Atmospheric and Oceanic Fluid Dynamics}.
  AMS, 2022.

\bibitem{zhu2022testing}
Ying Zhu, Boris Semisalov, Giorgio Krstulovic, and Sergey Nazarenko.
\newblock Testing wave turbulence theory for the gross-pitaevskii system.
\newblock {\em Physical Review E}, 106(1):014205, 2022.

\bibitem{banks2022direct}
JW~Banks, T~Buckmaster, AO~Korotkevich, G~Kova{\v{c}}i{\v{c}}, and J~Shatah.
\newblock Direct verification of the kinetic description of wave turbulence for
  finite-size systems dominated by interactions among groups of six waves.
\newblock {\em Physical review letters}, 129(3):034101, 2022.

\bibitem{michel2022statistics}
Guillaume Michel, F{\'e}licien Bonnefoy, Guillaume Ducrozet, and Eric Falcon.
\newblock Statistics of rogue waves in isotropic wave fields.
\newblock {\em Journal of Fluid Mechanics}, 943:A26, 2022.

\bibitem{falcon2020saturation}
Eric Falcon, Guillaume Michel, Gaurav Prabhudesai, Annette Cazaubiel,
  Micha{\"e}l Berhanu, Nicolas Mordant, S{\'e}bastien Auma{\^\i}tre, and
  F~Bonnefoy.
\newblock Saturation of the inverse cascade in surface gravity-wave turbulence.
\newblock {\em Physical Review Letters}, 125(13):134501, 2020.

\bibitem{garrido2023large}
Miguel~Angel Garrido, Ricardo Grande, Kristin~M Kurianski, and Gigliola
  Staffilani.
\newblock Large deviations principle for the cubic nls equation.
\newblock {\em Communications on Pure and Applied Mathematics},
  76(12):4087--4136, 2023.

\bibitem{dematteis2018rogue}
Giovanni Dematteis, Tobias Grafke, and Eric Vanden-Eijnden.
\newblock Rogue waves and large deviations in deep sea.
\newblock {\em Proceedings of the National Academy of Sciences},
  115(5):855--860, 2018.

\bibitem{dematteis2019experimental}
Giovanni Dematteis, Tobias Grafke, Miguel Onorato, and Eric Vanden-Eijnden.
\newblock Experimental evidence of hydrodynamic instantons: the universal route
  to rogue waves.
\newblock {\em Physical Review X}, 9(4):041057, 2019.

\bibitem{majda1997one}
Andrew~J Majda, David~W McLaughlin, and EG1431687 Tabak.
\newblock A one-dimensional model for dispersive wave turbulence.
\newblock {\em Journal of Nonlinear Science}, 7(1):9--44, 1997.

\bibitem{cai2001dispersive}
David Cai, Andrew~J Majda, David~W McLaughlin, and Esteban~G Tabak.
\newblock Dispersive wave turbulence in one dimension.
\newblock {\em Physica D: Nonlinear Phenomena}, 152:551--572, 2001.

\bibitem{chibbaro2017weak}
Sergio Chibbaro, F~De~Lillo, and M~Onorato.
\newblock Weak versus strong wave turbulence in the majda-mclaughlin-tabak
  model.
\newblock {\em Physical Review Fluids}, 2(5):052603, 2017.

\bibitem{newell2012spontaneous}
Alan~C Newell, Benno Rumpf, and Vladimir~E Zakharov.
\newblock Spontaneous breaking of the spatial homogeneity symmetry in wave
  turbulence.
\newblock {\em Physical review letters}, 108(19):194502, 2012.

\bibitem{bertini2015macroscopic}
Lorenzo Bertini, Alberto De~Sole, Davide Gabrielli, Giovanni Jona-Lasinio, and
  Claudio Landim.
\newblock Macroscopic fluctuation theory.
\newblock {\em Reviews of Modern Physics}, 87(2):593, 2015.

\bibitem{bodineau2008long}
T~Bodineau, B~Derrida, V~Lecomte, and F~Van~Wijland.
\newblock Long range correlations and phase transitions in non-equilibrium
  diffusive systems.
\newblock {\em Journal of Statistical Physics}, 133(6):1013--1031, 2008.

\bibitem{illien2024deankawasakiequationstochasticdensity}
Pierre Illien.
\newblock The dean-kawasaki equation and stochastic density functional theory,
  2024.

\bibitem{bertini2009towards}
L~Bertini, Alberto De~Sole, D~Gabrielli, Giovanni Jona-Lasinio, and
  C25485962009JSP Landim.
\newblock Towards a nonequilibrium thermodynamics: a self-contained macroscopic
  description of driven diffusive systems.
\newblock {\em Journal of Statistical Physics}, 135:857--872, 2009.

\bibitem{DLS2}
B.~{Derrida}, J.~L. {Lebowitz}, and E.~R. {Speer}.
\newblock {Large Deviation of the Density Profile in the Steady State of the
  Open Symmetric Simple Exclusion Process}.
\newblock {\em Journal of Statistical Physics}, 107:599--634, May 2002.

\bibitem{BOUCHET:2016:D}
Freddy Bouchet, Krzysztof Gawedzki, and Cesare Nardini.
\newblock {Perturbative Calculation of Quasi-Potential in Non-equilibrium
  Diffusions: A Mean-Field Example}.
\newblock {\em {Journal of statistical physics}}, {163}({5}):{1157--1210},
  {JUN} {2016}.

\bibitem{rezakhanlou1998large}
F.~Rezakhanlou.
\newblock Large deviations from a kinetic limit.
\newblock {\em The Annals of Probability}, 26(3):1259--1340, 1998.

\bibitem{leonard1995large}
Christian L{\'e}onard.
\newblock On large deviations for particle systems associated with spatially
  homogeneous boltzmann type equations.
\newblock {\em Probability theory and related fields}, 101(1):1--44, 1995.

\bibitem{bouchet2020boltzmann}
Freddy Bouchet.
\newblock Is the boltzmann equation reversible? a large deviation perspective
  on the irreversibility paradox.
\newblock {\em Journal of Statistical Physics}, 181:515--550, 2020.

\bibitem{bodineau2022long}
Thierry Bodineau, Isabelle Gallagher, Laure Saint-Raymond, and Sergio
  Simonella.
\newblock Long-time derivation at equilibrium of the fluctuating boltzmann
  equation, 2022.

\bibitem{heydecker2023large}
Daniel Heydecker.
\newblock Large deviations of kac’s conservative particle system and energy
  nonconserving solutions to the boltzmann equation: A counterexample to the
  predicted rate function.
\newblock {\em The Annals of Applied Probability}, 33(3):1758--1826, 2023.

\bibitem{basile2024asymptotic}
Giada Basile, Dario Benedetto, Lorenzo Bertini, and Emanuele Caglioti.
\newblock Asymptotic probability of energy increasing solutions to the
  homogeneous boltzmann equation.
\newblock {\em The Annals of Applied Probability}, 34(4):3995--4021, 2024.

\bibitem{feliachi2021dynamical}
Ouassim Feliachi and Freddy Bouchet.
\newblock Dynamical large deviations for plasmas below the debye length and the
  landau equation.
\newblock {\em Journal of Statistical Physics}, 183(3):1--58, 2021.

\bibitem{feliachi2022dynamical}
Ouassim Feliachi and Freddy Bouchet.
\newblock Dynamical large deviations for homogeneous systems with long range
  interactions and the balescu--guernsey--lenard equation.
\newblock {\em Journal of Statistical Physics}, 186(2):1--29, 2022.

\bibitem{mielke2014relation}
Alexander Mielke, Mark~A Peletier, and DR~Michiel Renger.
\newblock On the relation between gradient flows and the large-deviation
  principle, with applications to markov chains and diffusion.
\newblock {\em Potential Analysis}, 41(4):1293--1327, 2014.

\bibitem{nazarenko2010statistics}
Sergey Nazarenko, Sergei Lukaschuk, Stuart McLelland, and Petr Denissenko.
\newblock Statistics of surface gravity wave turbulence in the space and time
  domains.
\newblock {\em Journal of Fluid Mechanics}, 642:395--420, 2010.

\bibitem{lvov_noisy_2004}
Yuri~V. Lvov and Sergey Nazarenko.
\newblock Noisy spectra, long correlations, and intermittency in wave
  turbulence.
\newblock {\em Physical Review E}, 69(6):066608, June 2004.
\newblock Number: 6.

\bibitem{campagne2019energy}
Antoine Campagne, Roumaissa Hassaini, Ivan Redor, Joel Sommeria, and Nicolas
  Mordant.
\newblock The energy cascade of surface wave turbulence: toward identifying the
  active wave coupling.
\newblock In {\em Turbulent Cascades II: Proceedings of the Euromech-ERCOFTAC
  Colloquium 589}, pages 239--246. Springer, 2019.

\bibitem{deike2013etudes}
Luc Deike.
\newblock {\em Etudes exp{\'e}rimentales et num{\'e}riques de la turbulence
  d'ondes de surface}.
\newblock PhD thesis, Paris 7, 2013.

\bibitem{aubourg20173}
Quentin Aubourg, Antoine Campagne, Charles Peureux, Fabrice Ardhuin, Joel
  Sommeria, Samuel Viboud, and Nicolas Mordant.
\newblock 3-wave and 4-wave interactions in gravity wave turbulence.
\newblock {\em arXiv preprint arXiv:1710.11372}, 2017.

\bibitem{deng2311long}
Y~Deng and Z~Hani.
\newblock Long time justification of wave turbulence theory (2023).
\newblock {\em arXiv preprint arXiv:2311.10082}.

\bibitem{deng2024long}
Yu~Deng, Zaher Hani, and Xiao Ma.
\newblock Long time derivation of the boltzmann equation from hard sphere
  dynamics.
\newblock {\em arXiv preprint arXiv:2408.07818}, 2024.

\bibitem{connaughton_non-stationary_2003}
Colm Connaughton, Alan~C. Newell, and Yves Pomeau.
\newblock Non-stationary spectra of local wave turbulence.
\newblock {\em Physica D: Nonlinear Phenomena}, 184(1-4):64--85, October 2003.

\bibitem{cramer1938nouveau}
Harald Cram{\'e}r.
\newblock Sur un nouveau th{\'e}oreme-limite de la th{\'e}orie des
  probabilit{\'e}s.
\newblock {\em Actual. Sci. Ind.}, 736:5--23, 1938.

\bibitem{Don}
M.D. {Donsker} and S.R.S. {Varadhan}.
\newblock {Asymptotic evaluation of certain markov process expectations for
  large time, I,II,III,IV}.
\newblock {\em Communications on Pure and Applied Mathematics.},
  28,28,29,36:1--47,279--301,389--461,183--212, 1975,1975,1976,1983.

\bibitem{Freidlin_Wentzel_1984_book}
M.~I. {Freidlin} and A.~D. {Wentzell}.
\newblock {\em {Random perturbations of dynamical systems}}.
\newblock Springer - New York, Berlin, 1984.

\bibitem{Gardiner_1994_Book_Stochastic}
C.~W. {Gardiner}.
\newblock {\em {Handbook of stochastic methods for physics, chemistry and the
  natural sciences}}.
\newblock Springer Series in Synergetics, Berlin: Springer, |c1994, 2nd
  ed.~1985.~Corr.~3rd printing 1994, 1994.

\bibitem{touchette2009large}
Hugo Touchette.
\newblock The large deviation approach to statistical mechanics.
\newblock {\em Physics Reports}, 478(1-3):1--69, 2009.

\bibitem{spohn2006phonon}
Herbert Spohn.
\newblock The phonon boltzmann equation, properties and link to weakly
  anharmonic lattice dynamics.
\newblock {\em Journal of statistical physics}, 124(2):1041--1104, 2006.

\bibitem{dyachenko1992optical}
S~Dyachenko, AC~Newell, A~Pushkarev, and VE1169619 Zakharov.
\newblock Optical turbulence: weak turbulence, condensates and collapsing
  filaments in the nonlinear schr{\"o}dinger equation.
\newblock {\em Physica D: Nonlinear Phenomena}, 57(1-2):96--160, 1992.

\bibitem{zakharov_collapse_nodate}
V~E Zakharov.
\newblock Collapse of {Langmuir} {Waves}.
\newblock page~7.

\bibitem{zakharov1967weak}
Vladimir~Evgen'evich Zakharov and NN~Filonenko.
\newblock Weak turbulence of capillary waves.
\newblock {\em Journal of applied mechanics and technical physics},
  8(5):37--40, 1967.

\bibitem{pushkarev2000turbulence}
AN~Pushkarev and VE~Zakharov.
\newblock Turbulence of capillary waves—theory and numerical simulation.
\newblock {\em Physica D: Nonlinear Phenomena}, 135(1-2):98--116, 2000.

\bibitem{deike2014direct}
Luc Deike, Daniel Fuster, Michael Berhanu, and Eric Falcon.
\newblock Direct numerical simulations of capillary wave turbulence.
\newblock {\em Physical review letters}, 112(23):234501, 2014.

\bibitem{10.1145/361573.361582}
R.~H. Bartels and G.~W. Stewart.
\newblock Algorithm 432 [c2]: Solution of the matrix equation ax + xb = c [f4].
\newblock {\em Commun. ACM}, 15(9), September 1972.

\bibitem{golub1979hessenberg}
Gene Golub, Stephen Nash, and Charles Van~Loan.
\newblock A hessenberg-schur method for the problem ax+ xb= c.
\newblock {\em IEEE Transactions on Automatic Control}, 24(6):909--913, 1979.

\bibitem{lu1991solution}
An~Lu and Eugene~L Wachspress.
\newblock Solution of lyapunov equations by alternating direction implicit
  iteration.
\newblock {\em Computers \& Mathematics with Applications}, 21(9):43--58, 1991.

\bibitem{li2002low}
Jing-Rebecca Li and Jacob White.
\newblock Low rank solution of lyapunov equations.
\newblock {\em SIAM Journal on Matrix Analysis and Applications},
  24(1):260--280, 2002.

\bibitem{lanchon2023energy}
Nicolas Lanchon and Pierre-Philippe Cortet.
\newblock Energy spectra of nonlocal internal gravity wave turbulence.
\newblock {\em Physical Review Letters}, 131(26):264001, 2023.

\bibitem{vladimirova2021turbulence}
Natalia Vladimirova, Ivan Vointsev, Alena Skoba, and Gregory Falkovich.
\newblock Turbulence of capillary waves on shallow water.
\newblock {\em Fluids}, 6(5):185, 2021.

\bibitem{wu2023breaking}
Jiarong Wu, St{\'e}phane Popinet, and Luc Deike.
\newblock Breaking wave field statistics with a multi-layer model.
\newblock {\em Journal of Fluid Mechanics}, 968:A12, 2023.

\bibitem{krasitskii_reduced_1994}
Vladimir~P. Krasitskii.
\newblock On reduced equations in the {Hamiltonian} theory of weakly nonlinear
  surface waves.
\newblock {\em Journal of Fluid Mechanics}, 272:1--20, August 1994.
\newblock Publisher: Cambridge University Press.

\end{thebibliography}

\appendix
\setcounter{secnumdepth}{3}   


\section{Appendices}

\subsection{Contraction principle\label{subsec:Contraction-principle}}

The contraction principle is a useful result in large deviation theory.
Suppose we have a sequence of random variables $X_{\epsilon}$ satisfying
a large deviation principle with rate function $I$ : $\mathbb{P}_{\epsilon}\left(x\right)\underset{\epsilon\to0}{\asymp}e^{-\frac{I\left(x\right)}{\epsilon}}$,
and we transform those random variables in a new sequence $Y_{\epsilon}=f\left(X_{\epsilon}\right)$.
Then there is a new LDP holding for variable $Y_{\epsilon}$ such
that the associated rate function is 
\[
J\left(y\right)=\underset{x}{\sup}\left\{ I\left(x\right)|f\left(x\right)=y\right\} .
\]

\subsection{Properties of the dynamics and equilibrium quasipotentials\label{subsec:Properties-dyn-a,d-eq-quasipot}}

In this paragraph, we explain the main results that the large deviation
frameworks presented in Sections \ref{subsec:Large-deviation-Hamiltonian-general-WT},
\ref{subsec:LD-isotropic-WT} entail. We present first the general
wave turbulence, then the isotropic wave turbulence. The isotropic
local case used for the end of the paper is left in the core of the
article (Section \ref{subsec:LD-isotropic-local-WT}). 

\subsubsection{General wave turbulence\label{subsec:Properties-dyn-a,d-eq-quasipot-general}}

In this paragraph, we explain the main properties derived from that
the large deviation framework presented in Section \ref{subsec:Large-deviation-Hamiltonian-general-WT}.
All these properties can be found in \cite{guioth_path_2022}.

\paragraph{Fluctuating dynamics}

The large deviation Hamiltonian (\ref{eq:LD-hamiltonian-general})
is quadratic in $\lambda$: formally it can be written
\[
H\left[n,\lambda\right]=b.\lambda+\lambda.a\lambda
\]
 where $.$ denotes canonical scalar product in the space of functions
of $\bm{k}$: $f.g=\int_{\mathbb{V}_{L}^{d}}\text{d}^{d}\bm{k}f\left(\bm{k}\right)g\left(\bm{k}\right)$.
We denote $\mathcal{R}_{3}^{d}\left(\bm{k}_{1}\right)=\left\{ \bm{k}_{2},\bm{k}_{3},\bm{k}_{4}\;|\;\bm{k}_{1}+\bm{k}_{2}=\bm{k}_{3}+\bm{k}_{4}\ \&\ \omega_{\bm{k}_{1}}+\omega_{\bm{k}_{2}}=\omega_{\bm{k}_{3}}+\omega_{\bm{k}_{4}}\right\} $
and we can write explicitly 
\begin{equation}
b\left[n\right]\left(\bm{k}_{1}\right)=4\int_{\mathcal{R}_{3}^{d}\left(\bm{k}_{1}\right)}|W_{1234}|^{2}n_{1}n_{2}n_{3}n_{4}\left[\frac{1}{n_{1}}+\frac{1}{n_{2}}-\frac{1}{n_{3}}-\frac{1}{n_{4}}\right]\label{eq:b-general}
\end{equation}
which gives the deterministic equation $\partial_{t}\bar{n}=\frac{\delta H}{\delta\lambda}\left[\bar{n},\lambda=0\right]=b\left[\bar{n}\right]$
verified by the average state $\bar{n}$, and $a$ is a nonnegative-definite,
self-adjoint operator which may be interpreted as a noise covariance,
associated to the quadratic form
\[
\lambda.a\lambda=\int_{\mathcal{R}_{4}^{d}}|W_{1234}|^{2}n_{1}n_{2}n_{3}n_{4}\left(\lambda_{1}+\lambda_{2}-\lambda_{3}-\lambda_{4}\right)^{2}
\]
if we denote $a_{1234}=\delta\left(\bm{k}_{1}+\bm{k}_{2}-\bm{k}_{3}-\bm{k}_{4}\right)\delta\left(\omega_{\bm{k}_{1}}+\omega_{\bm{k}_{2}}-\omega_{\bm{k}_{3}}-\omega_{\bm{k}_{4}}\right)|W_{1234}|^{2}n_{1}n_{2}n_{3}n_{4}$
we can express $a$ (using the symmetries of $W$ and those of $a_{1234}$
which is invariant under exchanges $1\leftrightarrow2,3\leftrightarrow4\text{ and }\left(1,2\right)\leftrightarrow\left(3,4\right)$):
\begin{align}
a\left[n\right]\left(\bm{k}_{1},\bm{k}_{2}\right) & =4\delta\left(\bm{k}_{1}-\bm{k}_{2}\right)\int_{\mathbb{V}_{d}^{d}}\text{d}^{d}\bm{k}_{3}\text{d}^{d}\bm{k}_{4}\text{d}^{d}\bm{k}_{5}a_{1534}\label{eq:a-general}\\
 & +4\int_{\mathbb{V}_{d}^{d}}\text{d}^{d}\bm{k}_{3}\text{d}^{d}\bm{k}_{4}a_{1234}-2\int_{\mathbb{V}_{d}^{d}}\text{d}^{d}\bm{k}_{3}\text{d}^{d}\bm{k}_{4}\left(a_{1432}+a_{4231}\right)\nonumber 
\end{align}
The statistics of $n$ is the same as one following a stochastic Langevin
equation with small noise:
\begin{equation}
\text{d}n=b\left[n\right]\text{d}t+\sqrt{2\left(\frac{2\pi}{L}\right)^{d}}\sigma\left[n\right]\text{d}W\label{eq:fluctuating-dyn-general}
\end{equation}
where $a=\sigma\sigma^{T}$ and $W$ is a Wiener process. This equivalent
fluctuating dynamics shows that the kinetic equation $\partial_{t}n=b\left[n\right]$
should be regarded as the average dynamics of a fluctuating spectrum
where the fluctuations are of order $\mathcal{O}\left(L^{-d/2}\right)$
and the covariance of this noise is given by $a$.

\paragraph{Conserved quantities}

The symmetries of the Hamiltonian enable to find the quantities which
are conserved by the dynamics. As known from general large deviation
theory, if a quantity $C\left[n\right]$ is conserved for any evolution
$n$, then for all functions $\lambda$ and $\alpha$ we have the
property
\[
H\left[n,\lambda+\alpha.\frac{\delta C}{\delta n}\right]=H\left[n,\lambda\right]
\]
We can show that the wave action number $\mathcal{N}=\int\text{d}^{d}\bm{k}n\left(\bm{k}\right)$,
the energy $\mathcal{E}=\int\text{d}^{d}\bm{k}\omega_{\bm{k}}n\left(\bm{k}\right)$
and the momentum $\bm{K}=\int\text{d}^{d}\bm{k}\bm{k}n\left(\bm{k}\right)$
are conserved.

\paragraph{Equilibrium quasipotential}

The stationary action for large times is called quasipotential and
denoted $Q_{\mathcal{N},\mathcal{E},\mathbf{K}}\left[n\right]$ and
is a functional of the stationary distribution $n$. In other words
in the stationary regime, we expect the following LDP
\[
P_{S}\left[\hat{n}=n\right]\underset{L\to\infty}{\asymp}e^{-\left(\frac{2\pi}{L}\right)^{d}Q_{\mathcal{N},\mathcal{E},\mathbf{K}}\left[n\right]}.
\]
In general, there exist a quasipotential as soon as there exist a
stationary distribution, and it is difficult to compute. However at
equilibrium, things are easier and we can show that
\[
Q_{\mathcal{N},\mathcal{E},\mathbf{K}}\left[n\right]=\begin{cases}
\int\text{d}^{d}\bm{k}\left(\frac{n}{n^{*}}-1-\ln\left(\frac{n}{n^{*}}\right)\right) & \text{if }E=\int\text{d}^{d}\bm{k}\omega n,\mathcal{N}=\int\text{d}^{d}\bm{k}n\text{ and }\bm{K}=\int\text{d}^{d}\bm{k}\bm{k}n\\
+\infty & \text{otherwise}
\end{cases}.
\]
We can check that this functional solves the stationary Hamilton-Jacobi
equation $H[n,\frac{\delta Q}{\delta n}]=0$ which is a necessary
(but not sufficient) condition for a functional to be the quasipotential.

\paragraph{Out-of-equilibrium dynamics\label{par:Out-of-equilibrium-dynamics}}

As explained in \cite{zakharov_kolmogorov_1992}, real wave turbulence
is often brought out of equilibrium by a forcing at given scales,
which is dissipated at other scales. In this case, we generally add
forcing terms to amount for sources and dissipation of $n$. This
can be done by adding terms for source $s\left[n,\bm{k},t\right]\geq0$
and dissipation $d\left[n,\bm{k},t\right]\geq0$ to the fluctuating
dynamics:
\[
\text{d}n=b\left[n\right]\text{d}t+\sqrt{2\left(\frac{2\pi}{L}\right)^{d}}\sigma\text{d}W+\left(s-d\right).
\]
 The properties of such dynamics will be studied in the next parts.

\subsubsection{Isotropic wave turbulence\label{subsec:Properties-dyn-a,d-eq-quasipot-isotropic}}

In this paragraph, we explain the main properties derived from that
the large deviation framework presented in Section \ref{subsec:LD-isotropic-WT}
for isotropic wave turbulence.

\paragraph{Properties of the large deviation Hamiltonian}

The large deviation Hamiltonian (\ref{eq:LD-Hamiltonian-isotropic})
is quadratic in $\Lambda$: formally it can be written
\[
H_{iso}\left[N,\Lambda\right]=b_{iso}.\Lambda+\Lambda.a_{iso}\Lambda
\]
 where
\begin{equation}
b_{iso}\left[N\right]\left(\omega_{1}\right)=4{\displaystyle \int\text{d}\omega_{1}\text{d}\omega_{2}\text{d}\omega_{3}\text{d}\omega_{4}}S_{\omega_{1}\omega_{2}\omega_{3}\omega_{4}}\delta\left(\omega_{1}+\omega_{2}-\omega_{3}-\omega_{4}\right)n_{1}n_{2}n_{3}n_{4}\left[\frac{1}{n_{1}}+\frac{1}{n_{2}}-\frac{1}{n_{3}}-\frac{1}{n_{4}}\right]\label{eq:b-isotropic}
\end{equation}
and we can use the symmetries of $a_{iso,1234}:=S_{\omega_{1}\omega_{2}\omega_{3}\omega_{4}}\delta\left(\omega_{1}+\omega_{2}-\omega_{3}-\omega_{4}\right)n_{1}n_{2}n_{3}n_{4}$
under the exchanges $1\leftrightarrow2,3\leftrightarrow4\text{ and }\left(1,2\right)\leftrightarrow\left(3,4\right)$
to write the noise covariance $a_{iso}$ :
\begin{align}
a_{iso}\left[N\right]\left(\omega_{1},\omega_{2}\right) & =4\delta\left(\omega_{1}-\omega_{2}\right)\int\text{d}\omega_{3}\text{d}\omega_{4}\text{d}\omega_{5}a_{iso,1534}\label{eq:a-isotropic}\\
 & +4\int\text{d}\omega_{3}\text{d}\omega_{4}a_{iso,1234}-2\int\text{d}\omega_{3}\text{d}\omega_{4}\left(a_{iso,1432}+a_{iso,4231}\right).\nonumber 
\end{align}
This operator $a_{iso}$ is still a is a nonnegative-definite, self-adjoint
operator. Moreover the statistics of $n$ is the same as one following
a stochastic Langevin equation with small noise:
\begin{equation}
\text{d}N=b_{iso}\left[N\right]\text{d}t+\sqrt{2\left(\frac{2\pi}{L}\right)^{d}}\sigma_{iso}\left[N\right]\text{d}W\label{eq:fluctuating-dyn-isotropic}
\end{equation}
where $a_{iso}=\sigma_{iso}\sigma_{iso}^{T}$ and $W$ is a Wiener
process. In addition to isotropy, we will simplify further by assuming
a locality property in the next part.

\paragraph{Conserved quantities}

As in the general case, there are conserved quantities: the total
the wave action number $\mathcal{N}=\int\text{d}^{d}\bm{k}n\left(\bm{k}\right)$
and total energy $\mathcal{E}=\int\text{d}^{d}\bm{k}\omega_{\bm{k}}n\left(\bm{k}\right)$
are still conserved. However, since supposed the state to be isotropic,
the integral defining the total momentum $\bm{K}={\displaystyle \int\bm{k}n\left(k\right)\text{d}^{d}\bm{k}}$
vanishes. In the following, we will ignore the momentum which no longer
contains relevant information.

\subsection{Discussion on the generality of the 4-wave form and nondecay dispersion
law\label{subsec:Appendix-4Wave}}

In this appendix, we explain the generality of the Hamiltonian (\ref{eq:dynamicsHamiltonian-4Wave})
without proving everything (which was already done in \cite{zakharov_kolmogorov_1992}).
We just recall the main results in order to keep in mind the generality
of the present paper. In general, the formula for the dynamic's Hamiltonian
is 
\[
\mathcal{H}={\displaystyle \sum_{\bm{k}}\omega_{\bm{k}}a_{\bm{k}}a_{\bm{k}}^{*}+\sum_{\bm{k}_{1}\bm{k}_{2}\bm{k}_{3}\bm{k}_{4}}\sum_{\left(\sigma_{1}\sigma_{2}\sigma_{3}\sigma_{4}\right)\in\left(\pm1\right)^{4}}W_{\bm{k}_{1}\bm{k}_{2}\bm{k}_{3}\bm{k}_{4}}^{\sigma_{1}\sigma_{2}\sigma_{3}\sigma_{4}}a_{\bm{k}_{1}}^{\sigma_{1}}a_{\bm{k}_{2}}^{\sigma_{2}}a_{\bm{k}_{3}}^{\sigma_{3}}a_{\bm{k}_{4}}^{\sigma_{4}}\delta_{\sigma_{1}\bm{k}_{1}+\sigma_{2}\bm{k}_{2}+\sigma_{3}\bm{k}_{3}+\sigma_{4}\bm{k}_{4},0}}.
\]
where $W_{\bm{k}_{1}\bm{k}_{2}\bm{k}_{3}\bm{k}_{4}}^{\sigma_{1}\sigma_{2}\sigma_{3}\sigma_{4}}$
is the interaction kernel and $a_{\bm{k}_{2}}^{\sigma_{2}}$ are the
rescaled Fourier amplitudes with the notation $a_{\bm{k}}^{+}=a_{\bm{k}}$
and $a_{\bm{k}}^{-}=a_{\bm{k}}^{*}$. The result is that under certain
hypothesis developed in the next paragraph, we can keep terms with
interactions of the type $2\to2$ and discard third order terms $1\to2,2\to1,$
and fourth order terms of the form $3\to1$ or $4\to0$. Symmetry
properties of the coefficients enable to keep only one term with $\left(\sigma_{1}\sigma_{2}\sigma_{3}\sigma_{4}\right)=(+1,+1,-1,-1)$.
We can thus suppose this change of variables done and ignore the sum
$\sum_{\left(\sigma_{1}\sigma_{2}\sigma_{3}\sigma_{4}\right)\in\left(\pm1\right)^{4}}$. 

The principle is the following. As explained in \cite[part 1.1.3]{zakharov_kolmogorov_1992},
we can define new canonical variables $b_{\bm{k}}$ replacing $a_{\bm{k}}$:
\[
b_{\bm{k}}=a_{\bm{k}}+p_{\bm{k}}^{\left(1\right)}a_{\bm{k}}^{2}+p_{\bm{k}}^{\left(2\right)}a_{\bm{k}}a_{\bm{k}}^{*}+p_{\bm{k}}^{\left(3\right)}a_{\bm{k}}^{*2}+q_{\bm{k}}^{\left(1\right)}a_{\bm{k}}^{3}+q_{\bm{k}}^{\left(2\right)}a_{\bm{k}}^{2}a_{\bm{k}}^{*}+q_{\bm{k}}^{\left(3\right)}a_{\bm{k}}a_{\bm{k}}^{*2}+q_{\bm{k}}^{\left(4\right)}a_{\bm{k}}^{*3}
\]
 which by wisely choosing the $p_{\bm{k}}^{\left(i\right)},q_{\bm{k}}^{\left(i\right)}$,
enable to delete terms of third order and the fourth order terms of
signature different from $(++--)$. We can find explicit formulas
for these coefficients. However, this change of variables requires
some attention on a few points:
\begin{itemize}
\item By doing so, we should beware that the empirical Fourier spectrum
$n$ is defined as
\[
n(\bm{\xi})=\sum_{\bm{k}}a_{\bm{k}}a_{\bm{k}}^{*}\delta(\bm{k}-\bm{\xi})
\]
and thus its definition is affected by this change of variable. When
we change variables, this will give rise to terms such as 
\[
a_{\bm{k}}a_{\bm{k}}^{*}=b_{\bm{k}}b_{\bm{k}}^{*}+p_{\bm{k}}^{\left(3\right)}\left(b_{\bm{k}}^{*3}+b_{\bm{k}}^{3}\right)+(p_{\bm{k}}^{(2)}+p_{\bm{k}}^{(1)})(b_{\bm{k}}^{*}b_{\bm{k}}^{2}+b_{\bm{k}}b_{\bm{k}}^{*2})+(q_{\bm{k}}^{(4)}+p_{\bm{k}}^{(1)}p_{\bm{k}}^{(3)})b_{\bm{k}}^{4}+\ldots
\]
At first sight, these terms seem non-vanishing. However we can argue
that since $B_{\bm{k}}=\mathcal{O}\left(\epsilon=\left(\frac{2\pi}{L}\right)^{d}\right)$
the new $n$ is equal to the old $n$ up to next order in $\epsilon$:
$\frac{n_{\bm{k}}^{\text{new}}-n_{\bm{k}}^{\text{old}}}{n_{\bm{k}}^{\text{old}}}=\mathcal{O}\left(\epsilon\right)$,
which is a fortiori true for the hamiltonian with relative error $\sim\mathcal{O}(\epsilon)$.
In short, the most important point is to keep the same convention
when comparing formulas of different papers. 
\item As mentioned in \cite{zakharov_kolmogorov_1992} the change of variable
is singular (there are possibly vanishing denominators in the expressions
of $p_{\bm{k}}^{i},q_{\bm{k}}^{i}$) when we are on the resonant manifold
of a mode that we would like to remove. Those of type $0\to3,0\to4$
are never verified because of positiveness of $\omega_{i}$'s, but
for the case $1\to2,1\to3$, this is not always true. So as to be
sure it does not occur, we have to make an hypothesis on the concave
nature of the dispersion relation $\omega(\bm{k})$. For the so-called
\emph{nondecay} dispersion relations 
\[
\omega''=\frac{\partial^{2}\omega(\bm{k})}{\partial k^{2}}<0,
\]
these transitions $1\to2,1\to3$ are impossible (there is a graphical
argument in the 2D case), see \cite[part 1.1.2]{zakharov_kolmogorov_1992}.
For the isotropic case we are interested in, the dispersion relation
$\omega(\bm{k})=\gamma\left\Vert \bm{k}\right\Vert ^{\alpha}$ is
of the \emph{nondecay} type if and only if
\[
\alpha<1.
\]
The limiting case $\alpha=1$ (linear dispersion relation) enables
decay only for colinear wave vectors which is a particular case.
Moreover, anisotropic dispersion relations do not enter in this categorization
as well as dispersion relations which change convexity with the value
of $k$. 
\end{itemize}
\textbf{To conclude, within this paper, we will make this nondecay
hypothesis $\frac{\partial^{2}\omega(\bm{k})}{\partial k^{2}}<0$
for any $\bm{k}$. }It is a sufficient hypothesis to discard third
order terms and fourth order terms with signature different from $(++--)$,
up to a canonical change of variables. Otherwise, we should highlight
that the computations remain valid provided we stay in the general
way of writing with a sum $\sum_{\vec{\sigma}}.$ However, this expression
of the dynamic's Hamiltonian with only $2\to2$ interactions helps
to justify the local approximation, which otherwise has to be assumed.
The table \ref{tab:dispersion-relations} gathers values of $\frac{\partial^{2}\omega}{\partial k^{2}}$
for different physical situations.

\begin{table}[H]
\begin{tabular}{|>{\centering}p{0.35\textwidth}|>{\centering}p{0.23\textwidth}|>{\centering}p{0.15\textwidth}|>{\centering}p{0.25\textwidth}|}
\hline 
\rowcolor{lightgray!50}
\footnotesize\textbf{Wave type} & $\omega$ & $\frac{\partial^{2}\omega}{\partial k^{2}}$ & type\tabularnewline
\hline 
\hline 
Non-Linear Schrödinger equation $i\partial_{t}\psi=-\mu\nabla^{2}\psi+\epsilon\left|\psi\right|^{2}\psi$ & $\omega=-\mu k^{2}$ & $\frac{\partial^{2}\omega}{\partial k^{2}}=-2\mu<0$ & \textbf{nondecay}\tabularnewline
\hline 
deep water gravity waves ($h>\frac{1}{2}\lambda$) & $\omega=\sqrt{gk}$ & $\frac{\partial^{2}\omega}{\partial k^{2}}=-\frac{\sqrt{g}}{4k^{3/2}}<0$ & \textbf{nondecay}\tabularnewline
\hline 
intermediate depth water gravity waves (any $h,\lambda$) & $\omega=\sqrt{gk\tanh(kh)}$ & $\frac{\partial^{2}\omega}{\partial k^{2}}<0$ & \textbf{nondecay}\tabularnewline
\hline 
shallow water gravity waves ($h<0.05\lambda$) & $\omega=k\sqrt{gh}$ & $\frac{\partial^{2}\omega}{\partial k^{2}}=0$ & decay possible only for colinear wave vectors\tabularnewline
\hline 
general gravity-capillary waves & $\omega=\sqrt{k(g+\sigma k^{2}/\rho_{0})\tanh(kh)}$ & $\frac{\partial^{2}\omega}{\partial k^{2}}$ changes sign & no defined type\tabularnewline
\hline 
shallow water gravity-capillary waves ($k\ll\sqrt{\rho_{0}g/\sigma}$) & $\omega=\sqrt{gh}k\left(1+\frac{k^{2}}{2\rho_{0}g/\sigma}\right)$ & $\frac{\partial^{2}\omega}{\partial k^{2}}>0$ & decay\tabularnewline
\hline 
shallow water capillary waves ($\sqrt{\rho_{0}g/\sigma}\ll k\ll h^{-1}$) & $\omega=\sqrt{\frac{\sigma h}{\rho_{0}}}k^{2}$ & $\frac{\partial^{2}\omega}{\partial k^{2}}>0$ & decay\tabularnewline
\hline 
deep water capillary waves ($k\gg h^{-1}$) & $\omega=\sqrt{\frac{\sigma}{\rho_{0}}}k^{3/2}$ & $\frac{\partial^{2}\omega}{\partial k^{2}}>0$ & decay\tabularnewline
\hline 
acoustic waves & $\omega=ck\pm\Omega(k)$ where at first order $\Omega(k)=ck(a^{2}k^{2})\ll ck$ & $\frac{\partial^{2}\omega}{\partial k^{2}}=\pm6a^{2}ck$ changes sign & no defined type\tabularnewline
\hline 
waves on a string & $\omega=k\sqrt{\frac{T}{\mu}}$ & $\frac{\partial^{2}\omega}{\partial k^{2}}=0$ & decay possible only for colinear wave vectors\tabularnewline
\hline 
light in vacuum & $\omega=ck$ & $\frac{\partial^{2}\omega}{\partial k^{2}}=0$ & decay possible only for colinear wave vectors\tabularnewline
\hline 
light in medium of optical index $n$ & $\omega=\frac{ck}{n(k)}$with $n(k)\simeq n_{0}+Bk^{2}$ & $\frac{\partial^{2}\omega}{\partial k^{2}}=-2cBk\frac{3n_{0}-Bk^{2}}{(n_{0}+Bk^{2})^{3}}$ & \textbf{nondecay} for usual cases where $Bk^{2}<3n_{0}$\tabularnewline
\hline 
nonrelativistic particle & $\omega=\frac{\hbar k^{2}}{2m}$ & $\frac{\partial^{2}\omega}{\partial k^{2}}=\frac{\hbar}{m}>0$ & decay\tabularnewline
\hline 
relativistic particle & $\omega=\frac{mc^{2}}{\hbar}+\frac{\hbar k^{2}}{2m}$ & $\frac{\partial^{2}\omega}{\partial k^{2}}=\frac{\hbar}{m}>0$ & decay\tabularnewline
\hline 
vibrations of elastic plates (Föppl-von Karman equation) & $\omega=ck^{2}$ with $c>0$ in the linear limit & $\frac{\partial^{2}\omega}{\partial k^{2}}=2c>0$ & decay\tabularnewline
\hline 
Plasma & many different dispersion laws &  & both cases exist (\textit{e.g. }Langmuir waves: decay; O-waves of
electrons: \textbf{nondecay})\tabularnewline
\hline 
Rossby waves & $\omega\propto\frac{k_{x}}{k_{y}^{2}}$ where $k_{x}\ll k_{y}$ &  & anisotropic\tabularnewline
\hline 
Petviashvilli equation $\partial_{t}\psi=\nabla^{2}\partial_{x}\psi-\psi\partial_{x}\psi$
(non-isotropic) & $\omega=k_{x}k^{2}$ &  & anisotropic\tabularnewline
\hline 
Charney-Hassegawa-Mima Equation $\partial_{t}\left[\nabla^{2}\psi-\rho^{-2}\psi\right]+\beta\partial_{x}\psi+\left(\partial_{x}\psi\right)\left(\partial_{y}\nabla^{2}\psi\right)-\left(\partial_{y}\psi\right)\left(\partial_{x}\nabla^{2}\psi\right)=0$(model
for Rossby waves and plasma drift waves, see \cite{nazarenko_wave_2011}) & $\omega=-\frac{\beta\rho^{2}k_{x}}{1+\rho^{2}k^{2}}$ &  & anisotropic\tabularnewline
\hline 
magnetized ion sound & $\omega=ck_{z}(1-k_{\perp}c_{s}/2\Omega_{s}^{2})$ &  & anisotropic\tabularnewline
\hline 
Heisenberg ferromagnet & $\omega=\sqrt{\omega_{m}(\omega_{0}-\omega_{m}N_{z})}\left|\frac{k_{z}}{k_{\perp}}\right|$ &  & anisotropic\tabularnewline
\hline 
\end{tabular}\caption{{\footnotesize Some classical dispersion relations \label{tab:dispersion-relations}}}
\end{table}

\subsection{Discussion of the local approximation \label{subsec:Appendix:-Local-approx-computation}}

In this appendix, we discuss the range of validity of the ``strong''
local approximation leading to the demonstration given in \cite{dyachenko1992optical}.
In a second paragraph, we give the expression of the large deviation
Hamiltonian if the dynamics cannot be reduced to $2\to2$ interactions.

\subsubsection{Range of validity}

In this paragraph, we will make more explicit the approximation which
is done when considering ``local'' approximation. 

For compactness, we will denote $\vec{\sigma}=\left(+1,+1,-1,-1\right)$,
$\vec{\bm{k}}=\left(\bm{k}_{1},\bm{k}_{2},\bm{k}_{3},\bm{k}_{4}\right)$,
$\vec{\lambda}=\left(\lambda_{1},\lambda_{2},\lambda_{3},\lambda_{4}\right)$
and $\vec{\frac{1}{n}}=\left(\frac{1}{n_{1}},\frac{1}{n_{2}},\frac{1}{n_{3}},\frac{1}{n_{4}}\right)$.
As in \cite{dyachenko1992optical}, we write for $i=2,3,4$ : $\omega_{i}=\omega_{1}\left(1+p_{i}\right)$,
use the homogeneity of $S$ and Taylor expand the terms $\vec{\sigma}.\left(\vec{\lambda}+\vec{\frac{1}{n}}\right)$
and $\vec{\sigma}.\vec{\lambda}$ up to second order, but we will
keep track of the higher orders $\mathcal{O}\left(p_{i}^{3}\right)$
\begin{align*}
H_{iso}\left[N,\Lambda\right] & ={\displaystyle \int\text{d}\omega_{1}\text{d}\omega_{2}\text{d}\omega_{3}\text{d}\omega_{4}S_{\vec{\omega}}}\delta\left(\vec{\sigma}\cdot\vec{\omega}\right)\left(\vec{\sigma}\cdot\vec{\lambda}\right)n_{1}n_{2}n_{3}n_{4}\left[\vec{\sigma}\cdot\left(\frac{1}{\vec{n}}+\vec{\lambda}\right)\right]\\
 & ={\displaystyle \int\text{d}\omega_{1}\omega_{1}^{3}\text{d}p_{2}\text{d}p_{3}\text{d}p_{4}\omega_{1}^{\gamma_{S}}S_{\overrightarrow{1+p}}\frac{1}{\omega_{1}}}\delta\left(\vec{\sigma}\cdot\overrightarrow{(1+p)}\right)\left(\sum_{i}\sigma_{i}\left[\lambda+\omega_{1}\partial_{\omega}^ {}\lambda p_{i}+\frac{1}{2}\omega_{1}^{2}\partial_{\omega}^{2}\left(\lambda\right)p_{i}^{2}+\mathcal{O}\left(\omega_{1}^{3}p_{i}^{3}\right)\right]\right)\\
 & \times n_{1}^{4}\omega_{1}^{2}\left[\sum_{j}\sigma_{j}\left[\frac{1}{n}+\lambda+\omega_{1}\partial_{\omega}^ {}\left(\frac{1}{n}+\lambda\right)p_{j}+\frac{1}{2}\omega_{1}^{2}\partial_{\omega}^{2}\left(\frac{1}{n}+\lambda\right)p_{j}^{2}+\mathcal{O}\left(\omega_{1}^{3}p_{i}^{3}\right)\right]\right]
\end{align*}
In the expansion, we will find the same result as \cite{dyachenko1992optical}
for the part without higher orders, which we will not write and leave
as ``$\ldots$''. The zeroth orders of the expansion in $p_{i}$
will eliminate due to the null sum of the $\sigma_{i}$, and the first
orders will lead to terms $\vec{\sigma}\cdot\vec{p}=\vec{\sigma}\cdot\overrightarrow{\left(1+p\right)}$
since only $2\rightarrow2$ waves interactions remain in our choice.
Therefore, both vanish. 
\begin{align*}
H_{iso}\left[N,\Lambda\right] & =\ldots+\frac{1}{2}{\displaystyle \int\text{d}\omega_{1}n_{1}^{4}\omega_{1}^{\gamma_{S}+9}\int\text{d}p_{2}\text{d}p_{3}\text{d}p_{4}S_{\overrightarrow{1+p}}}\delta\left(\vec{\sigma}\cdot\overrightarrow{p}\right)\\
 & \times\left\{ \left[\sum_{i}\sigma_{i}\mathcal{O}\left(p_{i}^{3}\right)\right]\left[\sum_{j}\sigma_{j}\partial_{\omega}^{2}\left(\frac{1}{n}+\lambda\right)p_{j}^{2}\right]\right.\\
 & \left.+\left[\sum_{i}\sigma_{i}\partial_{\omega}^{2}\lambda p_{i}^{2}\right]\left[\sum_{j}\sigma_{j}\mathcal{O}\left(p_{i}^{3}\right)\right]\right\} \\
 & =\ldots+\frac{1}{2}{\displaystyle \int\text{d}\omega_{1}n_{1}^{4}\omega_{1}^{\gamma_{S}+9}\int\text{d}p_{2}\text{d}p_{3}\text{d}p_{4}S_{\overrightarrow{1+p}}}\delta\left(\vec{\sigma}\cdot\overrightarrow{p}\right)\\
 & \times\left[\sum_{i}\sigma_{i}p_{i}^{2}\right]\left\{ \partial_{\omega}^{2}\left(\frac{1}{n}+\lambda\right)\left[\sum_{j}\sigma_{j}\mathcal{O}\left(p_{j}^{3}\right)\right]+\partial_{\omega}^{2}\lambda\left[\sum_{j}\sigma_{j}\mathcal{O}\left(p_{i}^{3}\right)\right]\right\} \\
 & =\ldots+\mathcal{O}\left({\displaystyle \int\text{d}\omega_{1}n_{1}^{4}\omega_{1}^{\gamma_{S}+9}\int\text{d}p_{2}\text{d}p_{3}\text{d}p_{4}S_{\overrightarrow{1+p}}}\delta\left(\vec{\sigma}\cdot\overrightarrow{(1+p)}\right)\left[\sum_{i}\sigma_{i}p_{i}^{2}\right]\left[\sum_{j}\sigma_{j}p_{j}^{3}\right]\right)
\end{align*}
In other words, the ``strong'' local approximation is valid as soon
as the integral prefactors have different order of magnitude:
\[
\int\text{d}p_{2}\text{d}p_{3}\text{d}p_{4}S_{\overrightarrow{1+p}}\delta\left(\vec{\sigma}\cdot\overrightarrow{p}\right)\left(\sum_{i}\sigma_{i}p_{i}^{2}\right)^{2}\gg{\displaystyle \int\text{d}p_{2}\text{d}p_{3}\text{d}p_{4}S_{\overrightarrow{1+p}}}\delta\left(\vec{\sigma}\cdot\overrightarrow{p}\right)\left[\sum_{i}\sigma_{i}p_{i}^{2}\right]\left[\sum_{j}\sigma_{j}p_{j}^{3}\right]
\]
which can be true if $S$ is peaked enough around $\left(1,1,1,1\right)$.
To give a very rough idea of when this has good chances to be true
without exact calculation, we can do a simple reasoning. We denote
$\varepsilon$ the characteristic width of the distribution $S_{\overrightarrow{1+p}}$.
Since this integral is made of products of integrals of the type $\int\text{d}pS_{\overrightarrow{1+p}}p_{i}^{a}$
(\textit{i.e. }moments of the distribution $S_{\overrightarrow{1+p}}$),
the order of magnitude of $\int\text{d}p_{2}\text{d}p_{3}\text{d}p_{4}S_{\overrightarrow{1+p}}\delta\left(\vec{\sigma}\cdot\overrightarrow{p}\right)\left(\sum_{i}\sigma_{i}p_{i}^{2}\right)^{2}$
should be $\varepsilon^{4}$ whereas that of ${\displaystyle \int\text{d}p_{2}\text{d}p_{3}\text{d}p_{4}S_{\overrightarrow{1+p}}}\delta\left(\vec{\sigma}\cdot\overrightarrow{p}\right)\left[\sum_{i}\sigma_{i}p_{i}^{2}\right]\left[\sum_{j}\sigma_{j}p_{j}^{3}\right]$
should be $\varepsilon^{5}$. This means that for the local hypothesis
to be valid, an order of magnitude to keep in mind is that the characteristic
width $\varepsilon$ of the distribution $S_{\overrightarrow{1+p}}$
should check 
\[
\varepsilon\ll1.
\]
To our knowledge, this clear criterion, although often implicit, has
never been written as such in the literature.

\subsubsection{What happens when the dynamics cannot be reduced to $2\to2$ interactions
?}

We still suppose the leading interactions are 4-wave. For compactness,
we will denote $\vec{\sigma}=\left(\sigma_{1},\sigma_{2},\sigma_{3},\sigma_{4}\right)\in\left\{ \pm1\right\} ^{4}$,
$\vec{\bm{k}}=\left(\bm{k}_{1},\bm{k}_{2},\bm{k}_{3},\bm{k}_{4}\right)$,
$\vec{\lambda}=\left(\lambda_{1},\lambda_{2},\lambda_{3},\lambda_{4}\right)$
and $\vec{\frac{1}{n}}=\left(\frac{1}{n_{1}},\frac{1}{n_{2}},\frac{1}{n_{3}},\frac{1}{n_{4}}\right)$.

In the case where we have resonant terms of the type $1\to3$ (those
of type $4\to0$ are always null because we cannot have $\delta\left(\vec{\sigma}\cdot\vec{\omega}\right)\delta\left(\vec{\sigma}\cdot\vec{\bm{k}}\right)\neq0$),
all the computations are exactly similar up to the isotropic Hamiltonian
(\ref{eq:LD-Hamiltonian-isotropic}) except that we have to sum over
more terms, such that the interaction kernels $W_{\overrightarrow{\bm{k}}}^{\vec{\sigma}}$
and $S_{\omega_{1}\omega_{2}\omega_{3}\omega_{4}}^{\vec{\sigma}}$would
take an argument $\vec{\sigma}$. In addition, the hamiltonian takes
a sum over the signature $\vec{\sigma}$ inherited from the dynamic's
Hamiltonian $\mathcal{H}$. The only thing that we need to change
is to add a $\sum_{\vec{\sigma}}$ at every line. We end up with the
Hamiltonian (\ref{eq:LD-Hamiltonian-isotropic}) rewriting 
\[
H_{iso}\left[N,\Lambda\right]=\sum_{\vec{\sigma}}{\displaystyle \int\text{d}\omega_{1}\text{d}\omega_{2}\text{d}\omega_{3}\text{d}\omega_{4}S_{\omega_{1}\omega_{2}\omega_{3}\omega_{4}}^{\vec{\sigma}}}\delta\left(\vec{\sigma}.\vec{\omega}\right)\left(\vec{\sigma}.\vec{\lambda}\right)n_{1}n_{2}n_{3}n_{4}\left[\vec{\sigma}.\left(\vec{\lambda}+\vec{\frac{1}{n}}\right)\right].
\]
We proceed exactly as in \cite{dyachenko1992optical}, but do not
assume $2\to2$ interactions. We write for $i=2,3,4$ : $\omega_{i}=\omega_{1}\left(1+p_{i}\right)$,
use the homogeneity of $S$ and Taylor expand the terms $\vec{\sigma}.\left(\vec{\lambda}+\vec{\frac{1}{n}}\right)$
and $\vec{\sigma}.\vec{\lambda}$ up to second order.
\begin{align*}
H_{iso}\left[N,\Lambda\right] & \simeq\sum_{\vec{\sigma}}{\displaystyle \int\text{d}\omega_{1}\omega_{1}^{3}\text{d}p_{2}\text{d}p_{3}\text{d}p_{4}\omega_{1}^{\gamma_{S}}S_{\overrightarrow{1+p}}^{\vec{\sigma}}\frac{1}{\omega_{1}}}\delta\left(\vec{\sigma}\cdot\overrightarrow{(1+p)}\right)\left(\sum_{i}\sigma_{i}\left[\lambda+\omega_{1}\nabla\lambda p_{i}+\frac{1}{2}\omega_{1}^{2}\nabla^{2}\left(\lambda\right)p_{i}^{2}\right]\right)n_{1}^{4}\omega_{1}^{2}\\
 & \qquad\times\left[\sum_{j}\sigma_{j}\left[\frac{1}{n}+\lambda+\omega_{1}\nabla\left(\frac{1}{n}+\lambda\right)p_{j}+\frac{1}{2}\omega_{1}^{2}\nabla^{2}\left(\frac{1}{n}+\lambda\right)p_{j}^{2}\right]\right]
\end{align*}

every time we have a term like $\sum_{i}\sigma_{i}p_{i}=\vec{\sigma}\cdot\vec{p}$
we can use the $\delta\left(\vec{\sigma}\cdot\overrightarrow{(1+p)}\right)$
to write $-\sum_{i}\sigma_{i}$ instead. If we expand the product,
this yields:{\footnotesize
\begin{align*}
{\tiny {\scriptsize {\scriptsize {\footnotesize {\tiny {\footnotesize {\small H_{iso}\left[N,\Lambda\right]}}}}}}} & {\tiny {\scriptsize {\scriptsize {\footnotesize {\tiny {\footnotesize {\small =\int\text{d}\omega_{1}\omega_{1}^{\gamma_{S}+4}n_{1}^{4}\sum_{\vec{\sigma}}{\displaystyle \int\text{d}p_{2}\text{d}p_{3}\text{d}p_{4}S_{\overrightarrow{1+p}}^{\vec{\sigma}}}\delta\left(\vec{\sigma}\cdot\overrightarrow{(1+p)}\right)\left[\left(\sum_{i}\sigma_{i}\right)^{2}\left[\lambda-\omega_{1}\nabla\lambda\right]\left[\frac{1}{n}+\lambda-\omega_{1}\nabla\left(\frac{1}{n}+\lambda\right)\right]\right.}}}}}}}\\
{\tiny {\scriptsize {\scriptsize {\footnotesize {\tiny {\footnotesize {\small }}}}}}} & {\tiny {\scriptsize {\scriptsize {\footnotesize {\tiny {\footnotesize {\small +\sum_{i}\sigma_{i}\left(\left[\lambda-\omega_{1}\nabla\lambda\right]\left[\sum_{j}\sigma_{j}\frac{1}{2}\omega_{1}^{2}\nabla^{2}\left(\frac{1}{n}+\lambda\right)p_{j}^{2}\right]+\left[\frac{1}{n}+\lambda-\omega_{1}\nabla\left(\frac{1}{n}+\lambda\right)\right]\left[\sum_{i}\sigma_{i}\frac{1}{2}\omega_{1}^{2}\nabla^{2}\left(\lambda\right)p_{i}^{2}\right]\right)\hfill\hfill\left(*\right)}}}}}}}\\
{\tiny {\scriptsize {\scriptsize {\footnotesize {\tiny {\footnotesize {\small }}}}}}} & {\tiny {\scriptsize {\scriptsize {\footnotesize {\tiny {\footnotesize {\small \left.+\left(\sum_{j}\sigma_{j}\frac{1}{2}\omega_{1}^{2}\nabla^{2}\left(\frac{1}{n}+\lambda\right)p_{j}^{2}\right)\left(\sum_{i}\sigma_{i}\frac{1}{2}\omega_{1}^{2}\nabla^{2}\left(\lambda\right)p_{i}^{2}\right)\right]}}}}}}}\\
{\tiny {\scriptsize {\scriptsize {\footnotesize {\tiny {\footnotesize {\small }}}}}}} & {\tiny {\scriptsize {\scriptsize {\footnotesize {\tiny {\footnotesize {\small =\int\text{d}\omega_{1}\omega_{1}^{\gamma_{S}+4}n_{1}^{4}\sum_{\vec{\sigma}}{\displaystyle \int\text{d}p_{2}\text{d}p_{3}\text{d}p_{4}S_{\overrightarrow{1+p}}^{\vec{\sigma}}}\delta\left(\vec{\sigma}\cdot\overrightarrow{(1+p)}\right)\left[\left(\sum_{i}\sigma_{i}\right)^{2}\left[\lambda-\omega_{1}\nabla\lambda\right]\left[\frac{1}{n}+\lambda-\omega_{1}\nabla\left(\frac{1}{n}+\lambda\right)\right]\right.}}}}}}}\\
{\tiny {\scriptsize {\scriptsize {\footnotesize {\tiny {\footnotesize {\small }}}}}}} & {\tiny {\scriptsize {\scriptsize {\footnotesize {\tiny {\footnotesize {\small \left.+\left(\sum_{j}\sigma_{j}\frac{1}{2}\omega_{1}^{2}\nabla^{2}\left(\frac{1}{n}+\lambda\right)p_{j}^{2}\right)\left(\sum_{i}\sigma_{i}\frac{1}{2}\omega_{1}^{2}\nabla^{2}\left(\lambda\right)p_{i}^{2}\right)\right]}}}}}}}
\end{align*}
}where the second line $\left(*\right)$ term vanishes due to the
symmetry $\sigma\leftrightarrow-\sigma$ in the sum $\sum_{\vec{\sigma}}$,
but the other one stay because some terms $1\to3$ give factors $2^{2}$. 

Instead of (\ref{eq:LD-Hamiltonian-isotropic}), we would therefore
obtain
\begin{equation}
H_{l}\left[N,\Lambda\right]=S_{0}\int\text{d}\omega n^{4}\omega^{s}\left[\partial_{\omega}^{2}\lambda\partial_{\omega}^{2}\left(\frac{1}{n}+\lambda\right)\right]+S_{1}\int\text{d}\omega\omega^{\gamma_{S}+4}n^{4}\left(\lambda-\omega\partial_{\omega}\lambda\right)\left(\frac{1}{n}+\lambda-\omega\partial_{\omega}\left(\frac{1}{n}+\lambda\right)\right)\label{eq:LD-hamiltonian-local-generalization-decay-type}
\end{equation}
 where $S_{0},S_{1}$ are numbers given by the integrals
\begin{align*}
S_{0} & =\frac{1}{4}\sum_{\vec{\sigma}}\int\text{d}p_{2}\text{d}p_{3}\text{d}p_{4}\hat{S}_{\overrightarrow{1+p}}^{\vec{\sigma}}\delta\left(\vec{\sigma}\cdot\overrightarrow{(1+p)}\right)\left(\sum_{i}\sigma_{i}p_{i}^{2}\right)^{2}\\
S_{1} & =4\sum_{\vec{\sigma}}{\displaystyle \int\text{d}p_{2}\text{d}p_{3}\text{d}p_{4}\hat{S}_{\overrightarrow{1+p}}^{\vec{\sigma}}}\delta\left(\vec{\sigma}\cdot\overrightarrow{(1+p)}\right)
\end{align*}
Interestingly, we do can generalise our results when the dynamics
cannot be reduced to $2\to2$ interactions (for instance if the dispersion
relation is of the decay type). The local large deviation Hamiltonian
has a modified expression (\ref{eq:LD-hamiltonian-local-generalization-decay-type})
with an additional term. This term is however still quadratic in $\lambda$,
meaning that the statistics of $n$ is locally Gaussian. However,
this is easily checked that the second term does not necessarily conserve
mass and energy as in the classical case. 

\subsection{Review of some values for interaction kernels and exponents in different
physical cases\label{subsec:Appendix:values-W-exponents}}

Let us write a few values of these exponents which will appear regularly
in this manuscript. We recall their definition: $d$ is the {\small dimension,}
$W_{\vec{\bm{k}}}$ is the kernel of the wave dynamics, $\alpha$
is the exponent of the dispersion law, $\gamma_{S}=\frac{2\gamma_{W}+3d}{\alpha}-4$
and $s=6+\gamma_{S}$, and gather them in the table \ref{tab:exponents}.

\begin{table}[H]
\begin{centering}
\begin{tabular}{|>{\raggedright}p{0.3\textwidth}|>{\centering}p{0.03\textwidth}|>{\centering}p{0.2\textwidth}|>{\centering}p{0.1\textwidth}|>{\centering}p{0.1\textwidth}|>{\centering}p{0.09\textwidth}|>{\centering}p{0.08\textwidth}|}
\hline 
\rowcolor{lightgray!50}
\footnotesize\textbf{Model} & {\small$d$} & $W_{\vec{\bm{k}}}$ & $\alpha$ & $\gamma_{W}$ & $\gamma_{S}$ & $s$\tabularnewline
\hline 
\hline 
{\small Non-linear Schrödinger equation $i\partial_{t}\psi=\mu\nabla^{2}\psi+\epsilon\left|\psi\right|^{2}\psi$} & 3;2 & $W_{kk_{1}k_{2}k_{3}}=1$ & 2 & {\small 0} & {\small$1/2;-1$} & {\small$9/2;5$}\tabularnewline
\hline 
{\small Petviashvilli equation $\partial_{t}\psi=\nabla^{2}\partial_{x}\psi-\psi\partial_{x}\psi$} & 3 & {\small$W_{\bm{kk_{1}k_{2}}}=\frac{1}{2}\sqrt{\left|k_{x}k_{1x}k_{2x}\right|}$} & anisotropic &  &  & \tabularnewline
\hline 
{\small Charney-Hassegawa-Mima Equation $\partial_{t}\left[\nabla^{2}\psi-\rho^{-2}\psi\right]+\beta\partial_{x}\psi+\left(\partial_{x}\psi\right)\left(\partial_{y}\nabla^{2}\psi\right)-\left(\partial_{y}\psi\right)\left(\partial_{x}\nabla^{2}\psi\right)=0$} & 2 & {\footnotesize$W_{\bm{k}\bm{k}_{2}\bm{k}_{3}}=\frac{i\rho^{2}\beta^{1/2}}{2}\sqrt{\left|k_{x}k_{1x}k_{2x}\right|}\times\left(\frac{k_{1y}}{1+\rho^{2}k_{1}^{2}}+\frac{k_{2y}}{1+\rho^{2}k_{2}^{2}}-\frac{k_{y}}{1+\rho^{2}k^{2}}\right)$} & anisotropic &  &  & \tabularnewline
\hline 
{\small deep water gravity waves ($h>\frac{1}{2}\lambda$; no surface
tension)} & 2 & see general case in \cite{krasitskii_reduced_1994} & $1/2$ & {\scriptsize 3-wave term: $\frac{7}{4}$}{\scriptsize\par}

{\scriptsize and 4-wave term: $3$} & {\scriptsize 3-wave: $15$ ;}{\scriptsize\par}

{\scriptsize 4-wave: 20} & {\scriptsize 3-wave: 9 ;}{\scriptsize\par}

{\scriptsize 4-wave: 14}\tabularnewline
\hline 
{\small shallow water gravity waves ($h<0.05\lambda$, no surface tension)} & 2 & see general case in \cite{krasitskii_reduced_1994} & $1$ & {\scriptsize 3-wave term: $\frac{5}{2}$}{\scriptsize\par}

{\scriptsize and 4-wave term: $3$} & {\scriptsize 3-wave: 7 ;}{\scriptsize\par}

{\scriptsize 4-wave: 8} & {\scriptsize 3-wave: 9 ;}{\scriptsize\par}

{\scriptsize 4-wave: 10}\tabularnewline
\hline 
{\small intermediate depth water capillary-gravity waves (any $h,\lambda$,
possibly surface tension, see \cite{krasitskii_reduced_1994} for
computations)} & 2 & see part 4 of \cite{krasitskii_reduced_1994} for the complete expression & $\omega=\sqrt{gk\tanh(kh)}$ : not a power law &  &  & \tabularnewline
\hline 
acoustic waves & 3 &  & $1$ (first order in $k$) &  &  & \tabularnewline
\hline 
light in vacuum & 3 & $W_{\vec{\bm{k}}}=0$ & / &  &  & \tabularnewline
\hline 
\end{tabular}
\par\end{centering}
\caption{{\footnotesize Classical exponent values.}}
{\footnotesize\label{tab:exponents}}{\footnotesize\par}
\end{table}

\subsection{Computation of the equilibrium quasipotential\label{subsec:Appendix:-equil-quasipot}}

We use a method exactly analog to \cite{guioth_path_2022} and adopt
the same notations. We denote $\mathcal{H}$ the energy, which is
conserved since we follow a Hamiltonian dynamics. We also denote $\mathscr{N}=\left(\frac{2\pi}{L}\right)^{d}\sum_{\bm{k}}A_{\bm{k}}A_{\bm{k}}^{*}$
the (total) wave action, which is conserved by the microscopic dynamics.
Eventually, we denote $\bm{K}=\left(\frac{2\pi}{L}\right)^{d}\sum_{\bm{k}}\bm{k}A_{\bm{k}}A_{\bm{k}}^{*}$
the total momentum which, in theory, should be zero in the isotropic
hypothesis. 

We define the microcanonical distribution
\[
\text{d}\mu_{\mathcal{E},\mathcal{N},L,\epsilon}=\frac{1}{\Gamma_{\mathcal{E},\mathcal{N},L,\epsilon}}\delta\left(\mathcal{E}-\mathcal{H}\right)\delta\left(\mathscr{N}-\mathcal{N}\right)\delta\left(\bm{K}\right)\prod_{\bm{k}}\text{d}A_{\bm{k}}\text{d}A_{\bm{k}}^{*}
\]
with $\Gamma_{\mathcal{E},\mathcal{N},L,\epsilon}=\int\delta\left(\mathcal{E}-\mathcal{H}\right)\delta\left(\mathscr{N}-\mathcal{N}\right)\delta\left(\bm{K}\right)\prod_{\bm{k}}\text{d}A_{\bm{k}}\text{d}A_{\bm{k}}^{*}$
the volume of the phase space associated with the macrostate$(\mathcal{E},\mathcal{N})$.
We keep the indices $L,\epsilon$ in order to remember that the energy
$\mathcal{H}$ depends on $\epsilon$ and that the system has size
$L$. 

First, we introduce the expression of the (rescaled) energy
\[
\mathcal{H}={\displaystyle \left(\frac{2\pi}{L}\right)^{d}\underset{:=\mathcal{H}_{2}=\mathcal{O}\left(\left(\frac{2\pi}{L}\right)^{d}\right)}{\underbrace{\sum_{\bm{k}}\omega_{\bm{k}}A_{\bm{k}}A_{\bm{k}}^{*}}}+\underset{\mathcal{O}\left(\left(\frac{2\pi}{L}\right)^{2d}\right)}{\underbrace{\left(\frac{2\pi}{L}\right)^{2d}\sum_{\bm{k}_{1}\bm{k}_{2}\bm{k}_{3}\bm{k}_{4}}\sum_{\sigma_{1}\sigma_{2}\sigma_{3}\sigma_{4}}W_{\bm{k}_{1}\bm{k}_{2}\bm{k}_{3}\bm{k}_{4}}^{\sigma_{1}\sigma_{2}\sigma_{3}\sigma_{4}}A_{\bm{k}_{1}}^{\sigma_{1}}A_{\bm{k}_{2}}^{\sigma_{2}}A_{\bm{k}_{3}}^{\sigma_{3}}A_{\bm{k}_{4}}^{\sigma_{4}}\delta_{\sigma_{1}\bm{k}_{1}+\sigma_{2}\bm{k}_{2}+\sigma_{3}\bm{k}_{3}+\sigma_{4}\bm{k}_{4},0}}}}.
\]
In the $L\to\infty$ limit, we expect the large deviation principle
\[
Q_{\mathcal{E},\mathcal{N}}\left[n\right]=-\text{Kin Lim}\left(\frac{2\pi}{L}\right)^{d}\ln P_{\mathcal{E},\mathcal{N},L,\epsilon}\left[n\right]
\]
where $P_{\mathcal{E},\mathcal{N},L,\epsilon}\left[n\right]$ is the
equilibrium pdf of the empirical density $\hat{n}(\bm{k})=A_{\bm{k}}A_{\bm{k}}^{*}$,
which writes 
\begin{equation}
P_{\mathcal{E},\mathcal{N},L,\epsilon}\left[n\right]=\mathbb{E}_{\mathcal{E},\mathcal{N},L,\epsilon}\left[\delta\left(\hat{n}-n\right)\right]=\frac{\Gamma_{\mathcal{E},\mathcal{N},L,\epsilon}\left[n\right]}{\Gamma_{\mathcal{E},\mathcal{N},L,\epsilon}}\label{eq:P_E-L-eps-N}
\end{equation}
where 
\begin{equation}
\Gamma_{\mathcal{E},\mathcal{N},L,\epsilon}\left[n\right]=\int\prod_{k}\text{d}A_{\bm{k}}\text{d}A_{\bm{k}}^{*}\delta\left(\mathcal{E}-\mathcal{H}_{2}+\mathcal{O}\left(\epsilon^{2}\right)\right)\delta\left(\bm{K}\right)\delta\left(\mathscr{N}-\mathcal{N}\right)\cdot\delta\left(\hat{n}-n\right)\label{eq:Gamma-n}
\end{equation}
is the volume of phase space associated with the macrostate $(\mathcal{E},\mathcal{N})$. 

We define the entropy 
\[
s_{\mathcal{E},\mathcal{N}}\left[n\right]=\text{Kin Lim}_{L\to\infty}\left(\frac{2\pi}{L}\right)^{d}\ln\Gamma_{\mathcal{E},\mathcal{N},L,\epsilon}\left[n\right]
\]
which enable to rewrite 
\[
Q_{\mathcal{E},\mathcal{N}}\left[n\right]=-s_{\mathcal{E},\mathcal{N}}\left[n\right]+\text{Cst}
\]
where the constant is chosen to check the condition $\inf_{n}Q_{\mathcal{E},\mathcal{N}}\left[n\right]=0$.
The entropy $s_{\mathcal{E},\mathcal{N}}\left[n\right]$ can be computed
using the Legendre-Fenchel transform of the free energy:

\begin{align*}
f_{\beta,\alpha}\left[\lambda\right] & =-\underset{L\to\infty}{\text{Kin Lim}}\left(\frac{2\pi}{L}\right)^{d}\ln\left\langle e^{-\left(\frac{L}{2\pi}\right)^{d}\left(\beta\mathcal{H}+\alpha\mathscr{N}+\left(\frac{2\pi}{L}\right)^{d}\sum_{\bm{k}}\lambda\left(\text{\ensuremath{\bm{k}}}\right)A_{\bm{k}}A_{\bm{k}}^{*}\right)}\right\rangle \\
 & =-\underset{L\to\infty}{\text{Kin Lim}}\left(\frac{2\pi}{L}\right)^{d}\ln\left\langle e^{-\sum_{\bm{k}}\left(\beta\omega_{\bm{k}}\left|A_{\bm{k}}\right|^{2}+\mathcal{O}\left(\left(\frac{2\pi}{L}\right)^{d}\right)+\alpha\left|A_{\bm{k}}\right|^{2}+\lambda\left(\text{\ensuremath{\bm{k}}}\right)\left|A_{\bm{k}}\right|^{2}\right)}\right\rangle \\
 & =-\underset{L\to\infty}{\text{Kin Lim}}\left(\frac{2\pi}{L}\right)^{d}\ln\left[\prod_{\bm{k}}\int\text{d}A_{\bm{k}}\text{d}A_{\bm{k}}^{*}e^{-\left(\beta\omega_{\bm{k}}\left|A_{\bm{k}}\right|^{2}+\alpha\left|A_{\bm{k}}\right|^{2}+\lambda\left(\text{\ensuremath{\bm{k}}}\right)\left|A_{\bm{k}}\right|^{2}\right)}\right]\\
 & =-\underset{L\to\infty}{\text{Kin Lim}}\left(\frac{2\pi}{L}\right)^{d}\ln\left[\prod_{\bm{k}}\left(\int\text{d}\left|A_{\bm{k}}\right|^{2}\text{d}\varphi_{\bm{k}}\right)e^{-\left(\beta\omega_{\bm{k}}\left|A_{\bm{k}}\right|^{2}+\alpha\left|A_{\bm{k}}\right|^{2}+\lambda\left(\text{\ensuremath{\bm{k}}}\right)\left|A_{\bm{k}}\right|^{2}\right)}\right]\\
 & =-\underset{L\to\infty}{\text{Kin Lim}}\left(\frac{2\pi}{L}\right)^{d}\sum_{\bm{k}}\ln\left[2\pi\int\text{d}\left|A_{\bm{k}}\right|^{2}e^{-\left(\beta\omega_{\bm{k}}+\alpha+\lambda\left(\text{\ensuremath{\bm{k}}}\right)\right)\left|A_{\bm{k}}\right|^{2}}\right]\\
 & =-\underset{L\to\infty}{\text{Kin Lim}}\left(\frac{2\pi}{L}\right)^{d}\sum_{\bm{k}}\ln\left[\frac{2\pi}{\left(\beta\omega_{\bm{k}}+\alpha+\lambda\left(\text{\ensuremath{\bm{k}}}\right)\right)}\right]\\
 & =\int\text{d}\bm{k}\ln\left[\frac{\beta\omega_{\bm{k}}+\alpha+\lambda\left(\text{\ensuremath{\bm{k}}}\right)}{2\pi}\right]
\end{align*}
This limit being well-defined and differentiable, we use Gärtner-Ellis
theorem to write the entropy as the Legendre-Fenchel transform of
$f_{\beta,\alpha}\left[\lambda\right]$:
\begin{align*}
s_{\mathcal{E},\mathcal{N}}\left[n\right] & =\underset{\beta,\alpha,\lambda}{\inf}\left\{ \beta\mathcal{E}+\alpha\mathcal{N}+\int\text{d}\bm{k}n\left(\bm{k}\right)\lambda\left(\text{\ensuremath{\bm{k}}}\right)-f_{\beta,\alpha}\right\} 
\end{align*}
The stationarity condition at the minimum writes
\[
\begin{cases}
\mathcal{E} & =\frac{\partial f_{\beta,\alpha}}{\partial\beta}=\int\text{d}\bm{k}\frac{\omega_{\bm{k}}}{\beta\omega_{\bm{k}}+\alpha+\lambda\left(\text{\ensuremath{\bm{k}}}\right)}\\
\mathcal{N} & =\frac{\partial f_{\beta,\alpha}}{\partial\alpha}=\int\text{d}\bm{k}\frac{1}{\beta\omega_{\bm{k}}+\alpha+\lambda\left(\text{\ensuremath{\bm{k}}}\right)}\\
n(\bm{k}) & =\frac{\delta f_{\beta,\alpha}}{\delta\lambda(\bm{k})}=\frac{1}{\beta\omega_{\bm{k}}+\alpha+\lambda\left(\text{\ensuremath{\bm{k}}}\right)}
\end{cases}
\]
This system has 3 unknowns and 3 equations but they are not independent.
This means that we get a condition on the two values of $\mathcal{E},\bm{K},\mathcal{N}$
and $n$ : $\mathcal{E}=\int\text{d}\bm{k}\omega_{\bm{k}}n$,$\mathcal{N}=\int\text{d}\bm{k}n$.
In the supremum point, we have $\beta\mathcal{E}+\alpha\mathcal{N}+\int\text{d}\bm{k}n\left(\bm{k}\right)\lambda\left(\text{\ensuremath{\bm{k}}}\right)=\int\text{d}\bm{k}\left[\beta\omega_{\bm{k}}+\alpha+\lambda\left(\text{\ensuremath{\bm{k}}}\right)\right]n\left(\bm{k}\right)=1$,
so we deduce
\[
s_{\mathcal{E},\mathcal{N}}\left[n\right]=\begin{cases}
\int\text{d}\bm{k}\left(1+\ln\left(2\pi\right)+\ln\left(n\left(\bm{k}\right)\right)\right) & \text{if }\mathcal{E}=\int\text{d}\bm{k}n\omega_{\bm{k}}\text{ and }\mathcal{N}=\int\text{d}\bm{k}n\\
-\infty & \text{otherwise}
\end{cases}
\]
We deduce the quasipotential:

\begin{equation}
Q_{\mathcal{E},\mathcal{N}}\left[n\right]=\begin{cases}
-\int\text{d}\bm{k}\ln\left(n\left(\bm{k}\right)\right)+\text{Cst} & \text{if }\mathcal{E}=\int\text{d}\bm{k}n\omega_{\bm{k}}\text{ and }\mathcal{N}=\int\text{d}\bm{k}n\\
+\infty & \text{otherwise}
\end{cases}\label{eq:eq-quasipot-2}
\end{equation}
The constant should be computed using the condition $\text{ }\mathcal{E}=\int\text{d}\bm{k}n\omega_{\bm{k}}\text{ and }\mathcal{N}=\int\text{d}\bm{k}n$.
On the manifold defined by $\mathcal{E}=\int\text{d}\bm{k}n\omega_{\bm{k}}\text{ and }\mathcal{N}=\int\text{d}\bm{k}n$,
any function combination $f(\mathcal{E},\mathscr{N})$ is a constant
since the energy and number of waves are conserved. We have the variational
problem 
\begin{align}
0 & =\underset{\beta,\alpha,n}{\inf}\left\{ Q_{E\mathcal{E},\mathcal{N}}\left[n\right]+\beta\left(\mathcal{E}-\int\omega n\right)+\alpha\left(\mathcal{N}-\int n\right)\right\} \label{eq:varprobQeq}\\
 & =\underset{\beta,\alpha,n}{\inf}\left\{ Q_{\mathcal{E},\mathcal{N}}\left[n\right]-\int n\left[\beta\omega+\alpha\right]+\beta\mathcal{E}+\alpha\mathcal{N}\right\} 
\end{align}
which gives a value for $n$ corresponding to the post probable one
at equilibrium:
\[
n^{*}\left(\bm{k}\right)=\frac{1}{A\omega_{\bm{k}}+B}
\]
which is the Rayleigh-Jeans distribution. Then we have to fix the
value of the constant and $A,B$ so that (\ref{eq:varprobQeq}) is
true. We obtain

\begin{equation}
Q_{\mathcal{E},\mathcal{N}}\left[n\right]=\begin{cases}
\int\text{d}^{d}\bm{k}\left(\frac{n}{n^{*}}-1-\ln\left(\frac{n}{n^{*}}\right)\right) & \text{if }\mathcal{E}=\int\text{d}^{d}\bm{k}n\omega_{\bm{k}}\text{ and }\mathcal{N}=\int\text{d}^{d}\bm{k}n\\
+\infty & \text{otherwise}
\end{cases}\label{eq:eq-quasipot-1}
\end{equation}
This way of writing exhibits that the minimum of the quasipotential
is zero and reached in $n=n^{*}$ (indeed, the function $x\mapsto x-x-\ln x$
is convex and has a single minimum in $x=1$). It has the same expression
as in \cite{guioth_path_2022}. 

Then by contraction principle we obtain the quasipotential in the
isotropic case. We have the change of variable $N(\omega)=\Omega_{0}k^{d-1}\frac{dk}{d\omega}n(k(\omega))=\Omega_{0}\frac{\omega^{\frac{d}{\alpha}-1}}{\gamma^{d/\alpha}\alpha}n(k(\omega))$.
By contraction, the probability law of $N$ verifies a LDP with a
a large deviation functional at equilibrium (quasipotential) given
by
\begin{align*}
Q_{\mathcal{E},\mathcal{N}}\left[N\right] & =\inf\left\{ Q_{\mathcal{E},\mathcal{N}}\left[n\right]\left|N(\omega)=\Omega_{0}\frac{\omega^{\frac{d}{\alpha}-1}}{\gamma^{d/\alpha}\alpha}n(k(\omega))\right.\right\} 
\end{align*}
giving
\[
Q_{\mathcal{E},\mathcal{N}}\left[N\right]=\begin{cases}
\int\text{d}\omega\Omega_{0}\frac{\omega^{\frac{d}{\alpha}-1}}{\gamma^{d/\alpha}\alpha}\left(\frac{N}{N^{*}}-1-\ln\left(\frac{N}{N^{*}}\right)\right) & \text{if }\mathcal{E}=\int\text{d}\omega N\omega\text{ and }\mathcal{N}=\int\text{d}\omega N\\
+\infty & \text{otherwise}
\end{cases}
\]
where 
\[
N^{*}\left(\omega\right)=\Omega_{0}\frac{\omega^{\frac{d}{\alpha}-1}}{\gamma^{d/\alpha}\alpha}\frac{1}{a\omega+b}
\]
 is the Rayleigh-Jeans distribution and where $a,b$ are fixed so
that $\left(\mathcal{E}=\int\text{d}\omega N\omega\text{ and }\mathcal{N}=\int\text{d}\omega N\right)$
is verified. 

\subsection{Out-of-equilibrium dynamics induced by a forcing in the isotropic
local framework\label{subsec:Out-of-equilibrium-dynamics-induced-by-forcing}}

In this section, we introduce a forcing and a dissipation and we explain
their impact on the fluxes. For a situation where there is a separation
between forcing scales and dissipation scales, we compute explicitly
the values of the (constant) fluxes in the regions where forcing and
dissipation are absent (which we will call \textit{inertial ranges}).

As explained in paragraph \ref{par:Out-of-equilibrium-dynamics},
physical systems are often studies out-of-equilibirum by adding forcing
terms to amount for sources and dissipation of $N$. This can be done
by adding terms for source $s\left[N,\omega,t\right]\geq0$ and dissipation
$d\left[N,\omega,t\right]\geq0$ to the dynamics. Importantly, these
functions must be isotropic as well in order to keep isotropy of the
state $N$. We write the fluctuating dynamics:
\[
\partial_{t}N=\partial_{\omega}^{2}K+\left(s-d\right).
\]
This has the property to break the conservation laws since we create
and dissipate $N$ with the terms $s$ and $d$. We will now write
the expression of the fluxes. By integrating the conservation equations
between two frequencies $\omega$ and $\omega'$, we find that
\begin{align*}
\partial_{t}\int_{\omega_{}}^{\omega'}\text{d}\omega N+j_{N}\left(\omega'\right)-j_{N}\left(\omega\right) & =\int_{\omega}^{\omega'}\text{d}\omega\left(s-d\right)\\
\partial_{t}\int_{\omega_{}}^{\omega'}\text{d}\omega E+j_{E}\left(\omega'\right)-j_{E}\left(\omega\right) & =\int_{\omega}^{\omega'}\text{d}\omega\omega\left(s-d\right)
\end{align*}
If we are in stationary regime, it means that the flux differences
between two given frequencies are given by the integral of the forcing:
\begin{align}
j_{N}\left(\omega'\right)-j_{N}\left(\omega\right) & =\int_{\omega}^{\omega'}\text{d}\omega\left(s-d\right)\label{eq:je-jn-from-forcing-integrals}\\
j_{E}\left(\omega'\right)-j_{E}\left(\omega\right) & =\int_{\omega}^{\omega'}\text{d}\omega\omega\left(s-d\right)\nonumber 
\end{align}
Now we will study the paradigmatic case where there is a localized
source of amplitude $S_{f}$ around a given frequency $\omega_{f}$
and a dissipation at large scales $\omega\leq\omega_{d}^{-}$ and
at small scales $\omega\geq\omega_{d}^{+}$.  In the inertial ranges
(which are for the moment not supposed large) $\omega_{d}^{-}<\omega<\omega_{f}$
and $\omega_{f}<\omega<\omega_{d}^{+}$ the fluxes are constant. Let
us compute all fluxes in these both intervals. We denote $j_{N}^{(1)},j_{E}^{(1)}$
the (constant) value of the fluxes in the first inertial range, and
$j_{N}^{(2)},j_{E}^{(2)}$ the (constant) value of the fluxes in the
second one. The formulas (\ref{eq:je-jn-from-forcing-integrals})
give expressions for $j_{N}^{(1)},j_{E}^{(1)},j_{N}^{(2)},j_{E}^{(2)}$
(using that fluxes vanish in zero and at infinity). Then, it easy
to compare the value of $j_{E}$ and $j_{N}\omega$ in both inertial
ranges. In the first inertial range $\omega_{d}^{-}<\omega<\omega_{f}$
we have:
\begin{equation}
\left|j_{E}^{(1)}\right|=\int_{0}^{\omega_{d}^{-}}d\omega<\omega_{d}^{-}\int_{0}^{\omega_{d}^{-}}d=\omega_{d}^{-}\left|j_{N}^{(1)}\right|<\omega\left|j_{N}^{(1)}\right|\label{eq:ineq fluxes1}
\end{equation}
and in the second $\omega_{f}<\omega<\omega_{d}^{+}$: 
\begin{equation}
j_{E}^{(2)}=\int_{\omega_{d}^{+}}^{+\infty}d\omega>\omega_{d}^{+}\int_{\omega_{d}^{+}}^{+\infty}d=\omega_{d}^{+}j_{N}^{(2)}>\omega j_{N}^{(2)}\label{eq:ineq fluxes2}
\end{equation}
In addition, when the dissipation is made on a narrow range around
$\omega_{d}^{-}$ and $\omega_{d}^{+}$ (which we will suppose for
the rest of this part), the first inequalities written in (\ref{eq:ineq fluxes1},\ref{eq:ineq fluxes2})
become equalities and we actually have 
\begin{align}
j_{E}^{(1)} & =\omega_{d}^{-}j_{N}^{(1)}=\omega j_{N}^{(1)}\frac{\omega_{d}^{-}}{\omega}\label{eq:relations-je-jn-at-boundary-frequencyRatio}\\
j_{E}^{(2)} & =\omega_{d}^{+}j_{N}^{(2)}=\omega j_{N}^{(2)}\frac{\omega_{d}^{+}}{\omega}\nonumber 
\end{align}
So in the inertial ranges, the large inequalities remain valid, but
we have an exact correspondence between the fluxes $j_{e}$ and $\omega j_{N}$
up to frequency ratios given by $\frac{\omega_{d}^{-}}{\omega}$ and
$\frac{\omega_{d}^{+}}{\omega}$. Using equations (\ref{eq:je-jn-from-forcing-integrals}),
all fluxes can be obtained with the source amplitude $S_{f}$:
\begin{equation}
\begin{cases}
j_{N}^{(1)} & =-S_{f}\left(\frac{\omega_{d}^{+}-\omega_{f}}{\omega_{d}^{+}-\omega_{d}^{-}}\right)\\
j_{N}^{(2)} & =S_{f}\left(\frac{\omega_{f}-\omega_{d}^{-}}{\omega_{d}^{+}-\omega_{d}^{-}}\right)\\
j_{E}^{(1)} & =-S_{f}\omega_{d}^{-}\left(\frac{\omega_{d}^{+}-\omega_{f}}{\omega_{d}^{+}-\omega_{d}^{-}}\right)\\
j_{E}^{(2)} & =S_{f}\omega_{d}^{+}\left(\frac{\omega_{f}-\omega_{d}^{-}}{\omega_{d}^{+}-\omega_{d}^{-}}\right)
\end{cases}\label{eq:je-jn-with-forcingAmpllitude-and-frequencies}
\end{equation}
Classical argument (Fjørtoft argument, see \textit{e.g.} \cite{nazarenko_wave_2011})
state that in the first inertial range ($\omega_{d}^{-}\ll\omega\ll\omega_{f}$)
there is an inverse wave-action cascade with constant negative flux
and no energy flux, while in the second range ($\omega_{f}^ {}\ll\omega\ll\omega_{d}^{+}$)
there is a direct energy cascade with constant positive flux and no
wave-action flux. Here, we extended this by computing all flux values
even when the inertial ranges are not large. In the following, we
will mainly assume large inertial ranges ($\omega_{d}^{+}\gg\omega_{f}\gg\omega_{d}^{-}$)
yielding a direct energy cascade and an inverse wave-action cascade,
and we then derive the stationary solutions.

\subsection{Computation of the Kolmogorov constant for general wave turbulence\label{subsec:Appendix:-Computation-of-kolmogorov-constant}}

In this appendix, we compute the Kolmogorov constant in the complete
theory, where we suppose neither isotropy nor locality. The final
value is different from the one obtained in the isotropic local theory
(\ref{eq:n_N-sol-powerlaw}). This is a strictly analog result at
\cite{pushkarev2000turbulence}, adapted for 4-wave interactions.

We consider an isotropic four-wave kinetic equation written in frequency
variables as

\[
\partial_{t}n(\omega_{1})=b_{\mathrm{iso}}[n](\omega_{1})
\]
with collision integral

\[
b_{\mathrm{iso}}[n](\omega_{1})=4\int\mathrm{d}\omega_{2}\mathrm{d}\omega_{3}\mathrm{d}\omega_{4}S_{\omega_{1}\omega_{2}\omega_{3}\omega_{4}}\delta(\omega_{1}+\omega_{2}-\omega_{3}-\omega_{4})n_{1}n_{2}n_{3}n_{4}\left(\frac{1}{n_{1}}+\frac{1}{n_{2}}-\frac{1}{n_{3}}-\frac{1}{n_{4}}\right)
\]
We assume that the interaction kernel is homogeneous in frequency
space,
\[
S_{\lambda\omega_{1},\lambda\omega_{2},\lambda\omega_{3},\lambda\omega_{4}}=\lambda^{\gamma_{S}}S_{\omega_{1}\omega_{2}\omega_{3}\omega_{4}}
\]
where $\gamma_{S}$ is the homogeneity exponent. We look for stationary
Kolmogorov--Zakharov spectra of the form

\[
n(\omega)=A\,\omega^{-x}.
\]
Substituting this ansatz gives

\[
b_{\mathrm{iso}}[n](\omega_{1})=4A^{3}\int\mathrm{d}\omega_{2}\mathrm{d}\omega_{3}\mathrm{d}\omega_{4}\,S_{\omega_{1}\omega_{2}\omega_{3}\omega_{4}}\delta(\omega_{1}+\omega_{2}-\omega_{3}-\omega_{4})\Psi_{x},
\]
with

\[
\Psi_{x}=\omega_{1}^{-x}\omega_{2}^{-x}\omega_{3}^{-x}\omega_{4}^{-x}\left(\omega_{1}^{x}+\omega_{2}^{x}-\omega_{3}^{x}-\omega_{4}^{x}\right).
\]
Introducing $u_{i}=\omega_{i}/\omega_{1}$ for $i=2,3,4$, one finds

\[
b_{\mathrm{iso}}[n](\omega_{1})=A^{3}\omega_{1}^{\gamma_{S}+2-3x}I(x),
\]
with

\[
I(x)=4\int\mathrm{d}u_{2}\mathrm{d}u_{3}\mathrm{d}u_{4}\,S_{1,u_{2},u_{3},u_{4}}\delta(1+u_{2}-u_{3}-u_{4})u_{2}^{-x}u_{3}^{-x}u_{4}^{-x}\left(1+u_{2}^{x}-u_{3}^{x}-u_{4}^{x}\right).
\]
The Kolmogorov--Zakharov exponent $x_{KZ}$ is defined as the exponent
of a KZ solution, by

\[
I(x_{KZ})=0.
\]
To remove the indeterminate form in the flux, we use the L'Hospital
rule and expand
\[
I(x)=I'(x_{KZ})(x-x_{KZ})+O\big((x-x_{KZ})^{2}\big).
\]
The constant energy flux $J_{E}$ satisfies
\[
J_{E}=\mathcal{G}\,A^{3}\,I'(x_{KZ}),
\]
hence
\[
A=\left(\frac{J_{E}}{\mathcal{G}\,I'(x_{KZ})}\right)^{1/3}.
\]
For a dispersion relation as a power law $\omega(\mathbf{k})=\gamma k^{\alpha}$
in dimension $d$, we have $\mathrm{d}^{d}k=\frac{\Omega_{d}}{\alpha\,\gamma^{d/\alpha}}\omega^{\frac{d}{\alpha}-1}\mathrm{d}\omega$,
where $\Omega_{d}$ is the surface of the unit sphere in dimension
$d$. The geometric factor is in this context:
\[
\mathcal{G}=\frac{4\,\Omega_{d}^{3}}{\alpha^{4}\,\gamma^{3d/\alpha+1}},
\]
and

\[
x_{KZ}=\frac{\gamma_{S}+2}{3}.
\]

\subsection{Perturbative corrections to the spectrum. \label{subsec:perturbative corr-to-n}
}

In this section, we outline a simple perturbative method to obtain
solutions with two currents, or when both currents are non-negligible.
This serves only as a remark, since the analytical results in the
paper focus on single-current inertial ranges, while the numerical
studies in section \ref{sec:Application-to-WT} apply more generally
to any stationary spectrum. However, the final result of this appendix
is, to our knowledge, original. By introducing a small dimensionless
parameter, we derive perturbative corrections to the spectrum from
equation (\ref{eq:stationary-kinetic-eq-twofluxes}). For simplicity
in this part only, we set $m=1/n$ to obtain the simpler equation
\[
\omega^{s}\partial_{\omega}^{2}m=m^{4}\left(J_{E}-J_{N}\omega\right).
\]
 In the following paragraphs, we will solve this equation when either
of $J_{E}$ and $J_{N}\omega$ is small compared to the other. This
is the case when the separation of scales between source and dissipation
is large but not large enough to totally neglect the second current.

\subsubsection{First case: $J_{E}\gg\omega J_{N}$:}

We consider $\frac{J_{N}\omega_{d}^{+}}{J_{E}}$ a dimensionless parameter
(of order one if the dissipation is localised) such that $J_{E}-J_{N}\omega=J_{E}\left(1-\frac{\omega}{\omega_{d}^{+}}\frac{J_{N}\omega_{d}^{+}}{J_{E}}\right)$.
We compute perturbatively the solution of $J_{E}\gg\omega J_{N}$,
that is to say $\omega^{s}\partial_{\omega}^{2}m=m^{4}J_{E}\left(1-\zeta^{+}\left(\omega\right)\right)$
where $\zeta^{+}\left(\omega\right)=\frac{\omega}{\omega_{d}^{+}}\left(\frac{J_{N}\omega_{d}^{+}}{J_{E}}\right)$
is a small parameter. We write $m$ as a perturbative expansion in
$\zeta$:
\[
m=A_{E}^{-1}\omega^{x_{E}}\left(1+\sum_{i=1}^{+\infty}\left(\zeta^{+}\right)^{i}a_{i}\right)
\]
where $a_{i}$ are numbers and where we recall $x_{E}=\frac{s-2}{3},A_{E}=\left(\frac{J_{E}}{x_{E}\left(x_{E}-1\right)}\right)^{1/3}$.
We inject it in the kinetic equation and identify orders. The first
order in $\zeta^{+}$ gives $a_{1}=\frac{x_{E}-1}{3x_{E}-5}.$ This
leads to $n$ at first order in $\zeta^{+}$ in the form
\[
n=\frac{1}{m}\underset{\zeta^{+}\to0}{=}\frac{1}{A_{E}^{-1}\omega^{x_{E}}\left(1+a_{1}\zeta^{+}\right)}\underset{\zeta^{+}\to0}{=}A_{E}\omega^{-x_{E}}\left(1-\zeta^{+}\frac{x_{E}-1}{3x_{E}-5}\right).
\]

\textbf{Remark:} The correction is negative in general because $x_{E}$
is in general positive and bigger than $5/3$. This correction $-\zeta^{+}\frac{x_{E}-1}{3x_{E}-5}$
depends linearly on $\omega/\omega_{d}^{+}$ and is thus larger when
we consider large frequencies $\omega$ which are close to $\omega_{d}^{+}$.

Moreover we have a recursive formula for higher orders $i>1$:
\begin{align*}
a_{i}=\frac{x_{E}\left(x_{E}-1\right)}{i\left(i-1\right)+2ix_{E}-3x_{E}\left(x_{E}-1\right)}\Phi\left(\left\{ a_{j}\right\} _{j<i}\right)
\end{align*}
where we introduce the notation 
\begin{align}
\Phi\left(\left\{ a_{j}\right\} _{j<i}\right) & =\left[\left(6\left(\sum_{j=1}^{i-1}a_{i-j}a_{j}\right)+4\left(\sum_{j=1}^{i-2}\sum_{k=1}^{i-j-1}a_{j}a_{k}a_{i-j-k}\right)+\sum_{j,k,l,n\geq1,j+k+l+n=i}a_{j}a_{k}a_{l}a_{n}\right)\right.\label{eq:Phi-order-i}\\
 & \left.-\left(4a_{i-1}+6\left(\sum_{j=1}^{i-2}a_{i-1-j}a_{j}\right)+4\left(\sum_{j=1}^{i-3}\sum_{k=1}^{i-j-2}a_{j}a_{k}a_{i-j-k}\right)+\sum_{j,k,l,n\geq1,j+k+l+n=i-1}a_{j}a_{k}a_{l}a_{n}\right)\right]\nonumber 
\end{align}

\subsubsection{Second case: $J_{E}\ll\omega J_{N}$:}

In the reciprocal case $J_{E}\ll\omega J_{N}$, we can easily adapt
the previous reasoning to the equation $\omega^{s}\partial_{\omega}^{2}m=m^{4}\omega j_{n}\left(\zeta^{-}-1\right)$
with $\zeta^{-}=\frac{\omega_{d}^{-}}{\omega}\left(\frac{J_{E}}{J_{N}\omega_{d}^{-}}\right)$
another dimensionless small parameter. By the same way, we write $m$
as a perturbative expansion in $\zeta^{-}$: $m=A_{N}^{-1}\omega^{x_{N}}\left(1+\sum_{i=1}^{+\infty}\left(\zeta^{-}\right)^{i}b_{i}\right)$
where $b_{i}$ are numbers. One can compute $b_{1}=\frac{x_{N}}{2+3x_{N}}$,
such that at first order, we have 
\[
n\underset{\zeta^{-}\to0}{=}A_{N}\omega^{-x_{N}}\left(1-\zeta^{-}\frac{x_{N}}{2+3x_{N}}\right).
\]
\textbf{Remark:} The correction is negative in general because $x_{N}$
is in general positive. This correction $-\zeta^{-}\frac{x_{N}}{2+3x_{N}}$
depends linearly on $\omega_{d}^{-}/\omega$ and is thus larger when
we consider small frequencies $\omega$ which are closer to $\omega_{d}^{-}$.

We can also compute the recursive formula for the order $i>2$:
\begin{align*}
b_{i} & =\frac{x_{N}\left(x_{N}-1\right)}{\left(i\left(i+1\right)-2ix_{N}\right)-3x_{N}\left(x_{N}-1\right)}\Phi\left(\left\{ b_{j}\right\} _{j<i}\right)
\end{align*}
where $\Phi$ is defined in (\ref{eq:Phi-order-i}).

\subsection{Spatially inhomogeneous case\label{subsec:Appendix-inhomogeneous-WT}}

In this appendix only, we will discuss the inhomogeneous wave turbulence,
where the empirical spectrum $\hat{n}\left(\bm{k},\bm{x},t\right)$
can have a dependency in $\bm{x}$ the spatial coordinate. The theory
of large deviations was already derived in the inhomogeneous case
\cite{Onuki_2023,Guioth_2024}. These works generalised the theory
with inhomogeneity respectively due to linear wave scattering by a
random medium, and to nonlinear wave-wave scattering. The purpose
of this paragraph is not to demonstrate everything rigorously but
rather to justify that our isotropic local theory generalises to the
inhomogeneous case. Taking the example of inhomogeneous nonlinear
wave scattering \cite{Guioth_2024} (we could do the same kind of
computation with the theory developed for inhomogeneous random media
as in \cite{Onuki_2023}), the \emph{empirical }local spectrum $\hat{n}$
defined as a Wigner transform 
\begin{equation}
\hat{n}(\bm{x},\bm{k},\tau)=\frac{1}{(2\pi)^{d}}\int_{\mathbb{R}^{d}}\Psi(\bm{x}+\mu\tfrac{\bm{y}}{2},\mu^{-1}\tau)\Psi^{\ast}(\bm{x}-\mu\tfrac{\bm{y}}{2},\mu^{-1}\tau)\mathrm{e}^{-i\bm{k}\cdot\bm{y}}\label{eq:def_local_spectrum}
\end{equation}
involving $\Psi(\tilde{\bm{x}},\tilde{t})$ the nondimensional signal,
where we denote 
\begin{align*}
L & \text{ the macroscopic lengthscale}\\
\ell & \text{ the typical correlation length (typical wave length)}\\
\mu=\frac{\ell}{L} & \text{ a nondimensional parameter which will give the large deviation parameter }\varepsilon=\left(2\pi\mu\right)^{-d}.
\end{align*}
The paper \cite{Guioth_2024} computes the large deviation Hamiltonian
$H$ and shows that it can be splitted into two contributions:
\begin{equation}
H[n,\lambda]=H_{\mathrm{tr}}[n,\lambda]+H_{\text{int}}[n,\lambda]\label{eq:LD-Hamiltonian-inhomogeneous}
\end{equation}
with
\begin{align*}
H_{\mathrm{tr}}[n,\lambda] & =-\int_{\mathbb{R}^{2d}}\lambda(\bm{x},\bm{k})\nabla_{\bm{k}}\omega\cdot\nabla_{\bm{x}}n(\bm{x},\bm{k})\\
H_{\text{int}}[n,\lambda] & =4\pi\int\text{d}\bm{x}{\displaystyle \int\text{d}\bm{k}_{1}\text{d}\bm{k}_{2}\text{d}\bm{k}_{3}\text{d}\bm{k}_{4}|W_{\vec{\bm{k}}}^{\vec{\sigma}}|^{2}}\delta\left(\vec{\sigma}\cdot\vec{\bm{k}}\right)\delta\left(\vec{\sigma}\cdot\vec{\omega}_{\vec{\bm{k}}}\right)\left(\vec{\sigma}\cdot\vec{\lambda}\right)n_{1}\left(\bm{x}\right)n_{2}\left(\bm{x}\right)n_{3}\left(\bm{x}\right)n_{4}\left(\bm{x}\right)\left[\vec{\sigma}\cdot\left(\frac{1}{\vec{n}}+\vec{\lambda}\right)\right]\,.
\end{align*}
where an arrow denotes a list of 4 components evaluated in $(\bm{x},\bm{k}_{i})$
for $i=1,\ldots,4$. The part $H_{\text{int}}[n,\lambda]$ has a very
similar form as in the homogeneous case. The computation for the isotropic
case can be done in the exact same way, as well as the local approximation.
We still can perform angle integration and perturbative expansion
in $\omega$ as in the homogeneous case and the only difference is
that variables depend on $\bm{x}$ and the spatial integration $\int\text{d}\bm{x}$. 

Now let us have a look at $H_{\mathrm{tr}}[n,\lambda]$ (which is
linear in $\lambda$) and write it in the isotropic hypothesis. We
take the the new variables
\begin{align*}
N(\bm{x},\omega) & =\Omega_{d}k^{d-1}\frac{dk}{d\omega}n(\bm{x},k(\omega))\\
\Lambda(\bm{x},\omega) & =\lambda(\bm{x},k(\omega))
\end{align*}
 as in the previous sections (which this time depend on $\bm{x}$
in addition to $\omega$). Since the dispersion relation is assumed
isotropic $\omega\left(\bm{k}\right)=\omega\left(k\right)=\gamma k^{\alpha}$,
the group velocity writes in spherical coordinates $\left(\vec{e_{r}},\vec{e_{\theta}},\vec{e_{\varphi}}\right)$:
$\nabla_{\bm{k}}\omega=\partial_{k}\omega\vec{e_{r}}=v_{g}\vec{e_{r}}=\alpha\gamma^{1/\alpha}\omega^{1-1/\alpha}\vec{e_{r}}$.
We can write this transport term $H_{\mathrm{tr}}[n,\lambda]=-\int\text{d\ensuremath{\bm{x}\text{d}\bm{k}}}\lambda(\bm{x},\bm{k})\nabla_{\bm{k}}\omega\cdot\nabla_{\bm{x}}n(\bm{x},\bm{k})$
as 
\[
H_{tr}\left[N,\Lambda\right]=-\int\text{d\ensuremath{\bm{x}\text{d}\omega}}\Lambda(\bm{x},\omega)\partial_{k}\omega\vec{e_{r}}\cdot\nabla_{\bm{x}}n(\bm{x},\omega)
\]
Thus if we keep spherical coordinates and denote $r=\left\Vert \bm{x}\right\Vert $:
The only term remaining in the scalar product is the radial term $\left(\nabla_{\bm{k}}\omega\right)_{r}\left(\nabla_{\bm{x}}N(\bm{x},\omega)\right)_{r}$
\begin{align*}
H_{tr}\left[N,\Lambda\right] & =-\int\text{d\ensuremath{\bm{x}\text{d}\omega}}\Lambda(\bm{x},\omega)\partial_{k}\omega\partial_{r}N(\bm{x},\omega).
\end{align*}
 Now we simply notice that this expression is already local, so there
is no need for other hypothesis nor calculation to get the isotropic
local Hamiltonian:
\[
H\left[N,\Lambda\right]=\int\text{d\ensuremath{\bm{x}\text{d}\omega}}\left\{ S_{0}\left[\partial_{\omega}^{2}\Lambda\partial_{\omega}^{2}\left(\frac{1}{n}+\Lambda\right)\right]{\displaystyle n^{4}\omega^{s}}-\Lambda(\bm{x},\omega)\partial_{k}\omega\partial_{r}N(\bm{x},\omega)\right\} .
\]
 Its expression is identical to that in the homogeneous case (\ref{eq:LD-Hamiltonian-local})
up to a space dependency and the additional transport term. We still
have a Hamiltonian which is quadratic in $\Lambda$, with an unchanged
expression in the noise covariance (quadratic term) and a linear term
which has an additional transport term. Finally, we deduce the associated
fluctuating dynamics
\begin{equation}
\partial_{t}N\left(\bm{x},\omega,t\right)+\underset{\nabla_{\bm{k}}\omega\cdot\nabla_{\bm{x}}N(\bm{x},\bm{k},t)}{\underbrace{\partial_{k}\omega\partial_{r}N\left(\bm{x},\omega,t\right)}}=\partial_{\omega}^{2}\left[\mathcal{K}\left[N\right]+\sqrt{2\varepsilon n^{4}\omega^{s}}\frac{\text{d}W}{\text{d}t}(\bm{x},\omega,t)\right]+s\left(N,\bm{x},\omega,t\right)-d\left(N,\bm{x},\omega,t\right)\label{eq:stoch-kin-eq-inhomogeneous-isotropic-local}
\end{equation}
where $\varepsilon=\left(2\pi\frac{\ell}{L}\right)^{-d}$ is the small
large deviation parameter, $\mathcal{K}\left[N\right]$ is defined
in (\ref{eq:def-K-(deterministic part)}) and $W$ is a Wiener process
of $(\bm{x},\omega,t)$. Similarly, one can directly write the linearised
dynamics 
\[
\partial_{t}\delta N\left(\bm{x},\omega,t\right)=\mathcal{L}_{\text{in}}\delta N+\partial_{\omega}^{2}\left[\sqrt{2\bar{n}^{4}\omega^{s}}\frac{\text{d}W}{\text{d}t}(\bm{x},\omega,t)\right]
\]
where $\mathcal{L}_{\text{in}}=\mathcal{L}+\mathcal{L}_{\text{tr}}$
where
\[
\mathcal{L}_{\text{tr}}\cdot=-\nabla_{\bm{k}}\omega\cdot\nabla_{\bm{x}}\cdot
\]
is a transport term, and $\mathcal{L}$ stands for the classical expression
of the linearised operator for the homogeneous dynamics used in the
article (adding $\bm{x}$ dependencies to variables). Similarly, one
can write the same Lyapunov equation over the cumulant at the same
space point $\bm{x}$ (similar reasoning can be done for two different
space points $\bm{x}$, $\bm{y}$) $C_{2}\left(\omega_{1},\omega_{2},\bm{x},t\right)=\mathbb{E}\left[\delta N\left(\omega_{1},\bm{x},t\right)\delta N\left(\omega_{2},\bm{x},t\right)\right]$,
just replacing $\mathcal{L}$ by $\mathcal{L}_{\text{in}}$ and adding
$\bm{x}$-dependencies:
\[
\partial_{t}C_{2}\left(\omega_{1},\omega_{2},\bm{x},t\right)=\mathcal{L}_{\text{in}}C_{2}\left(\omega_{1},\omega_{2},\bm{x},t\right)+C_{2}\left(\omega_{1},\omega_{2},\bm{x},t\right)\mathcal{L}_{\text{in}}^{T}+2A\left(\omega_{1},\omega_{2},\bm{x},t\right)
\]
with $\mathcal{L}_{\text{in}}$ acting on $\omega_{1}$ and $\mathcal{L}_{\text{in}}^{T}$
acting on $\omega_{2}$ and $A\left(\omega_{1},\omega_{2},\bm{x},t\right)=\partial_{\omega_{1}}^{2}\partial_{\omega_{2}}^{2}\left[\mu\left(\omega_{1},\bm{x},t\right)\delta\left(\omega_{1}-\omega_{2}\right)\right]$.
The long-range part of the cumulant $B\left(\omega_{1},\omega_{2},\bm{x},t\right)=C_{2}\left(\omega_{1},\omega_{2},\bm{x},t\right)-C_{\text{eq}}\left(\omega_{1},\bm{x},t\right)\delta\left(\omega_{1}-\omega_{2}\right)$
is still a continuous function (where $C_{\text{eq}}=-\frac{1}{\bar{g}'}$)
because it verifies a slightly modified elliptic equation in the stationary
regime:
\begin{equation}
\mathscr{L}B=\left(\mathcal{K}^{*}\left(\omega_{1}\right)+\mathcal{K}^{*}\left(\omega_{2}\right)\right)\delta''\left(\omega_{1}-\omega_{2}\right)+2\kappa\left(\omega_{1}\right)\delta\left(\omega_{1}-\omega_{2}\right)\label{eq:Lyapunov-B-2CL-2}
\end{equation}
where $\mathscr{L}\cdot=\mathcal{L}_{\text{in}}+\mathcal{L}_{\text{in}}^{T}$,
$\mathcal{K}^{*}=\frac{\bar{\mathcal{\mu}}'\bar{\mathcal{K}}}{\bar{\mathcal{\mu}}\bar{g}'}$
and $\kappa\left(\omega,\bm{x},t\right)=-\nabla_{\bm{k}}\omega.\nabla_{\bm{x}}C_{\text{eq}}\left(\omega,\bm{x},t\right)$.

To conclude, all of the present work can be generalised to inhomogeneous
dynamics, up to an additional space dependency in $\bm{x}$ and an
additional transport term. In this paper however, we will keep it
in the homogeneous framework for simplicity.

\subsection{Linearised dynamics and Lyapunov equation\label{subsec:Intro-Linearised-dynamics+Lyapunov}}

In this section, we are interested in small fluctuations of the density
$N$. For a general Langevin equation with small noise, if $N$ deviates
by a small amount of its average value $\bar{N}$, the dynamics of
$N$ can be linearised around $\bar{N}$. We give the formal expression
of the linearised dynamics and we derive the equation for the rescaled
second order cumulant in a simple manner. This equation is called
Lyapunov equation and will be fundamental for the rest of this paper.
We consider $N$ following a Langevin equation with small noise
\begin{equation}
\text{d}N=b\left[N\right]\text{d}t+\sqrt{2\varepsilon}\sigma\left[N\right]\text{d}W\label{eq:dynamics-general-intro-Lyapunov}
\end{equation}
 where $b\left[N\right]$ is a functional of $N$ and possibly depends
on $\omega$ and $t$, $\sigma\left[N\right]$ is an operator, depending
on $N$, possibly on $\omega$ and $t$. It acts on the function $W$
which is a Wiener process depending on $\omega$ and $t$. $\varepsilon$
is the large deviation parameter. We denote $\delta N=\frac{N-\bar{N}}{\sqrt{\varepsilon}}$.
since $\bar{N}$ verifies the deterministic equation $\dot{N}=b$,
the deviation $N-\bar{N}$ will likely be small and we can expand
the equation (\ref{eq:dynamics-general-intro-Lyapunov}) around $\bar{N}$
in powers of $\sqrt{\varepsilon}$. we find that up to a $\mathcal{O}\left(\sqrt{\varepsilon}\right)$
correction, $\delta N$ verifies the linearised dynamics 
\begin{equation}
\text{d}\delta N=\mathcal{L}\delta N\text{d}t+\sqrt{2}\sigma\left[\bar{N}\right]\text{d}W.\label{eq:linearised-dynamics-intro}
\end{equation}
where $\mathcal{L}\cdot=\frac{\delta b}{\delta N}\left[\bar{N}\right]\cdot$
is the operator of the linearised dynamics. It is a linear operator
acting on $\delta N$. If $b$ is a simple function of $N$, $\mathcal{L}\delta N$
is given by $b'\left[N^{*}\right]\delta N$.

This linearised dynamics (\ref{eq:linearised-dynamics-intro}) will
enable us to derive an equation on the second order cumulant $C_{2}\left(\omega_{1},\omega_{2},t\right)=\mathbb{E}\left[\delta N\left(\omega_{1}\right)\delta N\left(\omega_{2}\right)\right]$.
We use Itô equation on the function $f:X\mapsto\mathbb{E}\left[X\left(\omega_{1}\right)X\left(\omega_{2}\right)\right]$,
with $\delta N$ satisfying the linearised dynamics. It gives $\frac{\text{d}f}{\text{d}t}\left(X\right)=\frac{\delta f}{\delta X\left(\omega\right)}.\frac{\text{d}X\left(\omega\right)}{\text{d}t}+\frac{\delta^{2}f}{\delta X\left(\omega\right)\delta X\left(\omega'\right)}:\left(\sigma\text{d}W\right)\left(\omega\right)\left(\sigma\text{d}W\right)\left(\omega'\right)$
where $.$ stands for the simple contraction in variable $\omega$
and $:$ stands for the double contraction in variables $\omega,\omega'$.
The derivatives are given by
\begin{align*}
\frac{\delta f}{\delta X\left(\omega\right)}\left(\omega_{1},\omega_{2}\right) & =\mathbb{E}\left[\delta\left(\omega_{1}-\omega\right)X\left(\omega_{2}\right)\right]+\mathbb{E}\left[X\left(\omega_{1}\right)\delta\left(\omega-\omega_{2}\right)\right]\\
\frac{\delta^{2}f}{\delta X\left(\omega\right)\delta X\left(\omega'\right)}\left(\omega_{1},\omega_{2}\right) & =\mathbb{E}\left[\delta\left(\omega_{1}-\omega\right)\delta\left(\omega_{2}-\omega'\right)\right]+\mathbb{E}\left[\delta\left(\omega_{1}-\omega'\right)\delta\left(\omega-\omega_{2}\right)\right].
\end{align*}
We then use the identity $\text{d}W\left(\omega\right)\text{d}W\left(\omega'\right)=\delta\left(\omega-\omega'\right)$
and the equation (\ref{eq:linearised-dynamics-intro}). We denote
$\mathcal{L}_{\omega}$ the operator $\mathcal{L}$ acting on functions
of $\omega$ and $\sigma_{\omega}$ the operator $\sigma\left[\bar{N}\right]$
acting on function of $\omega$. We have $\frac{\text{d}f}{\text{d}t}\left(X\right)\left(\omega_{1},\omega_{2}\right)=\frac{\delta f}{\delta X\left(\omega\right)}.\left(\mathcal{L}_{\omega}\delta N\left(\omega\right)\text{d}t+\sqrt{2}\sigma_{\omega}\frac{\text{d}W}{\text{d}t}\left(\omega\right)\right)+\frac{\delta^{2}f\left(\omega_{1},\omega_{2}\right)}{\delta X\left(\omega\right)\delta X\left(\omega'\right)}:\left(\sigma_{\omega}\sigma_{\omega'}\delta\left(\omega-\omega'\right)\right)$
and, when replacing the derivatives, we obtain
\[
\frac{\text{d}f}{\text{d}t}\left(X\right)\left(\omega_{1},\omega_{2}\right)=\mathcal{L}_{\omega_{1}}\mathbb{E}\left[X\left(\omega_{1}\right)X\left(\omega_{2}\right)\right]+\mathcal{L}_{\omega_{2}}\mathbb{E}\left[X\left(\omega_{1}\right)X\left(\omega_{2}\right)\right]+2\sigma_{\omega_{1}}\sigma_{\omega_{2}}\delta\left(\omega_{1}-\omega_{2}\right).
\]
If we evaluate it in $\delta N$ we obtain the so-called Lyapunov
equation
\[
\frac{\text{d}C_{2}}{\text{d}t}\left(\omega_{1},\omega_{2}\right)=\mathcal{L}_{\omega_{1}}C_{2}\left(\omega_{1},\omega_{2}\right)+\mathcal{L}_{\omega_{2}}C_{2}\left(\omega_{1},\omega_{2}\right)+2A\left(\omega_{1},\omega_{2}\right)
\]
where 
\begin{equation}
A\left(\omega_{1},\omega_{2}\right)=\sigma_{\omega_{1}}\sigma_{\omega_{2}}\delta\left(\omega_{1}-\omega_{2}\right).\label{eq:A-intro}
\end{equation}
Which is classically written in the formal way
\begin{equation}
\frac{\text{d}C_{2}}{\text{d}t}=\mathcal{L}C_{2}+C_{2}\mathcal{L}^{T}+2A\label{eq:Lyapunov-equation-intro-formal}
\end{equation}
where $\cdot^{T}$ stands for the adjoint operator. Based on the
expression of the linear operator $\mathcal{L}\cdot=\frac{\delta b}{\delta N}\left[\bar{N}\right]\cdot$,
we can remark that the out-of-equilibrium dynamics influences the
correlations by two means. First, where the net forcing $s-d$ (implicitly
contained in $b$) does not vanish and depends on the density $n$,
the operator $\mathcal{L}\left[n\right]$ for the linearised dynamics
contains a source term $\frac{\delta}{\delta n}\left(s-d\right)$.
Second, if we choose an interval without forcing with a dynamics which
is brought out of equilibrium by an external forcing inducing a flux,
the Lynapunov equation is modified by the shape of the most probable
state $\bar{n}$ and by the boundary conditions, which become boundary
conditions of the flux type. We will see in the next sections that
imposing a flux at the boundary has a significant impact on the shape
of the equation and the boundary conditions, thus leading to very
different solutions for $C_{2}$. 

\subsection{Computation of the stationary spectrum for wave turbulence\label{subsec:Computation-spectrum-WT}}

In this section, we compute the most probable stationary spectrum
$\bar{N}$ for a dynamics 
\[
\text{d}N=\partial_{\omega}^{2}\left(\mathcal{K}\left[N\right]\text{d}t+\sqrt{2\varepsilon\mu\left[N\right]}\text{d}W\right)+\left(s-d\right)\text{d}t.
\]
 It verifies the kinetic equation
\begin{equation}
\partial_{t}\bar{N}=\partial_{\omega}^{2}\bar{\mathcal{K}}+\left(s-d\right).\label{eq:kinetic-equation-numerical-part}
\end{equation}
As shown in Section \ref{subsec:Out-of-equilibrium-solutions-kinetic-eq-wide-inertial-ranges},
in the absence of sources and dissipation and with a single flux,
the solutions are the KZ spectra, valid over wide inertial ranges.
Since our ultimate goal is to solve the Lyapunov equation numerically,
we now introduce a method to compute stationary spectra under more
general conditions, including source and dissipation terms. This approach
highlights the qualitative properties of the resulting correlations.
The section focuses on the methodology and illustrative examples with
different forcing and boundary conditions, using arbitrarily chosen
physical parameters rather than aiming for quantitative predictions.In
all this section, we try to minimize numerical complexity to give
numerical methods running on a simple laptop. 

\paragraph{Numerical method}

We can imagine many ways of solving the kinetic equation and has been
done multiple times in the literature. In this section, we will not
review all solutions, but we will just give a possibility. We choose
the resolution of the kinetic equation (\ref{eq:kinetic-equation-numerical-part})
by discretizing the space of $\omega$ in $I$ values $\left(\omega_{1},\ldots,\omega_{I}\right)$
with a regular spacing $\Delta\omega$. We integrate the equation
in time until convergence. The equation being a differential equation,
the operators are local in $\omega$ and can be represented as sparse
matrices. This enables to considerably speed up the evaluation of
the second member compared to nonlocal kinetic equations, because
no full matrix products are necessary. 

If we denote with an index $i\in\left\llbracket 1,I\right\rrbracket $
the functions evaluated in $\omega=\omega_{i}$, the discretised kinetic
equation reads 
\[
\partial_{t}N_{i}=\frac{\mathcal{K}_{i-1}-2\mathcal{K}_{i}+\mathcal{K}_{i+1}}{\Delta\omega^{2}}+s_{i}-d_{i}
\]
The function $\mathcal{K}=\omega^{s}n\left[2\left(\partial_{\omega}n\right)^{2}-n\partial_{\omega}^{2}n\right]$
can be computed by numerically differentiating $n$. So, for $i\in\left\llbracket 2,I-1\right\rrbracket $,
\begin{align}
\mathcal{K}_{i} & =-\omega_{i}^{s}n_{i}\left(n_{i}\frac{n_{i-1}-2n_{i}+n_{i+1}}{\Delta\omega^{2}}-2\left(\frac{n_{i+1}-n_{i-1}}{2\Delta\omega}\right)^{2}\right).\label{eq:K-discretised-with-n}
\end{align}
Furthermore, we should take care of implementing the correct boundary
conditions with fixed fluxes at the boundary, so that we have two
conserved quantities at the discretised level. It is equivalent to
prescribing the values for $\mathcal{K}_{0},\mathcal{K}_{1},\mathcal{K}_{I},\mathcal{K}_{I+1}$.
Since the wave action current $\mathcal{J}_{N}$ and the energy current
$\mathcal{J}_{E}$ can be expressed in terms of $\mathcal{K}$ as
$\mathcal{J}_{N}=-\partial_{\omega}\mathcal{K};\mathcal{J}_{E}=\mathcal{K}-\omega\partial_{\omega}\mathcal{K}=2\mathcal{K}-\partial_{\omega}\left(\omega\mathcal{K}\right)$,
we can discretise and invert these relations to find $\mathcal{K}_{0},\mathcal{K}_{1},\mathcal{K}_{I},\mathcal{K}_{I+1}$
as a function of the boundary currents. We denote $j_{N}^{(1)},j_{E}^{(1)}$
the currents imposed on the left side and $j_{N}^{(2)},j_{E}^{(2)}$
the currents imposed on the right side. After a straightforward computation,
we obtain an appropriate way to fix the boundary conditions:
\begin{equation}
\begin{cases}
\mathcal{K}_{0} & =j_{E}^{(1)}-\omega_{0}j_{N}^{(1)}\\
\mathcal{K}_{-1} & =j_{E}^{(1)}-\left(\omega_{0}-\Delta\omega\right)j_{N}^{(1)}\\
\mathcal{K}_{I+1} & =j_{E}^{(2)}-\left(\omega_{I}+\Delta\omega\right)j_{N}^{(2)}\\
\mathcal{K}_{I} & =j_{E}^{(2)}-\omega_{I}j_{N}^{(2)}
\end{cases}.\label{eq:K-discretized-BCs}
\end{equation}
In addition, we have two conservation laws with two currents, defined
on a shifted grid:
\begin{align*}
\begin{cases}
\partial_{t}n_{i} & =-\frac{j_{n,i+1/2}-j_{n,i-1/2}}{\Delta\omega}\\
j_{n,i+1/2} & =\frac{K_{i+1}-K_{i}}{\Delta\omega}
\end{cases}\quad;\quad & \begin{cases}
\partial_{t}\omega_{i}n_{i} & =-\frac{j_{e,i+1/2}-j_{e,i-1/2}}{\Delta\omega}\\
j_{e,i+1/2} & =-\omega_{i}\frac{K_{i+1}-K_{i}}{\Delta\omega}+K_{i}
\end{cases}
\end{align*}
 So it naturally comes that if we choose consistently the values of
the imposed currents $j_{N}^{(1)},j_{E}^{(1)},j_{N}^{(2)},j_{E}^{(2)}$
at the boundary, there are indeed two conserved quantities at the
discretised level: the mass (or wave action number) $\mathcal{N}=\sum_{i=1}^{I}n_{i}$
and the energy $\mathcal{E}=\sum_{i=1}^{I}\omega_{i}n_{i}$. Different
choices are possible depending on the source and dissipation configuration
in the resolved domain $\mathcal{D}$. We will mainly be focused on
the following situations:
\begin{enumerate}
\item No source and dissipation in the domain; we impose a wave action current
or an energy current alone, the same value at both sides. In theory
the stationary result should be the KZ solutions.
\item Only the source is included in the domain. The source is narrow, located
around frequency $\omega_{f}$ and there are large inertial ranges
on both sides of it. The inertial ranges are not fully resolved and
we impose fluxes at both sides. We imagine dissipation scales far
away from the forcing and out of the domain. The imposed currents
therefore should respect relations (\ref{eq:relations-je-jn-at-boundary-frequencyRatio},\ref{eq:je-jn-with-forcingAmpllitude-and-frequencies}).
\item Source and dissipation are fully included in the domain. If correctly
chosen such as the sources compensate the sinks inside the domain,
the currents are null at the boundary.
\end{enumerate}
We can make an important remark regarding the configurations 2) and
3). Due to the relations between the boundary currents (\ref{eq:relations-je-jn-at-boundary-frequencyRatio}),
it is quite a difficult task to implement wide inertial ranges leading
to KZ solutions, or even to branches of KZ solutions. This kind of
solution appears when only one constant current is present, or more
generally when the current ratio $\frac{\mathcal{J}_{E}}{\omega\mathcal{J}_{N}}$
is very small or very large. For a current to be negligible compared
to the other in the inertial ranges (where the currents are constant),
we need the ratios
\begin{align*}
\frac{j_{E}^{(1)}}{\omega j_{N}^{(1)}} & =\frac{\omega_{d}^{-}}{\omega}\\
\frac{\omega j_{N}^{(2)}}{j_{E}^{(2)}} & =\frac{\omega}{\omega_{d}^{+}}
\end{align*}
to be very small in a large part of the resolved domain $\mathcal{D}=\left[\omega_{1},\omega_{I}\right]$.
In other words, to observe KZ branches, we should have very large
resolved domains, thus increasing the size of the space of frequencies.
This size linearly increases with the current ratios $\frac{j_{E}^{(1)}}{\omega j_{N}^{(1)}},\frac{\omega j_{N}^{(2)}}{j_{E}^{(2)}}$.
This often causes that we are unable to observe KZ branches, unless
either taking configuration 2) with the dissipation which is virtually
very far away (and thus unresolved), or configuration 1) with no resolved
source and dissipation at all.

To sum up, the numerical evaluation of the second member is summarized
in the following scheme
\[
\begin{array}{ccc}
\left(n_{i}\left(t\right)\right)_{1\leq i\leq I} & \rightarrow & \text{numerical differentiation }\partial_{\omega}^{2}n,\partial_{\omega}n\\
\downarrow &  & \downarrow\\
\text{remove indices }0\text{ and }I & \rightarrow & \text{compute }\left(\mathcal{K}_{i}\right)_{2\leq i\leq I-1}\text{ with eq. (\ref{eq:K-discretised-with-n})}\\
 &  & \downarrow\\
 &  & \text{use BCs (\ref{eq:K-discretized-BCs}) for }\mathcal{K}\\
 &  & \downarrow\\
 &  & \left(\mathcal{K}_{i}\right)_{-1\leq i\leq I+1}\\
 &  & \downarrow\\
 &  & \text{numerical differentiation }\partial_{\omega}^{2}\mathcal{K}\\
 &  & \downarrow\\
s-d & \rightarrow & \left(\partial_{\omega}^{2}\mathcal{K}_{i}+s_{i}-d_{i}\right)_{1\leq i\leq I}\\
 &  & \downarrow\\
 &  & \boxed{n_{i}\left(t+\text{d}t\right)}
\end{array}
\]

In order to increase the time step and make convergence faster, we
use implicit integration methods such as BDF, LSODA or LADAU (see
python documentation of the function \textit{scipy.integrate.solve\_ivp})
with an adapted convergence criterion.  This method enables to compute
stationary state for the kinetic equation with any forcing and dissipation,
but widening the inertial range between forcing scales and dissipation
scales remains challenging with limited computing capacities. 

\paragraph{Results for the stationary spectra}

\begin{figure}[h]
\centering{}%
\begin{tabular}{@{}ccc}
\includegraphics[totalheight=3.5cm]{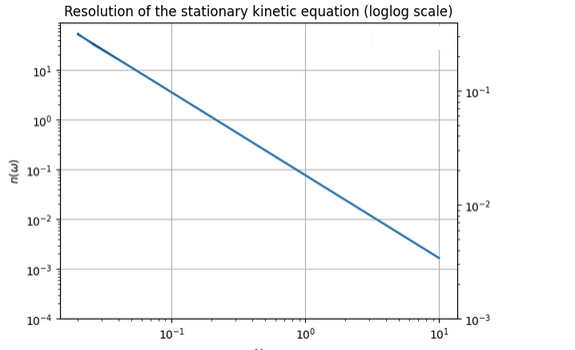} & \includegraphics[totalheight=3.5cm]{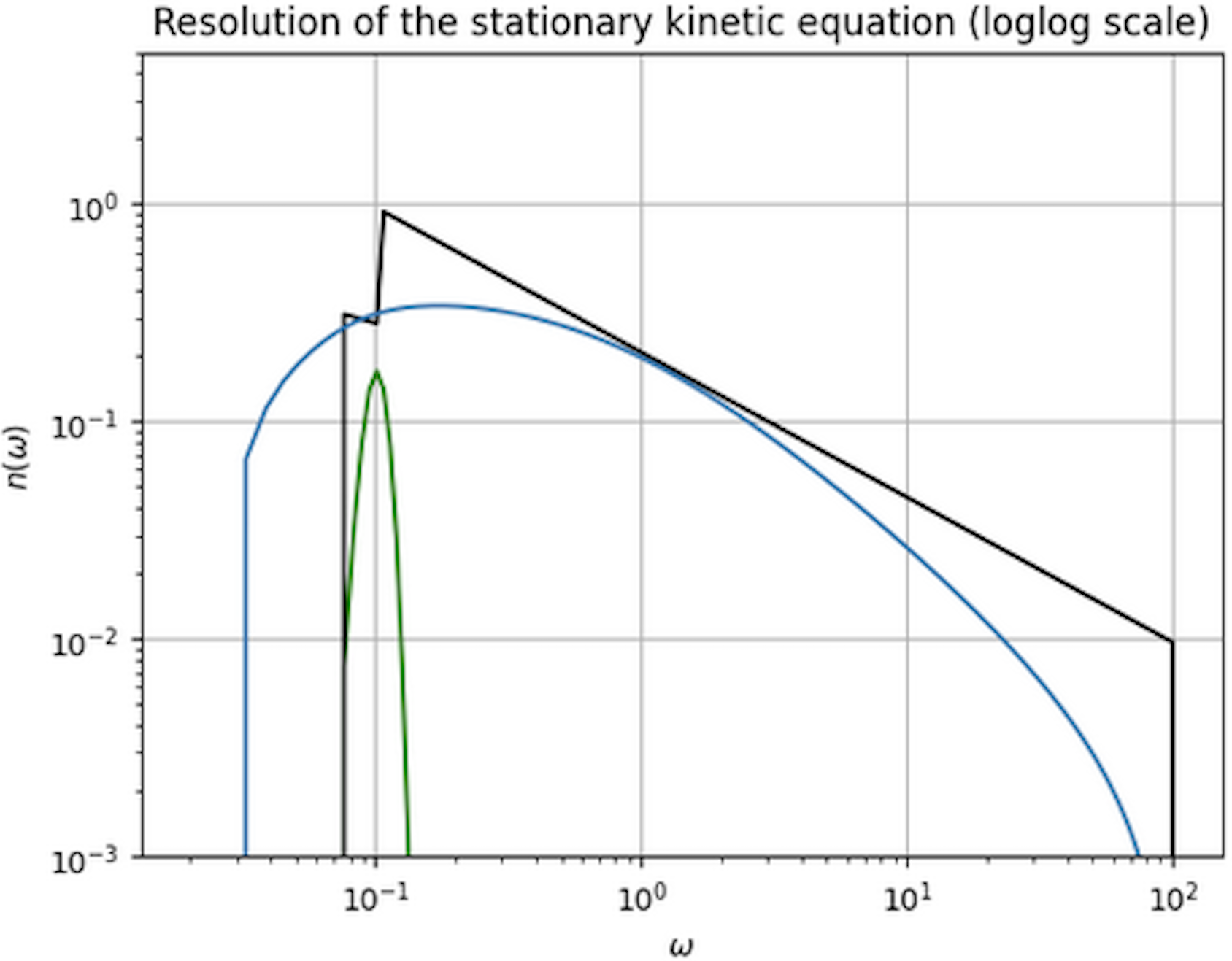} & \includegraphics[totalheight=3.5cm]{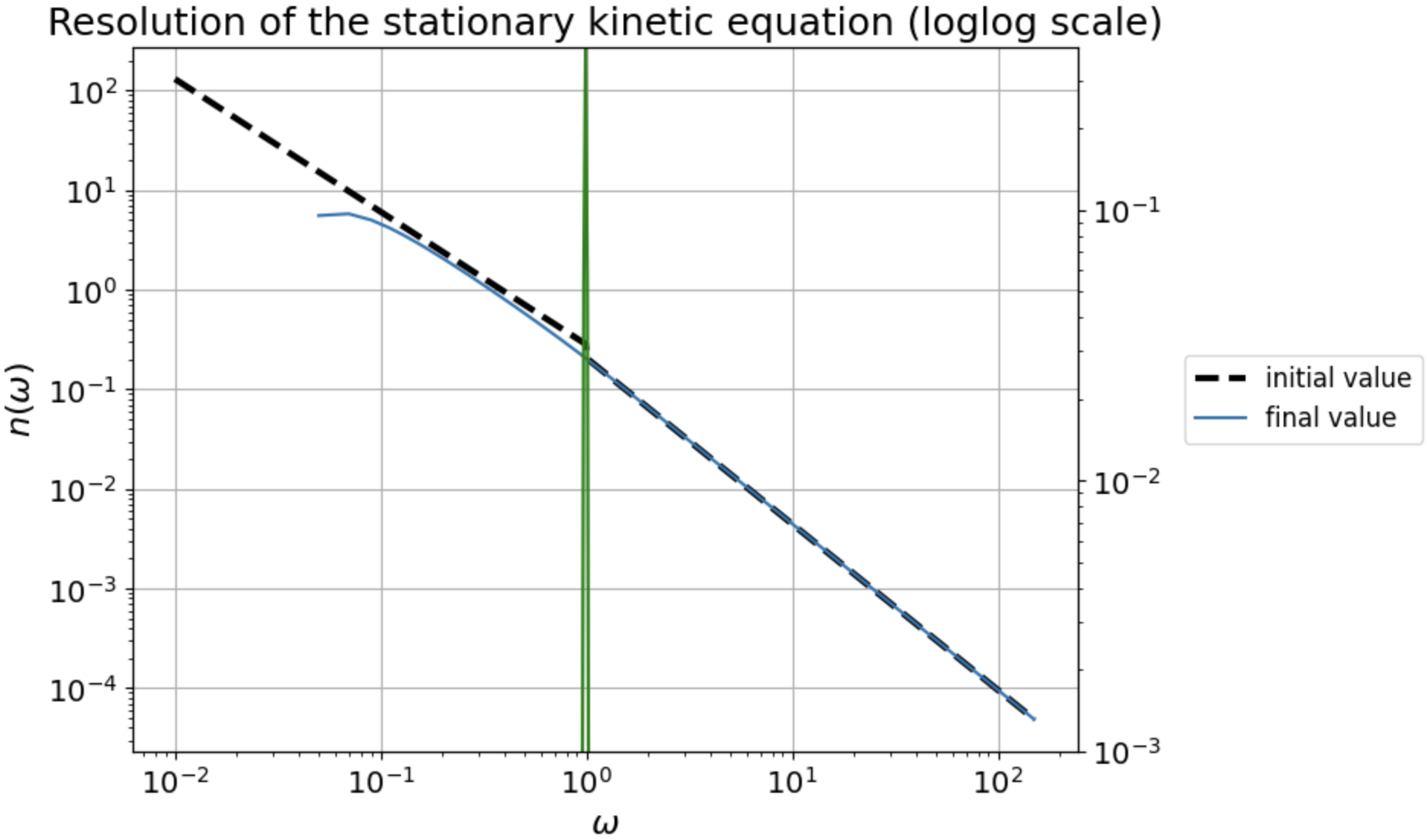}\tabularnewline
\textbf{(a)} & \textbf{(b)} & \textbf{(c)}\tabularnewline
\end{tabular}\caption{{\small\textbf{Stationary spectra}}{\small{} }\protect \\
{\small (a) infinite inertial range with only energy current $j_{E}$,
corresponding to configuration 1)}\protect \\
{\small (b) forcing in the bulk and dissipation near boundaries, with
zero boundary currents, corresponding to configuration 3)}\protect \\
{\small (c) forcing in the bulk with nonzero boundary currents $j_{N},j_{E}$,
corresponding to configuration 2). }\protect \\
{\small On all graphs, the blue curve represents the stationary solution
for the spectra, the green curve represents the forcing (if present)
and the black curve represents the branches of }KZ{\small{} solutions
in inertial ranges. A dissipation is present where the black curve
falls to $0$. We observe that excluding the dissipation from the
resolution interval and adopting suitable flux-type boundary conditions
facilitates the observation of branches of }KZ{\small{} solutions. \label{fig:stationary-spectra}}}
\end{figure}

In this paragraph, we expose the different spectra we computed. They
are presented in Figure \ref{fig:stationary-spectra}.

\begin{figure}[H]
\centering{}\includegraphics[totalheight=6cm]{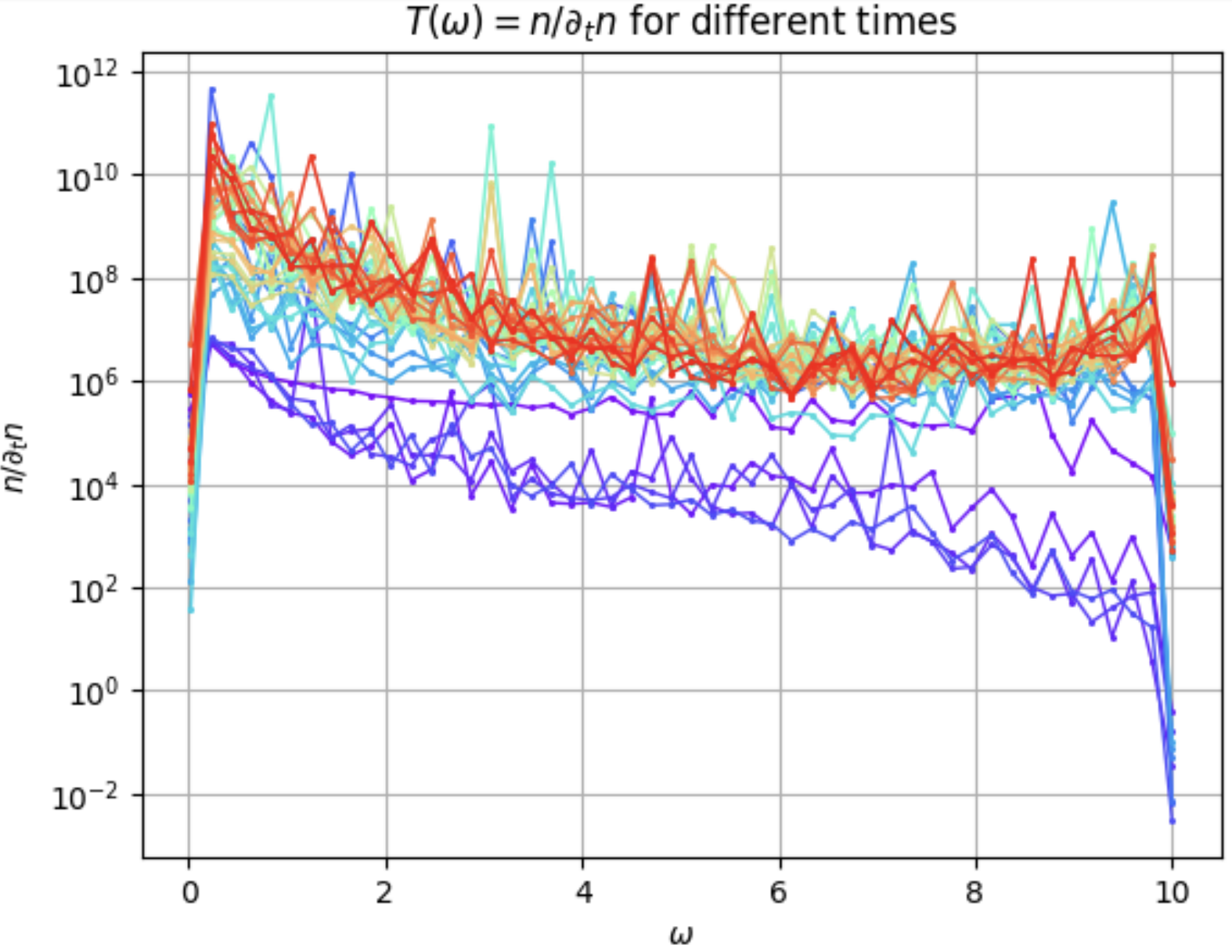}\caption{{\small\textbf{Convergence monitoring for configuration 2)}}{\small{}
}\protect \\
{\small In this figure, we show the successive time scale $\frac{N}{\partial_{t}N}$
as a function of $\omega$. The different curves correspond to different
times in the simulation. It starts in the purple-blue colours and
turn orange-red at the end. We observe that the time scale $\frac{N}{\partial_{t}N}$
grows and becomes large for all the values of $\omega$. It stabilizes,
probably due to numerical errors causing that the solution cannot
be perfect. \label{fig:convergence-monitoring}}}
\end{figure}

\begin{figure}[H]
\centering{}\includegraphics[totalheight=6cm]{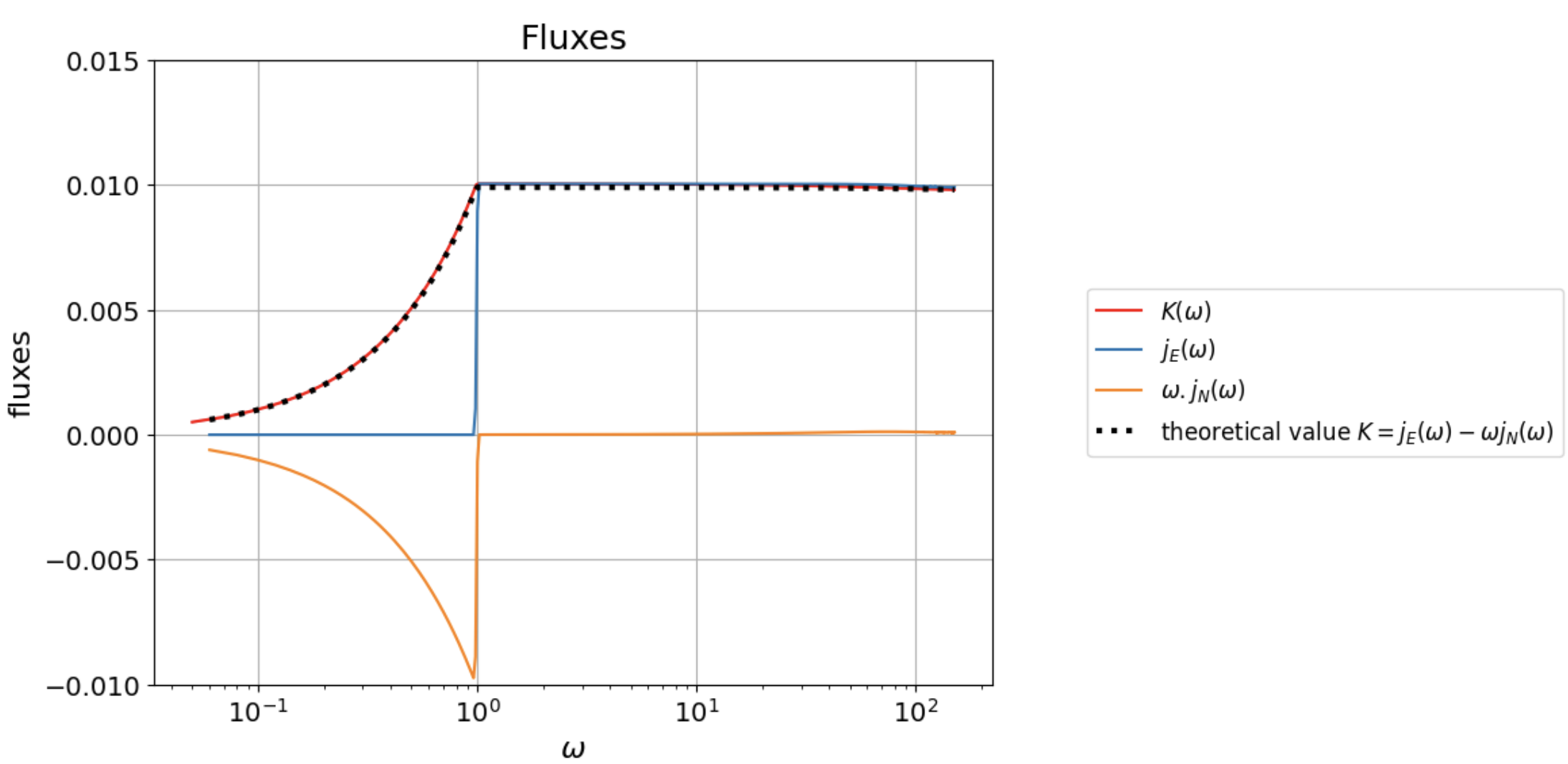}\caption{{\small\textbf{Fluxes}}\protect \\
{\small In this figure, we show the currents $j_{N}$, $j_{E}$ computed
numerically as a function of $\omega$ at the end of the simulation.
The the orange curve is the wave action current $j_{N}$ multiplied
by $\omega$, the blue one represents the energy current $j_{E}$.
The red curve is $\mathcal{K}=j_{E}-\omega j_{N}$ (numerically computed).
Eventually, the dotted black curve represents the theoretical value
for $\mathcal{K}=j_{E}-\omega j_{N}$, estimated by the fact that
in inertial ranges the currents should be constant. We observe a very
good agreement. }\protect \\
{\small\label{fig:fluxes}}}
\end{figure}

We observe that excluding the dissipation from the resolution interval
and adopting suitable flux-type boundary conditions facilitates the
observation of branches of KZ solutions. This highlights the interest
of the whole framework we developed in the precedent parts to clearly
set a problem which is adapted to flux-type boundary conditions. With
this technique, we will be able to compute the Gaussian fluctuations
close to KZ spectra in the next parts. To check that the algorithm
converges, we monitor the value of the second member and check its
value decreases and gets close to 0. This is consistent with the fact
that we observe a convergence of the spectra towards the final curves
represented in Figure \ref{fig:stationary-spectra} over time. An
example is given in Figure \ref{fig:convergence-monitoring}, for
configuration 2). We can check that the solution is physically relevant
by plotting the value of both currents and compare them to the expected
values. The wave action current and energy current are monitored and
converge to the expected value. We illustrate it in Figure \ref{fig:fluxes}.
All these verifications enable us to validate the spectrum profiles
we obtained in Figure \ref{fig:stationary-spectra}. 

Note that our resolution is quite incomplete and limited to some simple
cases on a frequency domain which is quite limited. We chose to focus
on the method and we do not pretend to give something more complete
and quantitative, this is not the intent of this article. This is
due to several issues, that we will now briefly discuss. First, for
physical cases, the value of the exponent $s$ can be quite large,
causing $\omega^{s}$ varying in many orders of magnitude. This issue
can possibly be solved by changing the variable to $\rho=\omega^{u}n$
with $u$ a judiciously chosen exponent, so that large powers of $\omega$
vanish in the new kinetic equation over $\rho$. Second, in order
to enlarge the resolution domain, this could be a promising idea to
solve in log variable $x=\log\omega$. We have not done it in this
work because it demands a proper re-writing of the kinetic equation
and we have chosen to focus our numerical effort on the resolution
of the Lyapunov equation. This will be the object of the next parts,
where we will use a numerical method developed in \cite{article2_brice_freddy}
to solve numerically the correlations for isotropic local wave turbulence. 

\subsection{Computation of the discretised Lyapunov equation for two conservation
laws and its declination for wave turbulence\label{subsec:Appendix:-computation-discretized-Lyapunov-2CL}}

In this appendix, we give more details on the result given in the
discretization used in the numerical method to solve Lyapunov equation
with two conservation laws.

For two conservation laws, the Lyapunov system reads 
\begin{equation}
\begin{cases}
\mathscr{L}B=\left(\mathcal{K}^{*}\left(\omega_{1}\right)+\mathcal{K}^{*}\left(\omega_{2}\right)\right)\delta''\left(\omega_{1}-\omega_{2}\right)\\
\begin{cases}
LB\left(\omega_{1},\omega_{2}\right) & =\mathcal{K}^{*}\left(\omega_{1}\right)\delta\left(\omega_{1}-\omega_{2}\right)\\
\partial_{\omega_{1}}\left(LB\left(\omega_{1},\omega_{2}\right)\right) & =\partial_{\omega_{1}}\left(\mathcal{K}^{*}\delta\left(\omega_{1}-\omega_{2}\right)\right)
\end{cases} & \text{ for }\omega_{1}\in\partial\mathcal{D}\\
\text{ same symmetrized boundary conditions } & \text{ for }\omega_{2}\in\partial\mathcal{D}
\end{cases}\label{eq:Lyapunov-B-2CL-1}
\end{equation}
where $\mathscr{L}\cdot=\partial_{\omega_{1}}^{2}\left(L\cdot\right)+\partial_{\omega_{2}}^{2}\left(\cdot L^{T}\right)$
with $L\cdot=\frac{\delta\mathcal{K}}{\delta N}\left[\bar{N}\right]\cdot=\bar{\mathscr{D}}'\partial_{\omega_{1}}^{2}\bar{g}\cdot+\bar{\mathscr{D}}\partial_{\omega_{1}}^{2}\left(\bar{g}'\cdot\right)+\bar{\mu}'E$
and $\mathcal{K}^{*}=\frac{\bar{\mathcal{\mu}}'\bar{\mathcal{K}}}{\bar{\mathcal{\mu}}\bar{g}'}$.
$g$ can be any function of the density $N$. We will follow the same
idea as for SEP, by subtracting a part related to the nonzero boundary
conditions, which will entail a slight modification of the second
member. Then we will discretise the operator $\partial_{\omega_{1}}^{2}L$
using null flux boundary conditions, that is to say we will prescribe
the values of $LB$ and $\partial_{\omega_{1}}\left(LB\right)$ at
the boundary. The operator $L$ reads
\begin{equation}
L=u\left(\omega_{1}\right)\partial_{\omega_{1}}^{2}+v\left(\omega_{1}\right)\partial_{\omega_{1}}+w\left(\omega_{1}\right)\label{eq:L-operator-continuous}
\end{equation}
where for example in wave turbulence,
\begin{align*}
u\left(\omega\right) & =-\omega^{s}n^{2}\\
v\left(\omega\right) & =\left(4\omega^{s}n\partial_{\omega}n\right)\\
w\left(\omega\right) & =\left(2\omega^{s}\left(n\partial_{\omega}^{2}n-\left(\partial_{\omega}n\right)^{2}\right)\right)
\end{align*}
so we can discretise using the central finite difference coefficients:
\begin{align*}
\left(\partial_{\omega_{1}}B\right)_{ij} & =\frac{B_{i+1,j}-B_{i-1,j}}{2\Delta\omega}\\
\left(\partial_{\omega_{1}}^{2}B\right)_{ij} & =\frac{B_{i+1,j}-2B_{i,j}+B_{i-1,j}}{\Delta\omega^{2}}\\
\left(\partial_{\omega_{1}}^{3}B\right)_{ij} & =\frac{B_{i+2,j}-2B_{i+1,j}+2B_{i-1,j}-B_{i-2,j}}{2\Delta\omega^{3}}
\end{align*}
Therefore, we get the discretised expressions of $LB$ and $\partial_{\omega_{1}}LB$
that we should fix at the boundary:
\begin{align}
\left(LB\right)_{ij} & =u_{i}\frac{B_{i+1,j}-2B_{i,j}+B_{i-1,j}}{\Delta\omega^{2}}+v_{i}\frac{B_{i+1,j}-B_{i-1,j}}{2\Delta\omega}+w_{i}B_{i,j}\nonumber \\
\left(\partial_{\omega_{1}}LB\right)_{ij} & =\left(\partial_{\omega_{1}}u\right)_{i}\frac{B_{i+1,j}-2B_{i,j}+B_{i-1,j}}{\Delta\omega^{2}}+u_{i}\frac{B_{i+2,j}-2B_{i+1,j}+2B_{i-1,j}-B_{i-2,j}}{2\Delta\omega^{3}}\label{eq:discretized-LB-and-dLB}\\
 & +\left(\partial_{\omega_{1}}v\right)_{i}\frac{B_{i+1,j}-B_{i-1,j}}{2\Delta\omega}+v_{i}\frac{B_{i+1,j}-2B_{i,j}+B_{i-1,j}}{\Delta\omega^{2}}\nonumber \\
 & +\left(\partial_{\omega_{1}}w\right)_{i}B_{i,j}+w_{i}\frac{B_{i+1,j}-B_{i-1,j}}{2\Delta\omega}\nonumber 
\end{align}
These relations can be written on the boundary, by virtually adding
ghost points $B_{-1,j},B_{0j},B_{I+1j},B_{I+2,j}$. Using the boundary
conditions, we have additional relations:
\begin{align}
 & \begin{cases}
\left(LB\right)_{i=1,j} & =g_{j}^{\left(0\right)}\\
\left(LB\right)_{i=I,j} & =g_{j}^{\left(1\right)}\\
\left(\partial_{\omega_{1}}LB\right)_{i=1j} & =h_{j}^{\left(0\right)}\\
\left(\partial_{\omega_{1}}LB\right)_{i=I,j} & =h_{j}^{\left(1\right)}
\end{cases}\label{eq:LB boundary 2CL}
\end{align}
where $g_{j}^{\left(0\right)},g_{j}^{\left(1\right)}$ and $h_{j}^{\left(0\right)},h_{j}^{\left(1\right)}$
are the discretizations of the boundary functions $\mathcal{K}^{*}\left(\omega_{1}\right)\delta\left(\omega_{1}-\omega_{2}\right)$
and $\partial_{\omega_{1}}\left(\mathcal{K}^{*}\left(\omega_{1}\right)\delta\left(\omega_{1}-\omega_{2}\right)\right)=\mathcal{K}^{*}\left(\omega_{2}\right)\delta'\left(\omega_{1}-\omega_{2}\right)$
respectively:
\begin{align*}
g_{j}^{\left(0\right)} & =\mathcal{K}^{*}\left(\omega_{\text{min}}\right)\frac{\delta_{j,1}}{\Delta\omega}\\
g_{j}^{\left(1\right)} & =\mathcal{K}^{*}\left(\omega_{\text{max}}\right)\frac{\delta_{j,I}}{\Delta\omega}\\
h_{j}^{\left(0\right)} & =\mathcal{K}^{*}\left(\omega_{j}\right)\frac{\delta'_{j,1}}{\Delta\omega^{2}}\\
h_{j}^{\left(1\right)} & =\mathcal{K}^{*}\left(\omega_{j}\right)\frac{\delta'_{j,1}}{\Delta\omega^{2}}
\end{align*}
where $\frac{\delta_{j,1}}{\Delta\omega}$ is the discretization of
the Dirac delta function $\delta$ and $\frac{\delta'_{j,1}}{\Delta\omega^{2}}$
is the discretization of $\delta'$, namely $\delta'_{j,1}=\left(\begin{array}{ccccc}
-1 & 1 & 0 & \ldots & 0\end{array}\right)_{j}$ and $\delta'_{j,I}=\left(\begin{array}{ccccc}
0 & \ldots & 0 & -1 & 1\end{array}\right)_{j}$. These relations (\ref{eq:LB boundary 2CL}) give $B_{-1,j},B_{0j},B_{I+1j},B_{I+2,j}$
as functions of $B_{1,j},B_{2j},B_{3,j},B_{Ij},B_{I-1,j},B_{I-2,j}$.
We could write them explicitly but the computation is not necessary.
Indeed, now we will perform a change of variable $B\mapsto\tilde{B}$
such that $B=\tilde{B}$ in the bulk but slightly differ on the boundaries,
such that
\begin{equation}
\begin{cases}
\left(L\tilde{B}\right)_{i=1,j} & =0\\
\left(L\tilde{B}\right)_{i=I,j} & =0\\
\left(\partial_{\omega_{1}}L\tilde{B}\right)_{i=1j} & =0\\
\left(\partial_{\omega_{1}}L\tilde{B}\right)_{i=I,j} & =0
\end{cases}.\label{eq:BCs-LBtilde-discretized}
\end{equation}
 Namely for $-1\leq j\leq I+2$:
\begin{equation}
\tilde{B}_{ij}=\begin{cases}
B_{ij} & \text{if }1\leq i\leq I\\
B_{-1,j}-\frac{2\Delta\omega^{3}h_{j}^{\left(0\right)}}{u_{-1}} & \text{if }i=-1\\
B_{0,j}+\frac{\Delta\omega^{2}g_{j}^{\left(0\right)}}{u_{0}-\frac{\Delta\omega v_{0}}{2}} & \text{if }i=0\\
B_{I+1,j}+\frac{\Delta\omega^{2}g_{j}^{\left(1\right)}}{u_{I+1}+\frac{\Delta\omega v_{I+1}}{2}} & \text{if }i=I+1\\
B_{I+2,j}+\frac{2\Delta\omega^{3}h_{j}^{\left(1\right)}}{u_{I+2}} & \text{if }i=I+2
\end{cases}\label{eq:Btilde-def}
\end{equation}
such that $\tilde{B}$ has null flux boundary conditions. We can check
that the relations (\ref{eq:BCs-LBtilde-discretized}) are verified
by inserting the expression of $\tilde{B}$ given in (\ref{eq:Btilde-def})
into the discretised expressions (\ref{eq:discretized-LB-and-dLB}).
The operator $\mathfrak{L}$ acting on $\tilde{B}$ can be written
the same way as for SEP. In the discretised space, we can write $\partial_{\omega_{1}}^{2}\left(L\tilde{B}\right)\in\mathscr{M}_{I,I}\left(\mathbb{R}\right)$
from $\partial_{\omega_{1}}^{2}\left(L\tilde{B}\right)$ with additional
terms corresponding to the difference: $\partial_{\omega_{1}}^{2}\left(LB\right)+\partial_{\omega_{1}}^{2}\left(L\left(\tilde{B}-B\right)\right)$
so in the discretised space, we have for $1\leq i,j\leq I$:
\[
\left[\partial_{\omega_{1}}^{2}\left(L\left(\tilde{B}-B\right)\right)\right]_{ij}=\frac{\left[L\left(\tilde{B}-B\right)\right]_{i+1,j}-2\left[L\left(\tilde{B}-B\right)\right]_{i,j}+\left[L\left(\tilde{B}-B\right)\right]_{i-1,j}}{\Delta\omega^{2}}
\]
where the terms $\left[L\left(\tilde{B}-B\right)\right]_{ij}$ can
be written using (\ref{eq:discretized-LB-and-dLB}) and the boundary
conditions (\ref{eq:LB boundary 2CL}, \ref{eq:BCs-LBtilde-discretized}).
This quantity vanishes, except for terms involving boundary terms
$\left(\tilde{B}-B\right)_{-1,j}$, $\left(\tilde{B}-B\right)_{0j}$,
$\left(\tilde{B}-B\right)_{I+1j}$, $\left(\tilde{B}-B\right)_{I+2,j}$.
Actually, we obtain {\small
\begin{align*}
\left(L\left(\tilde{B}-B\right)\right)_{ij} & =u_{i}\frac{\left(\tilde{B}-B\right)_{i+1,j}-2\left(\tilde{B}-B\right)_{i,j}+\left(\tilde{B}-B\right)_{i-1,j}}{\Delta\omega^{2}}+v_{i}\frac{\left(\tilde{B}-B\right)_{i+1,j}-\left(\tilde{B}-B\right)_{i-1,j}}{2\Delta\omega}+w_{i}\left(\tilde{B}-B\right)_{i,j}\\
 & =\begin{cases}
0 & \text{if }2\leq i\leq I-1\\
\left(-LB\right)_{ij} & \text{if }i=0,1,I,I+1
\end{cases}\\
 & =\begin{cases}
0 & \text{if }2\leq i\leq I-1\\
\left(-LB\right)_{i=0j}=-g_{j}^{\left(0\right)}+\Delta\omega h_{j}^{\left(0\right)} & \text{if }i=0\\
\left(-LB\right)_{i=1,j}=-g_{j}^{\left(0\right)} & \text{if }i=1\\
\left(-LB\right)_{i=I,j}=-g_{j}^{\left(1\right)} & \text{if }i=I\\
\left(-LB\right)_{i=I+1,j}=-g_{j}^{\left(1\right)}-\Delta\omega h_{j}^{\left(1\right)} & \text{if }i=I+1
\end{cases}.
\end{align*}
} We can write the discretised version of $\partial_{\omega_{1}}^{2}\left(L\left(\tilde{B}-B\right)\right)\in\mathscr{M}_{I,I}\left(\mathbb{R}\right)$
as a matrix product involving the discrete second derivative 
\[
D_{I\times I+2}^{2}=\frac{1}{\Delta\omega^{2}}\left[\begin{array}{cccccc}
1 & -2 & 1 &  & (0)\\
 & 1 & -2 & 1\\
 &  & \ddots & \ddots & \ddots\\
 & (0) &  & 1 & -2 & 1
\end{array}\right]\in\mathscr{M}_{I,I+2}\left(\mathbb{R}\right)
\]
 and a matrix with the boundary terms. We denote 
\[
S:=D_{I\times I+2}^{2}\cdot\underset{\in\mathscr{M}_{I+2,I}\left(\mathbb{R}\right)}{\underbrace{\left[\begin{array}{|ccccc|}
\hline g_{1}^{\left(0\right)}-\Delta\omega h_{1}^{\left(0\right)} &  & \ldots &  & g_{I}^{\left(0\right)}-\Delta\omega h_{I}^{\left(0\right)}\\
g_{1}^{\left(0\right)} &  & \ldots &  & g_{I}^{\left(0\right)}\\
\hline  &  & \begin{array}{ccc}
\\ & (0)\\
\\\end{array} &  & \\
\hline g_{1}^{\left(1\right)} &  & \ldots &  & g_{I}^{\left(1\right)}\\
g_{1}^{\left(1\right)}+\Delta\omega h_{1}^{\left(1\right)} &  & \ldots &  & g_{I}^{\left(1\right)}+\Delta\omega h_{I}^{\left(1\right)}
\\\hline \end{array}\right]}}
\]
such that $\partial_{\omega_{1}}^{2}\left(L\left(\tilde{B}-B\right)\right)=-S$.
Therefore, since $\mathscr{L}\tilde{B}=\mathscr{L}B+\mathscr{L}\left(\tilde{B}-B\right)$,
we deduce that $\tilde{B}$ verifies the discretised equation:
\begin{equation}
\mathscr{L}\tilde{B}=A+A^{T}+S+S^{T}\label{eq:Lyapunov-over-C}
\end{equation}
where $A$ is the discretization of $\mathcal{K}^{*}\left(\omega_{1}\right)\delta''\left(\omega_{1}-\omega_{2}\right)$,
the second member of (\ref{eq:Lyapunov-B-2CL-1}). Moreover, there
are additional source terms $S$ accounting for the change of variable.
This way, the equation verified by $\tilde{B}$ has homogeneous boundary
conditions. Let us write the linear operator $\partial_{\omega_{1}}^{2}\left(L\cdot\right)$
such that $\tilde{B}$ verifies equation (\ref{eq:Lyapunov-B-2CL-1})
with the boundary conditions being automatically verified. After some
easy algebra, one can obtain the following expression for $\mathfrak{L}\in\mathscr{M}_{I,I}\left(\mathbb{R}\right)$
the discretised version of $\partial_{\omega_{1}}^{2}\left(L\cdot\right)$:
\[
\mathfrak{L}=D_{I\times I-2}^{2}\cdot\tilde{L}
\]
where 
\[
D_{I\times I-2}^{2}=\frac{1}{\Delta\omega^{2}}\left[\begin{array}{cccc}
1\\
-2 & 1\\
1 & -2 & 1\\
 & \ddots & \ddots & \ddots\\
 & 1 & -2 & 1\\
 &  & 1 & -2\\
 &  &  & 1
\end{array}\right]\in\mathscr{M}_{I,I-2}\left(\mathbb{R}\right)
\]
and 
\begin{align*}
\tilde{L} & =\underset{\in\mathscr{M}_{I-2,I-2}\left(\mathbb{R}\right)}{\underbrace{\left[\begin{array}{cccc}
u_{2}\\
 & u_{3}\\
 &  & \ddots\\
 &  &  & u_{I-1}
\end{array}\right]}}\frac{1}{\Delta\omega^{2}}\underset{\in\mathscr{M}_{I-2,I}\left(\mathbb{R}\right)}{\underbrace{\left[\begin{array}{cccccc}
1 & -2 & 1 &  & (0)\\
 & 1 & -2 & 1\\
 &  & \ddots & \ddots & \ddots\\
 & (0) &  & 1 & -2 & 1
\end{array}\right]}}\\
 & +\underset{\in\mathscr{M}_{I-2,I-2}\left(\mathbb{R}\right)}{\underbrace{\left[\begin{array}{cccc}
v_{2}\\
 & v_{3}\\
 &  & \ddots\\
 &  &  & v_{I-1}
\end{array}\right]}}\frac{1}{2\Delta\omega}\underset{\in\mathscr{M}_{I-2,I}\left(\mathbb{R}\right)}{\underbrace{\left[\begin{array}{cccccc}
1 & 0 & -1 &  & (0)\\
 & 1 & 0 & -1\\
 &  & \ddots & \ddots & \ddots\\
 & (0) &  & 1 & 0 & -1
\end{array}\right]}}\\
 & +\underset{\in\mathscr{M}_{I-2,I}\left(\mathbb{R}\right)}{\underbrace{\left[\begin{array}{cccccc}
0 & w_{2}\\
 &  & w_{3}\\
 &  &  & \ddots\\
 &  &  &  & w_{I-1} & 0
\end{array}\right]}}.
\end{align*}
 One can easily check that this operator automatically conserves
the mass and the energy:
\begin{align*}
\left(1...1\right).\mathfrak{L} & =0\\
\left(\omega_{1},\ldots,\omega_{I}\right).\mathfrak{L} & =0
\end{align*}
Now we just have to correctly discretise the second member $A:=\mathcal{K}^{*}\left(\omega_{1}\right)\delta''\left(\omega_{1}-\omega_{2}\right)$.
We have $A=\mathcal{K}_{I-2\times I}^{*}\delta''_{I}\in\mathcal{M}_{I,I}\left(\mathbb{R}\right)$
where
\begin{align*}
\mathcal{K}_{I-2\times I}^{*} & =\left[\begin{array}{cccc}
0\\
\mathcal{K}^{*}\left(\omega_{2}\right) &  &  & (0)\\
 & \mathcal{K}^{*}\left(\omega_{3}\right)\\
 &  & \ddots\\
(0) &  &  & \mathcal{K}^{*}\left(\omega_{I-1}\right)\\
 &  &  & 0
\end{array}\right]\in\mathscr{M}_{I,I-2}\left(\mathbb{R}\right)\\
\delta''_{I} & =\frac{1}{\Delta\omega^{3}}\left[\begin{array}{cccccc}
1 & -2 & 1 &  & (0)\\
 & 1 & -2 & 1\\
 &  & \ddots & \ddots & \ddots\\
 & (0) &  & 1 & -2 & 1
\end{array}\right]\in\mathscr{M}_{I-2,I}\left(\mathbb{R}\right).
\end{align*}
 
\end{document}